\documentclass[12pt]{amsart} 
\usepackage{amsmath,amsthm,amsfonts,amscd,amsxtra,amsopn,verbatim,array}
\usepackage{eucal}
\usepackage{epsfig}
\usepackage{amsfonts}
\usepackage{amsmath}
\usepackage{amsthm}
\usepackage{pdfpages}

\usepackage{mathrsfs}
\usepackage[all]{xy}

\renewcommand{\mathfrak}{\mathcal}
\renewcommand{\mathcal}{\bold}

\newpage

\usepackage[numbers]{natbib}

\usepackage[colorlinks=false,linktocpage=true]{hyperref}

\newcommand{\beq}{\begin{equation}}
\newcommand{\eeq}{\end{equation}}
\newcommand{\beqarr}{\begin{eqnarray}}
\newcommand{\eeqarr}{\end{eqnarray}}
\newcommand{\beqa}{\begin{eqnarray*}}
\newcommand{\eeqa}{\end{eqnarray*}}

\begin{document}

\title[SEPAR -- Covid-19]{Studying the course of Covid-19 by a recursive delay approach}
\author[\small M. Kreck, E. Scholz]{\small  Matthias Kreck$^{\dag}$,  Erhard Scholz $^{\ddag}$}
\date{\today}

\begin{abstract} 
{\tiny In an earlier paper we proposed a recursive model for epidemics; in the present paper we generalize this model to include the asymptomatic  or unrecorded  symptomatic people,  which we call {\em  dark people} (dark sector). We call this the  SEPAR$_d$-model. A delay differential equation version of the model is added; it  allows a better comparison to other models. We carry this out by a comparison with the classical SIR model and indicate why we believe that the SEPAR$_d$ model may work better for Covid-19 than other approaches. 
 
 In the second part of the paper we explain how to deal with the data provided by the JHU, in particular we explain how to derive central model parameters from the data. Other  parameters, like the size of the dark sector, are less accessible and have to be estimated more roughly,  at best by  results of representative serological  studies which  are accessible, however, only for a few countries. 
We  start our country studies with  Switzerland where such data are available. 
Then  we apply the model to a collection of other countries, three European ones (Germany, France, Sweden), the three most stricken countries from three other continents (USA, Brazil, India). Finally we show that even the aggregated world data can be  well represented by our approach.
 
At the end of the paper we discuss the use of the model. Perhaps the most striking application is that it allows a quantitative analysis of the influence of the time until people are sent to quarantine or hospital. This suggests that imposing means to shorten this time is a powerful tool to flatten the curves. }
 \end{abstract}

\maketitle
\renewcommand{\thefootnote}{\fnsymbol{footnote}}

\footnotetext[2]{Mathematisches Institut der Universit\"at Bonn and Mathematisches Institut der Universit\"at Frankfurt, Germany, \quad kreck@math.uni-bonn.de}
\footnotetext[3]{University of Wuppertal, Faculty of  Math./Natural Sciences, and Interdisciplinary Centre for History and Philosophy of Science, \quad  scholz@math.uni-wuppertal.de}
\renewcommand{\thefootnote}{\arabic{footnote}}

\setcounter{tocdepth}{1}
\tableofcontents

\section*{Introduction} There is a flood of papers using the standard S(E)IR models for describing the outspread of Covid-19 and for forecasts.  Part of them is  discussed in \citep{Schaback:2020}. We propose alternative delay models and explain the differences. 

 In \cite{K-S} we have proposed a discrete delay model for an epidemic which we call SEPAR-model (in our paper we called it $SEPIR$ model). In this paper we explained  why and under which conditions the model is adequate for an epidemic. In the present note we add two new compartments reflecting asymptomatic or symptomatic, but not counted,  infected which we call the {\em dark sector}. We call this model the {\em generalized SEPAR-model}, abbreviated {\em SEPAR$_d$}, where $d$ stands for dark. This is our main new contribution. We will discuss the role of the dark sector in a theoretical comparison of the SEPAR$_d$-model with the SEPAR model. We will see, that -- as expected -- as long as the number  of susceptibles  is nearly constant, the difference of the two models is small, but in the long run it matters. 
  
 A second topic in this paper is a comparison with the standard SIR model. This comparison has two aspects, a purely theoretical one by comparing  the different fundaments on which the models are based, and a numerical one. For comparing two models it is helpful to derive them from similar inputs. For this we pass from the discrete model leading to difference equations to a  continuous model,  replacing difference equations by differential equations.  These differential equations fit into the general approach developed by Kermack/McKendrick in \cite{Kermack/McKendrick:1927}, as we learnt from O.  Diekmann. The analytic model resulting from our discrete model has been introduced independently by J. Mohring and coauthors  \citep{Mohring_ea:2020}   and, more recently, by B. Shayit  and M. Sharma  \citep{Shayit/Sharma:2020}. Also  F. Balabdaoui and D. Mohr work with a discrete delay approach with additional compartments and a stratification into different age layers adapted to the Swiss context    \citep{Balabdaoui/Mohr:2020}.
  Recently y R. Fe\ss ler has written a paper \citep{Fessler:2020}  in which  different differential equation models are discussed and compared,  including the classical S(E)IR-model and the analytic version of our model. 
Some hints  to earlier papers on the   analytic delay approach can be found there. 
  
In the second part of the present paper we apply the SEPAR model to selected countries and to the aggregated data of the world. To do so we first lay open how to pass from the data provided by the {\em Humdata} project of the JHU to the model parameters. The data themselves are obviously not reflecting the actual outspread correctly, which is most visible by the lower numbers of reported cases during weekends. But in addition there are aspects of the data which need to be corrected like for example a delay of reporting of recovered cases. All this is discussed carefully. Reliable data about the size of the dark sector are only available in certain countries where such studies were carried out. We found such studies for Germany and Switzerland, for other countries we estimate these numbers as good as we can. The case of Switzerland is particular interesting since  the effect of the dark sector which started to play a non-negligible role for the overall dynamics of the epidemic in the later part of 2020 for  the majority of the countries discussed  here (India, USA, Brazil, France)  can be studied there particularly well. For that reason we begin with this country and discuss the role of the dark sector in detail. 

 The paper closes with a discussion about what one can learn from the applications of the SEPAR model. We address three topics: The role of the constancy intervals, the role of the dark sector and the the influence of the time between infection and quarantine. The latter is perhaps the most striking application of our model offering a door for flattening the curves by sending people faster into quarantine, a restriction which imposes much less harm to the society than other means.  \\[3em]

\begin{center} {\bf PART I: Theoretical framework}
\end {center}


 \protect{\addcontentsline{toc}{chapter}{\protect \noindent Part I: Theoretical framework}  }

 \section{The SEPAR model and its comparison with other models \label{section model and comparison}}
 \subsection{The  SEPAR$_d$ model\label{subsection SEPAR}}
We begin by pointing out that we have changed our notation from \cite{K-S}. The  compartment  consisting of those who are isolated after sent to quarantine or hospital, which there was called $I$, is now being denoted by $A$ like {\em actually} infected,  in some places also described -- although a bit misleading -- as ``active'' cases (e.g. in Worldometer). This is why we speak now of the  SEPAR model  rather than of SEPIR.

Let us first recall the compartments introduced in \cite{K-S}. We observe 5 compartments which we call $S$, $E$, $P$, $A$, $R$, which people pass through in this order:  {\em Susceptibles} in compartment $S$ moving after infection to compartment $E$, where they are {\em exposed} but not infectious, after they are infected by people from compartment $P$ which comprises the   actively infectious people, those which {\em  propagate} the virus.  After $e$ days they move from $E$  to the compartment $P$, where they stay for $p$ days. After diagnosis they are sent into quarantine or hospital   and become members of the compartment   $A$,  where they no longer contribute to the spread of the virus although they are  then often counted as the actual cases  of the statistics. In order not to overload the model with too many details, we pass over the recording delay between diagnosis and the day of being recorded in the statistics. After another $q$ days the recorded infected move from compartment $A$ to the compartment $R$  of  {\em removed} (recovered or dead). 

We add two more compartments reflecting the role of the dark sector. There are two types of infected people, those who will at some moment be tested and  counted, and those who are never  tested, which we call people in the dark. This suggests to decompose compartment $P$ into two disjoint sub-compartments: $P_c$ of people who after $p_c$ days will  be tested and counted and move to compartment $A$,  and the collection  $P_d$ of people who after a longer period of $p_d$ days get immune and so move into a new compartment $R_d$ of removed people in the dark. To distinguish these removed people in the dark from those who come from compartment $P$ after recovery or death we introduce another new compartment $R_c$ of those removed people who occur in the statistics. Of course $R = R_c \cup R_d$.

The introduction of the dark sector in addition to the sector of counted people leads to the picture that  for the infected persons leaving compartment $E$ there is a branching process:  a certain fraction $\alpha(k)$ of people from $E$ moves to compartment $P_c$  at day $k$, whereas the fraction $ 1 - \alpha(k)$ of people moves to compartment $P_d$.

The existence of these compartments is a fundamental assumption which distinguishes the SEPAR$_d$ model from many other models including standard SIR.  The existence of these compartments is closely related to our picture of an epidemic like Covid-19. Of course this is a simplification. If one assumes that the passage from compartment $S$ to compartment $E$ takes place at a certain moment, the duration of the stay  in the next compartments varies from case to case. But it looks natural  to take the average of these durations leading for the different lengths $e$, $p_c$, $p_d$, and  $q$. All these have to be estimated from available information. 

Once one has agreed to this there is another fundamental assumption. This concerns the dynamics of the epidemic. Each person in compartment $P$ has a certain average number $\kappa (k)$ of contacts at day $k$. Depending on the strength of the infectious power of an individual the contacts will lead to newly exposed people. It is natural  to model  this development of the strength of infectiousness by a function $A(\tau)$, which measures the strength $\tau$ days after entry into compartment $P$. Again we simplify this very much, by replacing $A(\tau)$ by a constant $\gamma$, the average value of this assumed function. We will discuss this assumption later on in the light of information available for Covid-19. Given the parameters $\kappa (t)$ and $\gamma$ our next assumption is that, if we ignore the dark sector and set $\eta (k): = \gamma \kappa(k)$,  the dynamics of the infection can be described  by the following formula:

\beq
E_{new} (k) = \eta (k-1)  \frac{S(k-1)}{N}P(k-1).  \label{eq E-new}
\eeq

Here $E_{new}(k)$ is the number of additional members of compartment $E$ at day $k$ infected at day $(k-1)$ by people from compartment $P$ and $N$ the total number of the population. This is a very plausible formula. We call $\eta(k) $ the  {\em daily strength of infection}.  It is an integrated expression for the averaged contact behaviour of the population and the aggressiveness of the virus. 

This is the dynamics if we ignore the dark sector. But members of compartment $P_d$ also infect. We assume that the contacts are equal to those in compartment $P_c$. But  the average of the strength of infection of people from $P_d$ may be smaller than for those in compartment $P_c$, since in general they can be expected to stay longer in their compartment until they are immune and the strength of infection goes further down. Thus we introduce a separate measure $\gamma_c$ for those in compartment $P_c$ and $\gamma_d = \xi \gamma_c$, with $0\leq \xi \leq 1$,  for those in compartment $P_d$. 

Using this  the equation (1) has to be replaced by:
\beq
E_{new} (k) =  s(k-1)\eta (k-1)  [P_c(k-1) + \xi  P_d(k-1)]\label{eq E-newdark}.
\eeq 

Given this infection equation the rest of the model just  describes the time shifting passage of infected from one  compartment into the next  and counts  their cardinality at day $k$.  As usual we denote the latter by $S(k)$, $E(k)$, $P_c(k)$, $P_d(k)$, $A(k)$, $R_c(k)$ and $R_d(k)$. How such a translation is justified is  explained in \cite{K-S}. So we can just write down the self explaining formulas here: \\

\noindent
{\bf Introduction (definition) of the discrete SEPAR$_{d}$  model:} 
{\em  Let $e$, $p_c$, $p_d$, $q$ be integers standing for the duration of staying in the corresponding compartments, $0 \leq \alpha (k)  \leq 1$ be  branching ratios at day $k$ between later registered infected and those which are never counted,   $\eta (k)$  be  positive real numbers describing the daily strength of infection for $k\geq 0$, while $\eta(k)= 0 $ for $k<0$, and  $ \xi \le 1 $  a non-negative real number. Using $s(k)=\frac{S(k)}{N}$   the quantities $S(k),\, E(k)$, $E_{new}(k), \, P_c(k),\,P_d(k) , \, A(k),\, R_c(k),\,R_d(k) $ of the SEPAR$_{d}$ model  are given by  
\begin{itemize}
\item[a)] the start condition:\\
since the model is recursive we need an input for the first $e+p_d$ days (which we shift to negative values of $k$), i.e. start data 
 $E_{start}(k)$ for  \\$1-(e+p_d) \leq k \leq 0$, while $E_{start}(k)=0$ for all  $k>0$, \\ $P_c(k)=P_d(k)=A(k) = R_c(k)= R_d(k)=0$ (or some other well defined start values, cf. sec. \ref{subsection adaptation to the data})  for $ k < 1-(e+p_d)  $;  \\
 \item[b)]  the recursion scheme for $k \geq 1-(e+p_d)$:\\[-1em]
\beqarr 
E_{new}(k) &=&    s(k-1)\eta (k-1)  [P_c(k-1) +\xi  P_d(k-1)] + E_{start}(k)\nonumber \\
E(k) &=& E(k-1) + E_{new}(k) - E_{new}(k-e) \nonumber \\
P_c(k) &=& 
P_c(k-1) + \alpha(k) \,  E_{new}(k-e) - \alpha(k-p_c) \, E_{new}(k-e-p_c) \nonumber \\
P_d(k) &=& 
P_d(k-1) +(1- \alpha(k) )\,E_{new}(k-e) - (1- \alpha(k-p_d) )\, E_{new}(k-e-p_d) \nonumber \\
A(k) &=& A(k-1) + \alpha(k-p_c)\, E_{new}(k-e-p_c) - \alpha(k-p_c-q)\,E_{new}(k-e-p_c-q)] \nonumber \\
R_c(k) &=& R_c(k-1) + \alpha(k-p_c-q) \, E_{new}(k-e-p_c-q) \nonumber \\
R_d(k) &=& R_d(k-1) + (1-\alpha(k-p_d))\, E_{new}(k-e-p_d)\nonumber \\
S(k) &=& N - E(k) - P(k) - A(k) - R(k)  \nonumber \\
& & \hspace*{-4.5em} \mbox{with}  \quad  P(k)=P_c(k)+P_d(k) , \; R(k)=R_c(k)+R_d(k) \nonumber
\eeqarr
 \end{itemize}
}

The reason for  this definition  is easy to see. Additional people  in, for example, compartment $P_c$  at the day $k$ are $(P_c)_{new}(k) = \alpha(k) E_{new}  (k-e)$, while $\alpha(k) E_{new}(k-(e+p))$ move to the next compartment. Similar formulas hold for the compartments $P_d$, $A$, $R_c$ and $R_d$. 

An important parameter in an epidemic is the {\em reproduction number $\rho$}, the number of people infected by a single infectious person during its life time. If we assume that $\kappa (k) $  and $s(k)$ may be considered as  constant during $p_d$ days about $k$,  we can derive this number from the equations. It is 
\[\rho (k)  =(1+ \delta)^{-1}  \kappa(k)\,  s(k)  \big(  \gamma_c p_c + \delta \gamma _d  p_d \big)    \,, 
\]
where $\delta = \frac{1-\alpha}{\alpha}$. 
For $s(k)=1$   it is usually called the {\em basic} reproduction number,  in order to  distinguish it from the {\em effective} reproduction number with  $s(k) < 1$. In part I of this paper we usually mean the basic reproduction number when we speak of reproduction number, while in the part II the decreasing $s(k)$ hast to be taken into account and we usually speak of the effective reproduction number, also without use of the attribute ``effective''. 

If we set $\alpha(k) = 1$, $P_d = 0$ and $R_d =0$,  we obtain the  SEPAR model without dark sector as a special case of the SEPAR$_{d}$ model. Effects of vaccination can easily be implemented by sending the according number of persons directly from $S$ to $R$.

For later use the following observation is useful. The number of people in a given compartment at day $k$ is the sum of additional entries at previous days, for example
$$
E(k) = \sum_{j=0}^{e-1} E_{new}(k-j), 
$$
$$
P_c(k) = \alpha(k) \sum_{j=0}^{p_c-1} E_{new}(k-e-j), 
$$
 and so on for $P_d(k)$ and $A(k)$. 

We abbreviate 
\beq
H(k) =  E(k) + P(k) + A(k) + R(k) \, , \label{eq H(k)}
\eeq 
the number of {\em herd immunized} (without vaccination). Then the recursion scheme implies: 
$$
H(k) - H(k-1) = E_{new}(k) 
$$
Putting this into the  formula above: $
E(k) = \sum_{j=0}^{e-1} E_{new}(k-j) 
$, we obtain:
$$
E(k) = H(k) - H(k-e)
$$ 
and similarly
\beqarr
P_c(k) = \alpha(k)\,H(k-e) - \alpha(k-p_c)\,H(k-e-p_c)),  \label{eq P in terms of H} \\
P_d(k) = (1 - \alpha(k))\, H(k-e) - (1- \alpha(k-p_d))\,H(k-e-p_d)), \nonumber
\eeqarr
$$
A(k) = \alpha(k-p_c)\, (H(k-e-p_c) - \alpha(k-p_c-q)\, H(k-e-p_c-q)).
$$
Using $R_c(k) - R_c(k-1) = \alpha(k-p_c-q) E_{new} (k-e-p_c-q)= \alpha(k-p_c-q) H(k-e-p_c-q) - \alpha(k-p_c-q-1)\,H(k-e-p_c-q-1)$ we conclude:
$$
R_c(k) = \alpha(k-p_c-q) H(k-e-p_c-q) 
$$
and similarly
$$
R_d(k) =(1-\alpha(k-p_d)) H(k-e-p_d).
$$
This gives a very simple structure of the model in terms of a single recursion equation. \\

\noindent
{\bf 
 SEPAR$_d$-model:} {\em The recursion scheme of the SEPAR$_d$ model is given by a single recursion equation:
\beqarr
& & H(k)- H(k-1) =   \label{eq recursion for H(k)}\\
& &  s(k-1)\eta (k-1) \Big( \alpha(k-1)\, H(k-1-e) - \alpha(k-1-p_c)\,H(k-1-e-p_c)       \nonumber \\
&+& 
\xi \, \big[ (1 - \alpha(k-1))\, (H(k-1-e) -(1- \alpha(k-1-p_d))\,H(k-1-e-p_d)) \big]
\Big) \nonumber   
\eeqarr
 and
the functions $S(k)$,  $E(k)$, $P_c(k) $, $P_d(k) $, $A(k) $, $R_c(k)$ and $R_d(k)$ are given in terms of $H(k)$  by the equations above. 

If we pass from a daily recursion to a infinitesimal recursion,   replacing  the difference equation by a differential equation, we obtain the continuous recursion scheme, where now all functions are differentiable functions of the time $t$: 

\beq
\begin{split}
H'(t) = s(t)\eta (t) \Big( \alpha(t)\,H(t-e) - \alpha(t-p_c)\,H(t-e-p_c)) \\+ 
\xi \,\big[  (1 - \alpha(t))\, H(t-e) -(1- \alpha(t-p_d))\,H(t-e-p_d) \big]
\Big) \label{eq delay ODE for H(k)}
\end{split}
\eeq
}
\noindent
In both cases, discrete and continuous,  consistent  start conditions in an interval of length $e+p$ have to be added. For the discrete case see sec. \ref{subsection adaptation to the data}.
If we remove the dark sector, the continuous model was independently obtained in  \citep{Mohring_ea:2020}.

The branching ration $\alpha$ and with it the number  $\delta \approx \frac{1-\alpha}{\alpha}$ of unrecorded infected for each newly recorded one varies  drastically in space and time, roughly in the range  $1 \leq \delta \leq 50$.  For Switzerland and Germany  serological studies in late 2020 conclude $\delta \approx 2$, for the USA a recent study finds $\delta \approx 8$ and in part of India (Punjab)  a serological study found values indicating $\delta \approx 50$.\footnote{For Germany see \citep{Helmholtz-Zentrum-Muenchen:2020-12,RKI:2020SerologischeStudie}, for  Switzerland  \citep{Kuster:2021} announcing  a forthcoming study of {\em Corona-Immunitas}, for USA \citep{Reese_ea:2020USA} and for India  a report in ANI  {\tiny  \url{https://www.aninews.in/news/national/general-news/second-sero-survey-finds-2419-pc-of-punjab-population-infected-by-covid-1920201211181032/}} retrieved 12/21 2020. }
For our choice of the model parameter see below, section \ref{section countries}.

\begin{figure}[h]
\includegraphics[scale=0.7]{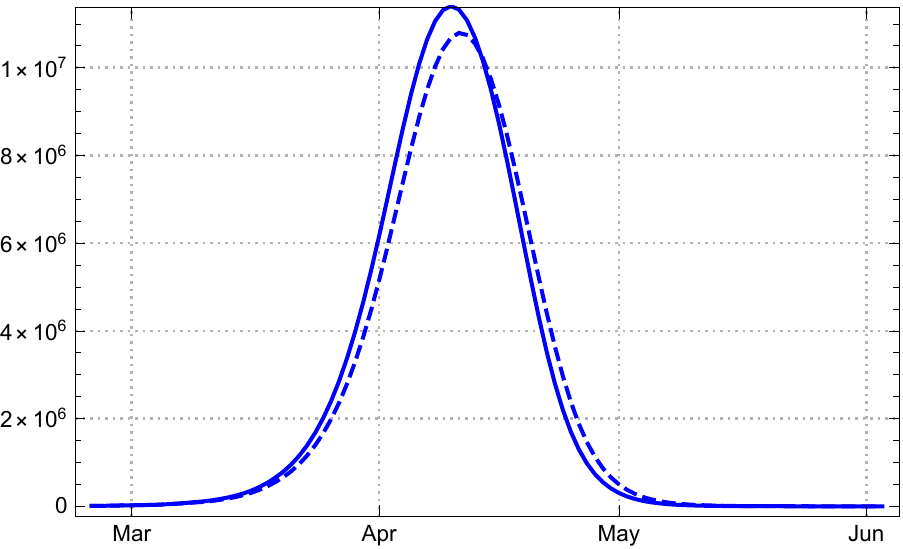}
\caption{Comparison of SEPAR$_d$ model for $A(k)$ between  dark sector with $p_c=7,\, p_d=10, \, \xi =0.9$ (solid blue) and simplified dark sector $p_c= p_d =p=7, \, \xi=1$ (dashed blue), assuming  constant  $\eta$.  \label{fig p-c and p-d} }
\end{figure}

 Besides the determination of $\alpha$ one needs to know the difference between $p_c$ and $p_d$ and between  $\eta_c (k)$ and $ \eta _d(k)$, if one wants to apply the $SEPAR_d$-model. As explained above we estimate $p_c = 7$.
 The mean time of active infectivity of people who are not quarantined seems to be not much longer, although in some cases it is.  According to  the study \cite[p. 466]{Woelfel_ea:2020} ``no isolates were obtained from samples taken after day 8 (after occurrence of symptoms) in spite of ongoing high viral loads''. 
 This allows to work with an estimate $p_d = 10$, and so it is not much larger than $p_c$. A comparison of the $SEPAR_d$ model  with a simplified version, where  we  assume $p_c = p_d=: p$ and  $\eta_c (k) = \eta _d(k) = :\eta(k)$ shows that with these values  the difference is very small (see figure \ref{fig p-c and p-d}).  In the following we therefore work  with the simplified $SEPAR_:d$ model setting   $p_c = p_d=: p$.
 
  Recent studies indicate that the  number of asymptomatic infected is often as low as about 1 in 5 symptomatic unrecorded and is thus  much smaller than originally expected \citep{Nogrady:Asymptomatic}. Although asymptomatic infected are there reported to be considerably less infective than the symptomatic ones, their relatively small number among all unreported cases justifies to  work in the  simplified dark model with the assumption  $ \eta _d(k) \approx \eta_c (k)  = :\eta(k)$.  If we set $P(k) = P_c(k) + P_d(k)$  as above, we see that $P_c(k) = \alpha(k) P(k)$ and $P_d(k) = (1-\alpha(k)) P(k)$. 
 
In part II we discuss how the time dependent   parameter $\eta(k)$ can be derived from the data and a rough estimate of the dark factor $\delta$ can be arrived at, although it lies in the nature of the dark sector that information is difficult to obtain. 

\begin{figure}[h]
\includegraphics[scale=0.7]{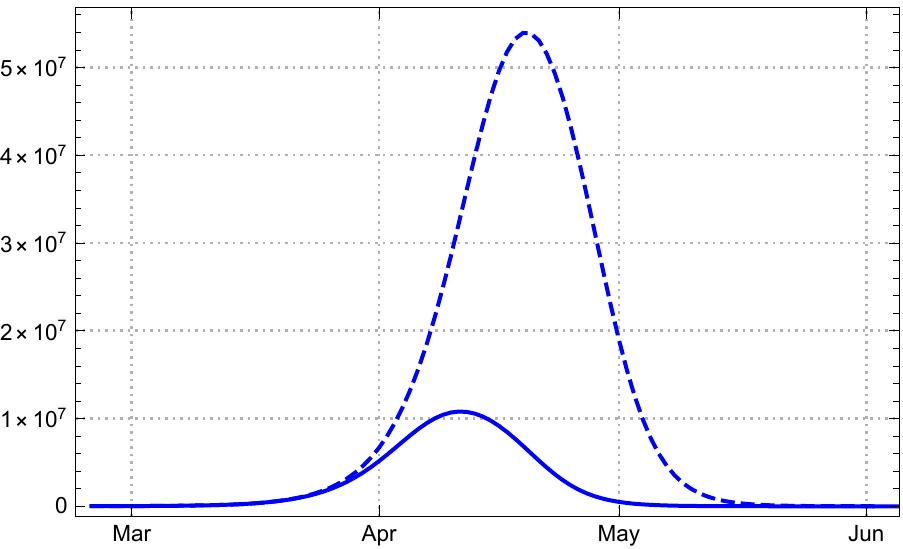}\includegraphics[scale=0.7]{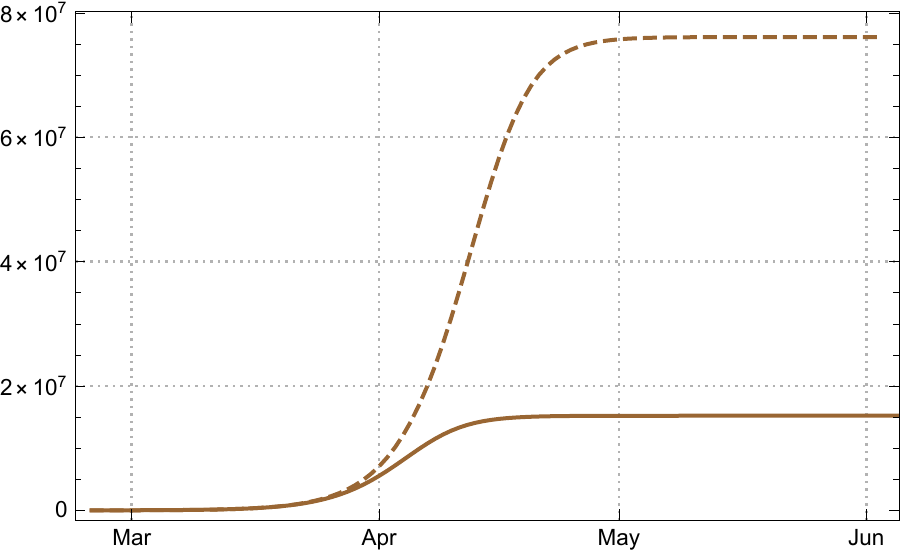}\\[0.3em]
\caption{Comparison of the course of an epidemic with constant reproduction number $\rho=3$  without dark sector (dashed), and with dark sector $\delta=4$ (solid lines): Left counted number of actual infected $A_c(k)$ (blue). Right: total number of confirmed infected $A_{tot \, c}(k)=A_c(k)+R_c(k)$ (brown).   \label{fig dark--no dark}}
\end{figure}

A comparison of the SEPAR model ($\delta=0$)  with the simplified SEPAR$_d$ model is given in  fig. \ref{fig dark--no dark}  for a constant parameter $\alpha = 0.2$, respectively dark factor $\delta=4$ and constant reproduction coefficient $\rho=3$. This illustrates the influence of the dark factor from the theoretical viewpoint. 

\subsection{The S(E)IR models and their assumptions\label{subsection SEIR}}

Whereas the derivation of the SEPAR$_d$-model is based on the idea of disjoint compartments, infected people pass through in time, there is a different approach with goes back to the seminal paper 
 \cite{Kermack/McKendrick:1927}. A special case is the standard SIR-model or SEIR model. It seems that most people use this as a black box without observing the assumptions on which it is built.  One should keep these assumptions in mind  whenever one applies a model. There is a modern and easy to understand paper by Breda, Diekmann, and de Graaf  with the title: {\em On the formulation of epidemic models (an appraisal of Kermack and McKendrick)} \cite{Breda/Diekmann:2012}, which explains the general derivation. In the introduction the authors state that the Kermack/McKendrick paper was cited innumerable times and  continue: "But how often is it actually read? Judging from an incessant misconception of its content one is inclined to conclude: hardly ever! If one observes the principles from which the S(E)IR models are derived one should be hesitant to apply it to Covid-19. 

Following \cite{Breda/Diekmann:2012} we shortly repeat the assumptions on which the general Kermack/McKendrick approach is based . The  general model considers a function 
$$
S(t) := \text {density (number per unit area) of susceptibles at time t}
$$
and related to this a function 
$$
F(t)
$$ 
called 
the force of infection at time t.
By definition, the force of infection is the probability per unit of time that a susceptible becomes infected. So, if numbers are large enough to warrant a deterministic description, we have
$$
I_{new}(t)  = F(t)S(t),
$$
where $I_{new}(t)$ is defined as the number of new cases per unit of time and area. The functions $S$ and $I$ are related by the equation:
$$
S'(t) = -F(t) S(t)
$$

Then the central modelling ingredient is introduced:
\[\begin{split} A(\tau  ) := \text{expected contribution  to the force of infection} \, \tau \, \text{units of time ago.}\end{split}
\]

Alone from this ingredient an integral differential equation is derived, which gives the model equations. For more details we also refer to a recent paper by Robert Fe\ss{}ler who derived the integral equation independently   \citep{Fessler:2020}. 

Already here we see a different  view of an epidemic. No compartments and their cardinality are mentioned; in their place the authors mention only certain functions. 

If the function $A$ is assumed to decay exponentially,
 $$
A(\tau) = \alpha e^{-\beta \tau}
$$
with constants $\alpha, \, \beta$.
The model derived from this input is  called the {\em standard SIR-model}. It leads  to two ordinary differential equations in the variable $t$:
$$
 I' = \alpha s   I - \beta I
 $$
 $$
 R' = \beta I
 $$
 $$S(t)  = N - I(t) - R(t),$$ 
 where $N$ as before is the number of the population and $s(t)=\frac{S(t)}{N}$. 
 
For the {\em standard SEIR-model}  there is an additional function $E(t)$ measuring the 
 {\em exposed}  and the input function is now 
 $$
A(\tau) = \alpha  \frac {\beta} {\beta - \gamma} (e^{- \gamma \tau } - e^{-\beta  \tau} )
 $$
 This leads to 3 ordinary differential equations in $t$
 $$
 E' = \alpha s I  - \beta E
 $$
$$
 I' = \beta E - \gamma I
 $$
 $$
 R' = \gamma I
 $$
 $$S(t)  = N - E(t) - I(t) - R(t),$$ 
 where $N$ as before is the number of the population.  The  infection function $A(\tau)$ considered here determines the convolution part of  an integral kernel in Fe\ss ler's approach mentioned above.
  
If Breda et al. are right, readers should be critical to papers applying the S(E)IR models without explaining  why the models, given their fundaments, are applicable. As far as we can see, the assumption of an exponential decay $A(t)$ is often not mentioned by authors applying it in situations where it would be necessary to discuss whether this assumption can be reasonably made. In a situation like Covid-19 where infectious people are  isolated  as soon as possible, it seems questionable whether this assumption holds. We are  surprised that in most of  the papers we have seen, which apply the S(E)IR model to an analysis of  Covid 19,   this problem is not even mentioned. This includes the papers of the group around Viola Priesemann which play an important part  in the discussion about how to deal with Covid 19 in Germany \cite{Dehning-ea:2020}, 
\citep{Contreras-ea:2020challenges}. 
 \subsection{Comparing  SIR with  SEPAR\label{subsection model comparison} }
 
When we want to compare the SIR models with the delay SEPAR model we have  to lower, in a first step, the number of compartments by removing $E$,  $P$ and $A$ and  to replace them by a single compartment, called  $I$, of infected people which are at the same time infectious.  For this model we assume  that infected susceptibles move right away to compartment $I$, where they stay for $p$ days. In contrast to the SEPAR model it is assumed that  these people are counted as actual infected people at the moment they are infected. After $p$ days they are counted as recovered or dead. So it is a strong simplification of the SEPAR model, but it follows the same pattern as the SEPAR model since it is a  delay model. We call it {\em d-SIR-model} (``d-''for delay) to distinguish it from the standard SIR-model. The equations for this model are based on the same principles as the SEPAR model: \\

\noindent
{\bf The continuous delay  d-SIR model:} {\em  Let $p$ be  a positive real number standing for the duration of staying in the compartment $I$ of infected and infectious people.  Let $\eta(t)$ be a differentiable function measuring the strength of infection (including the effects of social constraints).   The quantities $S(t),\, I(t),\, R(t) $ of the delay  SIR model are given by 
\begin{itemize}
\item[a)] the start condition:\\
 A differentiable function $  I(t)$ for $0\le t< p$ ,
 \item[b)]  and the delay differential equations:
\beqa 
I'(t) & = &\eta (t)   s(t)\, I(t)
- \eta (t-p) s(t-p)\, I(t-p) \\
R'(t) &=& \eta (t-p)s(t-p)\, I(t-p)  \\
S(t)  &=& N - I(t) -R(t)
\eeqa
\end{itemize}}

To compare this model with the standard SIR-model above we note that also the d-SIR model (like the continuous SEPAR$_d$ model) can be derived from the principles of Kermack/McKendrick, as explained in \citep{Fessler:2020}. 
One only has to take  the product of the characteristic function of the interval $[0,p]$ with $\eta  = \gamma \kappa$ as 
 the function $A(\tau)$. To compare the two models one has to relate the input parameters.   In the case of the d-SIR model they  are $\eta $ (for the comparison we assume that the contact rate is constant) and $p$, whereas for the SIR-model they are $\alpha$ and $\beta$. The role of $\eta$ is that of $\alpha$ in the SIR-model, so we set $\alpha = \eta$. There are several ways to relate the paramter $\beta$ of the SIR model with $p$ occurring in the d-SIR model. One is to assume that the total force of infection has to be the same if they describe the same developments, i.e. with the function $A(\tau)$ which is the product of the characteristic function of the interval $[0,p]$ with $\eta$ one has the condition:
$$
\int _0^ \infty \alpha e^{-\beta t} dt =  \int _0^ \infty A(t) dt
$$
Then the second relation:
$
\frac \alpha \beta = p\, \eta =  p \, \alpha$ and thus 
$$
\beta = \frac 1 p. 
$$
In both cases the reproduction number is $\frac \alpha \beta = p\, \alpha = p\, \eta$. 

If one applies this then there is a problem to find parameters so that at least at the beginning the two models are approximatively equal. Thus one can relate the tow models in a second way by choosing the parameters so that this is the case. For this we fix values for $\alpha $ and $\beta  $ and chose the start conditions of the d-SIR model so that they agree with the SIR-model during the first days. By construction of the SIR-model the function $I$ is nearly an exponential function as long as the function $S$ is nearly constant. Thus we chose the same exponential function as   start values for the d-SIR model.

Then the question is whether there are differences of the model curves in the long run and how large the differences are. 
 One should expect that the assumptions of  an exponential decay regulating the strength of infection of an infectious person in the case of the SIR-model versus a period of $p$ days, where the strength of infection is constant and after that goes immediately down to $0$ in the case of the delay d-SIR, should result in higher values for the functions $I(t)$ and $I_{tot}(t)= I(t) +R(t)$, the total number of infected until time $t$ of the  SIR model. The following graphics in which we assume a constant reproduction rate slightly above $1$ show, in fact, a dramatic difference supporting  the expectation. 
A similar observation can be found in \citep[fig. 5, 6]{Fessler:2020}

\begin{figure}[h]
\includegraphics[scale=0.7]{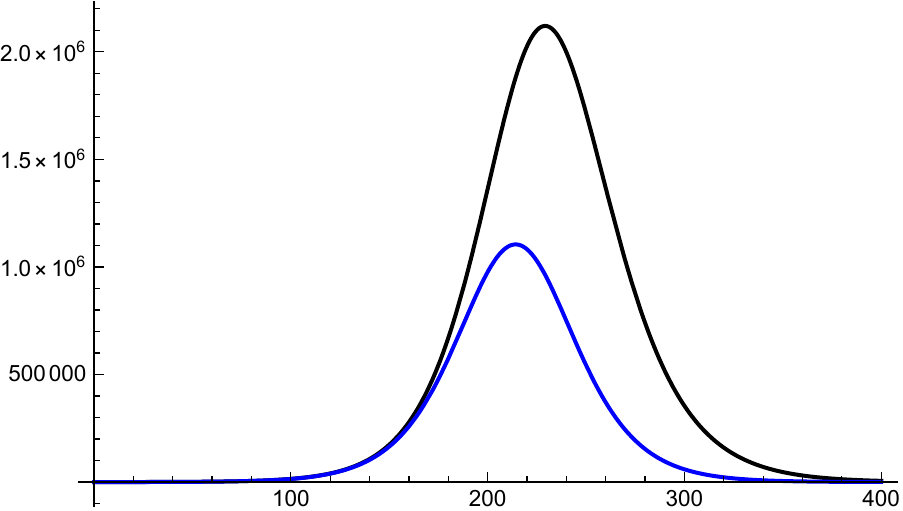}
\includegraphics[scale=0.7]{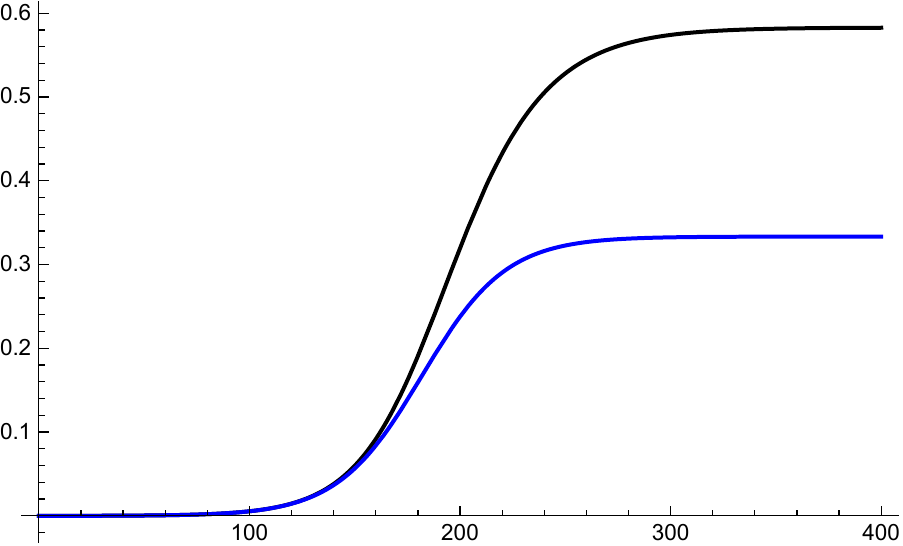}
\caption{Comparison of SIR (black) and dSIR (blue) for $s(t) \approx 1$ and constant coefficients  with  identical exponential increase and population size $N= 80$ M. Left: Number of infected $I(t)$. Right: $H(t)/N$ with $H(t)$ total number of  infected up to time $t$. Parameter values: $N=8$ M, $I_0=1 $ k,  $\alpha= \eta = 0.15, \, \beta= 0.1 $;  $p= 8.11$ for dSIR. \label{fig SIR -- dSIR}}
\end{figure}

This has an interesting consequence for a situation in which a high rate of immunity is achieved either by ``herd'' effects or by vaccination. According to a simple SIR  model with constant reproduction number $\rho=1.5$ and a population of  80 million people (like in Germany)  a little bit more than $0.6 \cdot 80 = 24$ million people would have to be infected or vaccinated to achieve herd immunity, whereas according to the delay model ``only'' about $0.3 \cdot 80=12$  million have to be infected (see fig. \ref{fig SIR -- dSIR}). 
The difference corresponds to the fact that equal initial exponential growth is related to different reproduction rates in the two models of  the example given: $\rho_{SIR}= \frac{\alpha}{\beta}=1.5$ and $\rho_{d-SIR}=p \, \eta  \approx 1.2$. Such a difference matters because, according to the plausibility arguments given above, the delay model may very well be more realistic than the SIR model for Covid-19.

Next we discuss  the differences between the SIR model and the delay SIR model during a time  when $s$ is still approximately equal to $1$,  both have constant reproduction numbers, and $\alpha = \eta$ like above. 
Moreover we assume  that the initial growths functions  of 
 both models are approximately identical to   the same exponential function (because both are designed to modelling the same growth process). 

 In reality one observes longer periods  in the data where the reproduction number is approximately constant until it changes in a short transition period to a new approximately 
 constant value. Such changes  may be due to containment measures (non-pharmaceutical interventions) imposed by governments, which influence the contact rate $\kappa (t)$. In the next graphics we show the effect of such a change for both models.
 \begin{figure}[h]
\includegraphics[scale=0.7]{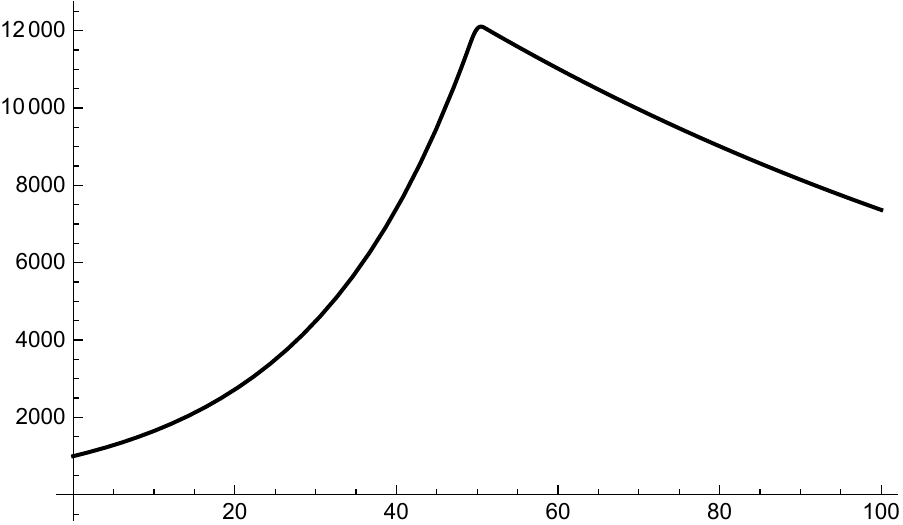}\includegraphics[scale=0.7]{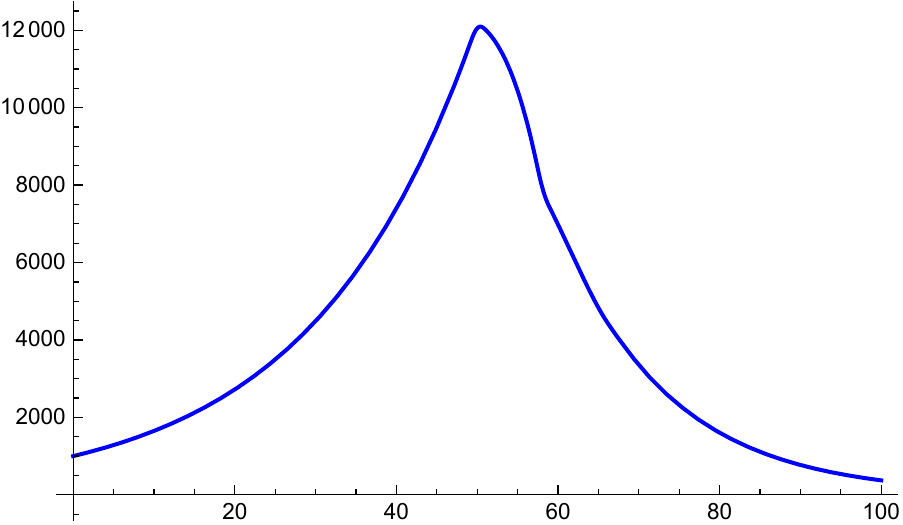}\\
\includegraphics[scale=0.7]{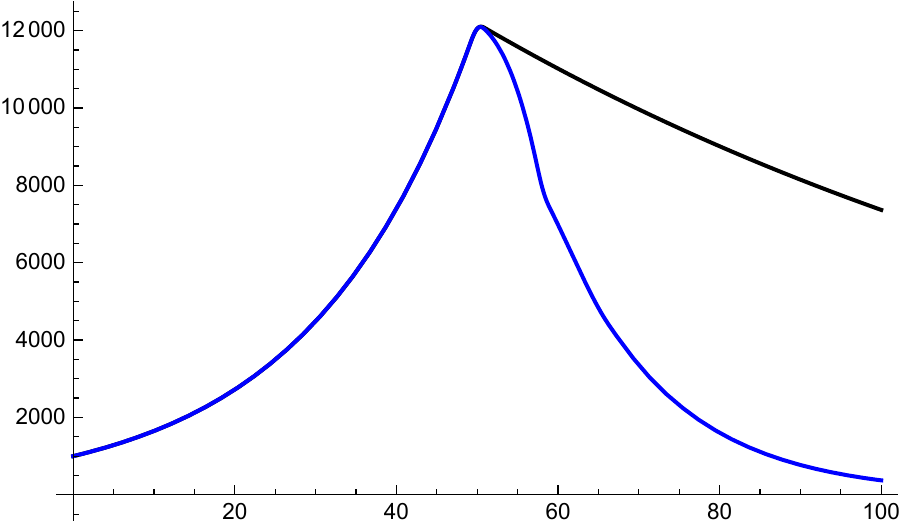}
\caption{Comparison of $I(t)$ for  SIR (top left) and dSIR (top right) for $s(t) \approx 1$ and constant coefficients for $t\leq 29$ and $t\geq 31$ with  identical exponential increase in the initial upswing. Reduction of reproduction rate by 40 \% in both cases. Bottom: SIR black, dSIR blue.  Parameter values: $N=8$ M, $I_0=1 $ k,  $\alpha_1= \eta_1 = 0.15, \, \beta_1 = \beta_2 =0.1, \; \alpha_2=\eta_2 =0.09, \, \beta_2 = \beta_1$; dSIR $p= 8.11$. \label{fig SIR--dSIR 2}}
\end{figure}
 
In figure \ref{fig SIR--dSIR 2}  we let the dSIR and the SIR curves start with identical exponential functions based on constant reproduction numbers. Then we lower the reproduction rate by 40 percent within three days. As expected the SIR curves have a cusp since one exponential function jumps into another, whereas the dSIR equation due to the delay character shows a slightly smoother transition. The second and more dramatic effect is that a similar phenomenon like in the long term comparison can be observed: the SIR solution is far above the dSIR solution. The reason seems to be  the same, the different assumptions made by the two approaches about the decay of the strength of infection. 
 
These considerations show that the choice of the model may result in important differences for the medium and long range development of the epidemic. We have given arguments why we consider the delay SIR model more realistic for Covid-19 than  the standard SIR approach. \\[0.5em]

But also the delay SIR model has defects when comparing it to the data. The reason is that in reality it is not the case that an infected person gets infectious the same day, and also it takes some time until an infectious person shows symptoms. This 
 speaks in favour of the delay SEPAR approach. 
 In part II we  apply the SEPAR$_d$ model to  data  of selected  countries. Here the model shows its high quality. Since data about the dark sector are  insecure we check how much the dark sector influences the overall dynamics of the epidemic in the discussed countries up to the  present  (until the end of 2020) and  choose  the dark factor of the model on the basis of the  analysis and given estimates for the respective countries. 
\\[3em]
 
 \begin{center} {\bf PART II: Applications of the  SEPAR model}
\end {center}
 \protect{\addcontentsline{toc}{chapter}{\protect \noindent Part II: Application}  }

\section{Determining empirical parameters for the model  \label{section data}} 
\subsection{JHU data\label{subsec JHU data}}
\subsubsection*{The basic data sets (JHU)} 
The worldwide  data 
 provided by the Humdata  project (Humanitarian Data Exchange) of the {\em Johns Hopkins University} provides   data on the development of the Covid-19 pandemic  for more than 200 countries and territories.\footnote{\url{https://data.humdata.org/dataset/novel-coronavirus-2019-ncov-cases}} The data are compressed into 3 basic data sets for each country/territory
 \[\mathit{Conf}(k), \; Rec(k), \; D(k) \, 
 \]
 where  $\mathit{Conf}(k)$ denotes the  total number of {\em confirmed cases}  until the  day $k$ (starting from January 22, 2020), $Rec (k)$ the number of reported {\em recovered} cases  and $D(k)$ the number of reported {\em deaths}  until the day $k$. 
  The last two entries can be combined to the number of {\em redrawn} persons of the  epidemic, captured  by the statistic,
\[ \hat{R}(k) = Rec(k)+ D(k)\, .
\] 
Empirical quantities derived from the JHU data set will be endowed with a hat, like $\hat{R}$, to distinguish  them from the corresponding model quantity, here $R$.

The (first) differences of $\mathit{Conf}(k)$ encode the daily numbers of {\em newly} reported and {\em acknowledged}   cases:
\beq \hat{A}_{new}(k) = \mathit{Conf}(k)- \mathit{Conf}(k-1) \,  \label{eq hat-A-new}
\eeq
The other way round, the  number of confirmed cases is the complete sum of newly reported ones, and may be considered as the total number of acknowledged cases
\beq \hat{A}_{tot}(k) = \sum_{j=1}^{k} \hat{A}_{new}(j) = \mathit{Conf}(k) \, , 
\label{eq hat-A-tot}
\eeq
while the difference 
\beq \hat{A}(k)= \mathit{Conf}(k)-\hat{R}(k) \, ,\label{eq hat-A}
\eeq  is  the empirical number  of  {\em acknowledged}, not yet redrawn,   {\em actual} cases. Some authors call it  the number of ``active cases'';\footnote{E.g. in \cite[p. 182]{Schaback:2020} \ldots A similar identification underlies the numbers for the active case in the {\em Worldometer} 
\url{https://www.worldometers.info/coronavirus/country/}.} 
but this  is misleading because
 the phase of  effective infectivity is usually over as soon as an infection  is diagnosed  and the person is quarantined. 

The number $Rec(k)$ of recovered people is often reported with much less care than the daily new cases and the deaths.
By this reason the recorded number of redrawn,  $\hat{R}(k)$, may be  heavily distorted, with the result   that neither itself  nor the derived  numbers $\hat{A}(k)$ can be  be taken at face value. The   most reliable {\em basic data}  remain therefore
\[ \hat{A}_{new}(k) \;, \quad  \mathit{Conf}(k) \qquad \mbox{and} \quad D(k) \; .
\]
Even $\hat{A}_{new}(k)$ has  its peculiarities due to the weekly cycle of reporting activities. 
In this paper we abstain from  discussing mortality rates and consider  the first two data sets of the mentioned three only.  $\hat{R}$ and $\hat{A}$ play an important  role for a complete image of an epidemic, but they are  reliable  only for a few countries;  for the majority of countries they have to be substituted or complemented by more adequate quantities derived from the basic data (see eq. \ref{eq hat-A_q}). 

\subsubsection*{Smoothing the weekly oscillations of $\hat{A}_{new}$}
For all countries  the reported number of daily new infections shows a characteristic 7-day oscillation  resulting from the reduction of tests  over weekends and the related  delay of transmission of data. A 3-day sliding average  suppresses  fluctuations on a  day-to-day scale and shows the weekly oscillations even more clearly.\\
\begin{figure}[h]
\includegraphics[scale=0.4]{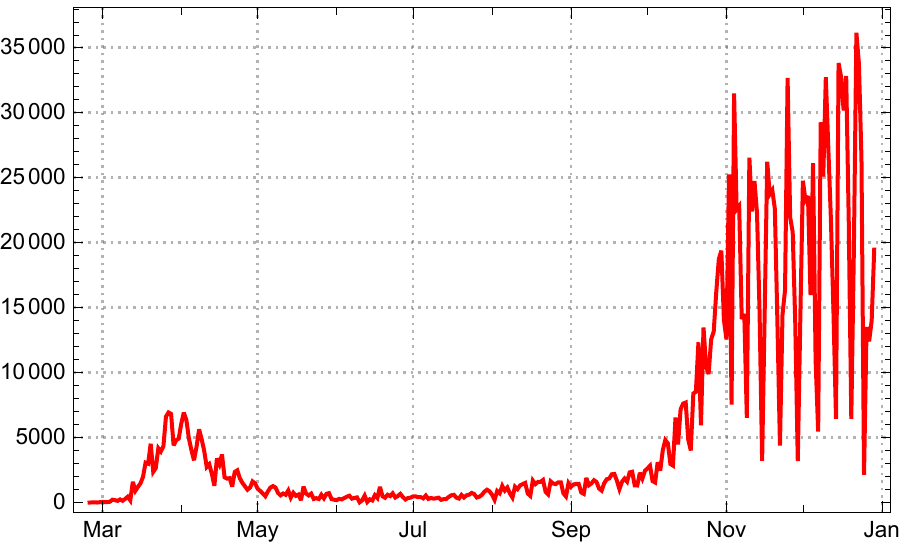} \hspace{1cm} \includegraphics[scale=0.4]{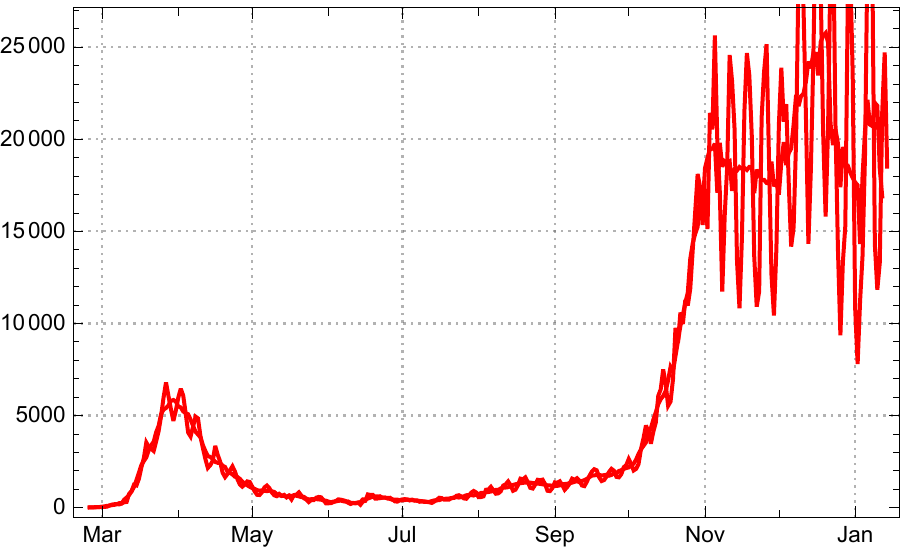}\hspace{1cm} \includegraphics[scale=0.4]{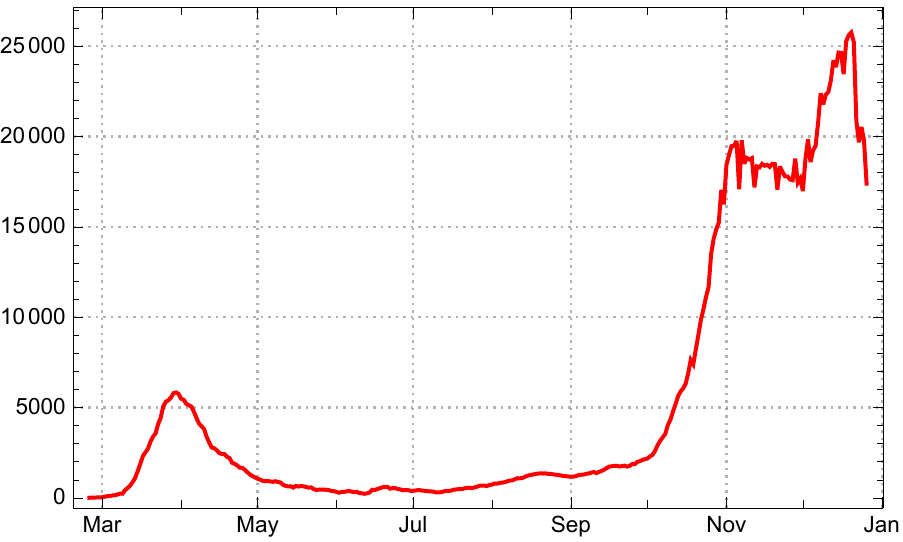}
\caption{\tiny Acknowledged cases $\hat{A}_{new}(k)$ for Germany (left),   sliding 3-day and 7-day averages $\hat{A}_{new,3}(k)$,  $\hat{A}_{new,3}(k)$ (middle)  and  $\hat{A}_{new,7}(k)$ (right)}
 \label{fig Inew D}
\end{figure}

In some countries these oscillations are  corrected for transmission delay by central institutions,\footnote{This is done by the {\em Robert Koch Institut} for the German case.}
but such  corrections are not implemented in the JHU data. A simple method for smoothing the weekly oscillations consists in using sliding {\em centred} 7-day averages:  
\beq \hat{A}_{new,7}(k) = \frac{1}{7} \sum_{j=-3}^3 \hat{A}_{new}(k+j) \, 
\eeq  
and similarly for the centred 3-day average $\hat{A}_{new,3}(k)$.
Note that in order to avoid a  time shift effect which would arise from using a purely backward sliding average, we  use a sliding average over 3 days forward and 3 days back.   For most countries this suffices for carving out the central tendency of the new infections quite clearly (fig. \ref{fig Inew D}).

For some countries already   the {\em daily} fluctuations  of $\hat{A}_{new}$ are extreme. 
The French data even indicate {\em negative} values for $\hat{A}_{new}$ for certain days, although this ought to be excluded by principle. Such effects indicate a highly unreliable system of data recording and transmission; they may be due to ex post  corrections of earlier exaggeration of transmitted numbers.
But even under such extreme conditions 
 the sliding 7-day average leads to reasonable information on the mean motion of the new infections, so that we don't have to exclude such countries from  further consideration (sec. \ref{subsection 4 European countries}).

\subsection{Data evaluation\label{subsec data evaluation}}
\subsubsection*{The  ``actual'' cases in the statistical sense}
\begin{figure}[h]
\includegraphics[scale=0.6]{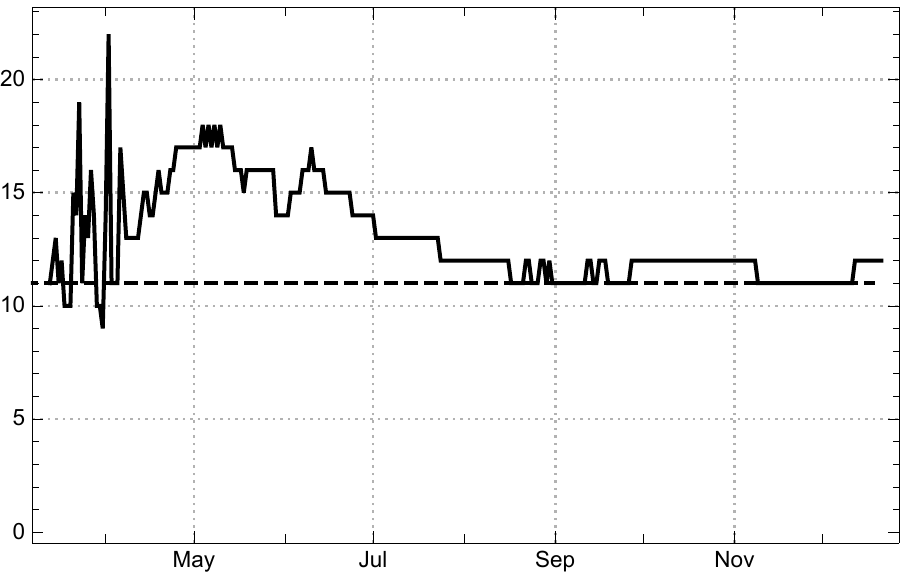}
\hspace{1cm}
\includegraphics[scale=0.6]{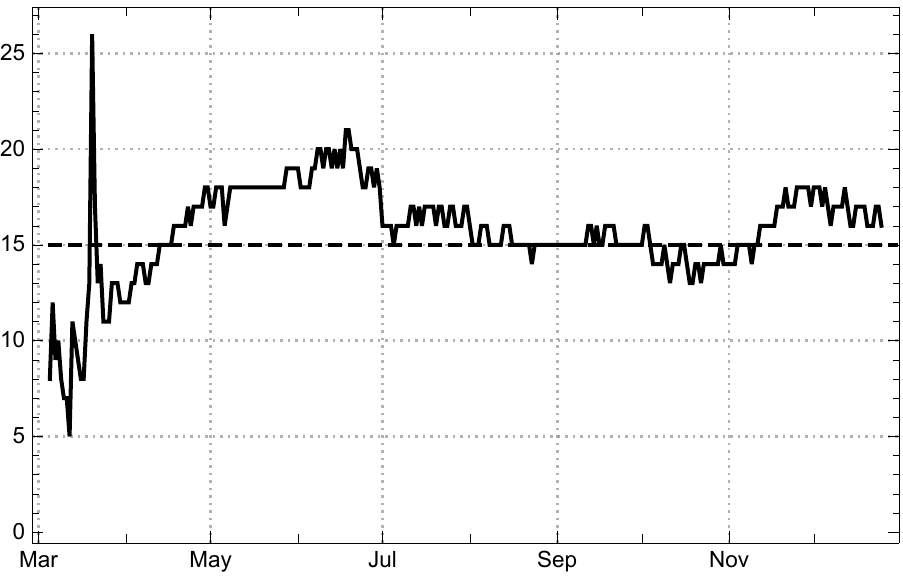} 
\caption{\tiny Daily values $\hat{q}(k)$ for the mean time of being statistically counted as  an  actual case  for  India (left) and Germany (right)  \label{fig q(k) D CH}}
\vspace{1em}
\includegraphics[scale=0.6]{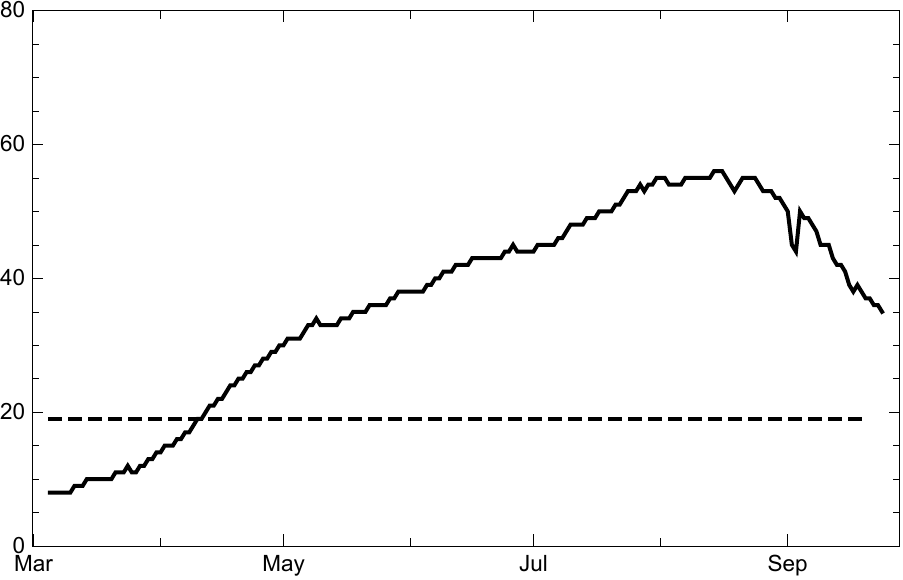} 
\hspace{1cm}
\includegraphics[scale=0.6]{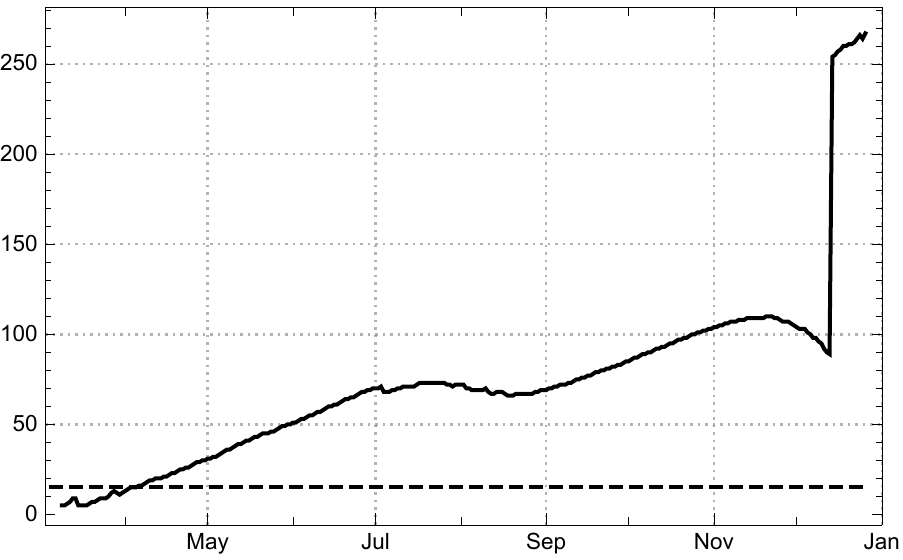}
\caption{\tiny Daily values of the mean time $\hat{q}(k)$  of being statistically counted as actual  infected for Italy (left) and USA (right)  \label{fig q(k) It US} }
\end{figure}

The difference  $\hat{A}(k)=\hat{A}_{tot}(k)-\hat{R}(k)$  (\ref{eq hat-A-tot}) can in principle be considered as an expression for the  number of {\em actual}  cases; but  
it is corrupted by the fact   that  the number of daily recovering $Rec(k)$ is irregularly reported.
For a critical investigation of this number we start from the truism that any actually infected person recorded at day $k$  
has been counted among the $\hat{A}_{new}(l)$ at some earlier day,  $l\leq k$. The smallest number $\hat{q}(k)$ of days preceding $k$ (including the latter) necessary for supplying  sufficient large numbers of infected  $\hat{A}(k)$,
\beq \hat{q}(k)= \min_{l} \,\Bigl[ \sum_{j=0}^l \hat{A}_{new}(k-j) \geq \hat{A}(k) \Bigr]\, , \label{eq hat q(k)}
\eeq  
 is  a good indicator for  the mean time of sojourn in the collective of infected which are recorded as ``actual cases''. As long as the number of severely ill among all infected  persons is relatively small and the time of severe illness well constrained, we may expect that the mean time of actual illness  does not deviate much from  the time of prescribed minimal time of isolation $q_{min}$ for infected persons. In the case of Covid-19  $q(k)$   surpasses  $q_{min}\approx 14$ only moderately  for  India, Germany,  Switzerland  etc. (fig. \ref{fig q(k) D CH}). For many other countries $\hat{q}(k)$ behaves differently. It starts near the time of quarantine or isolation but increases for a long time monotonically with the development of the epidemic, before often -- although not always -- it starts to decrease  again after the (local) peak of a  wave has been surpassed. This is the case, e.g., for Italy and the US; in the last case the deviation is extreme, $\hat{q}(k)$ surpasses  100  and shoots up a little later (see fig. \ref{fig q(k) It US})

   This effect cannot be attributed to medical reasons; the major contribution  rather  results from 
the unreliability of  the statistical book keeping: With the growing overload of the health system,  the time of recovery of registered infected persons is being  reported with an increasing time delay, sometimes not at all (e.g., Sweden, UK). The difference 
$ \hat{A}(k) = \hat{A}_{tot}(k) -\hat{R}(k) $
gets   increasingly confounded by the lack of correctness in the numbers  $Rec(k)$. In these countries it is an expression of  the number of {\em  ``statistically  actual''} cases only with, at best, an indirect relation to the real numbers of people in quarantine  or hospital. 

The information gathered for Covid-19  proposes the existence of a stable mean time $q$ of isolation  of infected persons (including  hospital) for long periods in each country. It is usually  a few days longer than the official duration of quarantine prescribed by the health authorities. 
Given $q$, the sum
\beq \hat{A}_q(k) = \sum_{j=0}^{q-1} \hat{A}_{new}(k-j) \label{eq hat-A_q}
\eeq
can be used as an  estimate of the number of infected recorded persons who are in isolation or hospital at the  day $k$. 
Here we do not use 7- or 3- day averages, because the summation compensates the daily oscillations anyhow. The  accordingly corrected number of redrawn   $\hat{R}_q$ is of course given by 
\beq \hat{R}_q = \hat{A}+ \hat{R}-\hat{A}_q \, .
\eeq

\begin{figure}[h]
 \includegraphics[scale=0.7]{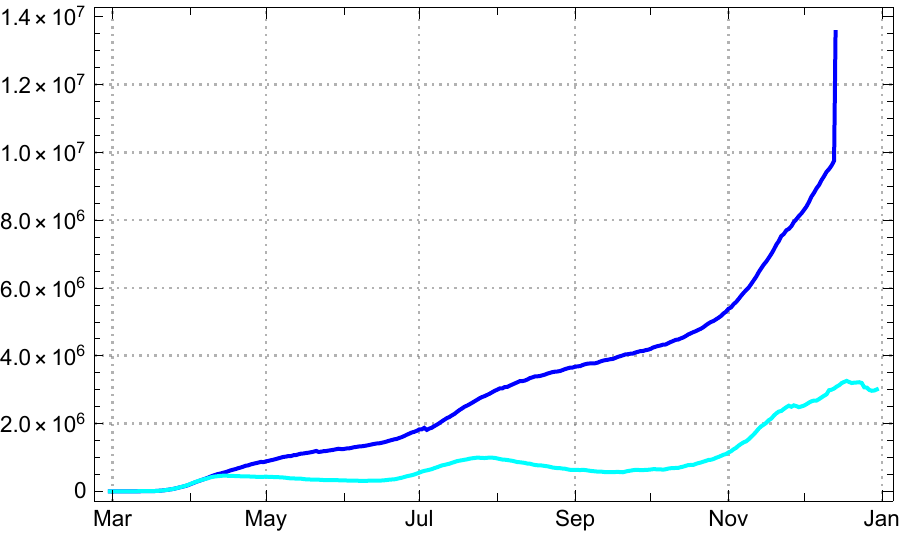}
\caption{\tiny $\hat{A}(k)$ (dark blue), $\hat{A}_q(k)$ (bright blue) for the USA ($q=15$)   \label{fig  US I Iq etc} }
\end{figure}

 Figure \ref{fig  US I Iq etc} shows $\hat{A}(k), \, \hat{A}_q(k)$ 
  for the USA (with $q=15$). It demonstrates  the difference between $\hat{A}(k)$ (dark blue) and $\hat{A}_q(k)$ (bright blue) and shows that $\hat{A}_q(k)$ is a more reliable estimate of actually infected than the numbers $\hat{A}(k)$    (the  ``active cases'' of the Worldometer).
  
For countries with reliable statistical recording of the recovered we find $\hat{q}(k) \approx const$. In this case we choose this constant as the value for the model $q$. For other countries one may use a default value, inferred from  comparable countries with a better status of recording the $Rec(k)$ data (i.e. $\hat{A}(k) \approx \hat{A}_q(k)$).

\subsubsection*{Simplifying assumptions on the duration $e$ of exposition  and the  duration $p$ of effective infectivity}
For Covid-19 it is known that there is a period of  duration say $e$  between the  exposition to the virus, marking the beginning of an infection, and the onset of active infectivity. Then a period of  propagation, i.e.   effective infectivity,  with duration  $p$  follows, before the infection is diagnosed,  the person is  isolated in quarantine or hospital and can  no longer  contribute to the further spread of the virus. 
 Although one might want to represent the mentioned durations by stochastic variables with their respective distributions and mean values, we use here the mean values only and make the simplifying assumption of constant $e$ and $p$ approximated by the nearest natural numbers. 

The Robert Koch Institute  estimates the mean time from infection to occurrence of symptoms to about 4 days \cite{RKI:Steckbrief}, (5.). This is divided into $e$ plus the time from getting infectious to the occurrence of symptoms. According to   studies already  mentioned above  the latter is estimated as $2$ days, so as a consequence we estimate $e=2$.   In section \ref{section countries} we generically use $p=7$.
We have checked the stability  of the model under a change of the conventions of parameter choice inside the mentioned intervals.

\subsubsection*{Estimate of the daily strength of infection}

As announced in part I we work with the simplified SEPAR$_d$ model. This means that the duration  in compartments $P_c$ and $P_d$ is equal, here  denoted by $p$,  and also the strength of infection is assumed to be equal: $\eta_c = \eta_d = \eta$. Furthermore, if $P(k) = P_c(k) + P_d(k)$ there is a branching ratio $\alpha$, which has to be estimated for each country, such that $P_c (k)= \alpha P(k)$ and $P_d(k) = (1-\alpha)P(k)$.  For every counted infected there are then 
\[ \delta= \frac{1-\alpha}{\alpha}
\]
uncounted ones. We call $\delta$ the {\em dark factor}.

Once $e$ and $p$ are  given (or fixed by convention inside their  intervals) one can  determine the empirical strength of infection $\eta(k)$ using the model equations. Namely 
$$
\eta(k) = \frac {E_{new}(k+1) }{s(k) P(k)}.
$$
In terms of the total number of infected $H(k)$ (see (\ref{eq P in terms of H})  and with constant $\alpha$ this is 
\beqa & & \eta(k) =  \\
& & \quad \frac{H(k+1)-H(k)}{s(k)\, \Big(\alpha \big(H(k-e) -H(k-(e+p_c))\big)  + \xi\, (1-\alpha)\big(H(k-e) -H(k-(e+p_d))) \big) \Big)} 
\eeqa

For the simplified SEPAR$_d$ model we have $\alpha E_{new}(k) = A_{new}(k+e+p)$ and $\alpha P(k) = \sum_{j=1}^p\hat{A}_{new}(k+j)$. Thus  $\alpha $ cancels and  we obtain:
\beq 
\eta(k) = \frac {A_{new}(k+e+p) }{s(k) \sum_{j=1}^p {A}_{new}(k+j)}    \label{eq eta}
\eeq 
Denoting, as before,  the values we obtain from the data by  $\hat \eta (k)$ etc.  we obtain
\beq  \hat{\eta}(k)  =  \frac {\hat A_{new}(k+e+p) }{\hat{s}(k) \sum_{j=1}^p{\hat A}_{new}(k+j)} \qquad \mbox{resp.} \quad \hat{\eta}_7(k)  =  \frac {\hat A_{new,7}(k+e+p) }{\hat{s}(k) \sum_{j=1}^p{\hat A}_{new}(k+j)}, \label{eq hat(eta)}
\eeq 
where in the second equation we work with the weekly averaged data.

An  estimation of the total number of new infections induced by an infected person during  the effective propagation time (and thus the whole time of illness) is then
\beq
\hat{\rho}(k) = \sum_{j=0}^{p-1} \hat{\eta} (k+j) \hat{s}(k+j) \, ; \label{eq repro number general}
\eeq
and similarly for $\hat{\rho}_7(k)$. In the following we generally use the latter but write just $\hat{\rho}(k)$. Note also that the determination of the strength of infection $\hat{\eta}(k)$ by (\ref{eq hat(eta)}) needs an estimation of the dark factor $\delta$ because the latter  enters into the ratio of susceptibles $\hat{s}(k)$,while it cancels in the calculation of the empirical reproduction rate $\hat{\rho}(k)$  (\ref{eq repro number empirical}). 

This is an empirical estimate for the {\em reproduction number} $\rho(k)$.
In periods of nearly constant daily strength of infection one may use the approximation
\beq
 \hat{\rho_7}(k) \approx p \, \hat{\eta_7}(k) \hat{s}(k) =   \frac{p \, A_{new,7}(k+e+p)}{\sum_{j=1}^p A_{new}(k+j)}\label{eq repro number empirical}
\eeq
Inspection of transition periods between constancy intervals for Covid 19 shows that this approximation is also feasible in such phases of change.  For  $p=7$
this  variant of the reproduction number stands in close relation to the reproduction numbers  used by the {\em Robert Koch Institut},   see appendix \ref{appendix generations}), which gives additional support to this  choice of the parameter. 
\\[1em]

  \subsection{SEPAR$_d$ parameters\label{subsection adaptation to the data}}
  
 For modelling Covid-19 in the simplified dark approach we use the parameter choice $e=2,\; p \,(=p_c=p_d) =7$  as explained in sec. \ref{subsec data evaluation}. Where we differentiate between $p_c$ and $p_d$ we usually use $p_c=7$ and $p_d=10$. 
 The value for  $q$ depends on the reported mean duration of reported infected  being counted as ``actual (active)'' case for each  country (sec. \ref{subsec JHU data}); in the following reports it usually lies between 10 and 17. For each country we let the recursion start at the first day $k_0$ at which the reported new infections become ``non-sporadic'' in the sense that no zero entries appear at least in the next $e+p_d$ days ($\hat{A}_{new}(k)\neq 0$ for $k_0 \leq k \leq k_0+(e+p_d)$).  For $k\geq k_0-1$ the  values $\hat{\eta}(k)$ are  calculated  in the simplified case, $p=p_c=p_d$ according to  (\ref{eq hat(eta)}). Otherwise  the formula above (\ref{eq eta}) has to be used. 
 
 \subsubsection*{Start values}
  For  the numerical calculations we use the recursion (\ref{eq recursion for H(k)}), with start values given by time shifted  numbers of the  statistically reported confirmed cases, expanded by the dark factor, for $k$ in the interval $J_{-1}= [k_{-1}, k_0 -1]$ where $k_{-1}= k_0-1-(e+p_d)$:
  \[ \hat{H}(k)=(1+\delta)\mathit{Conf}(k+e+p_c)= (1+\delta)\hat{A}_{tot}(k+e+p_c)
  \] 
Note that the time step parametrization in the introduction/definition of the SEPAR model in sec. \ref{subsection SEPAR}   works with $k_0= 1$.  
  
   If we set the model parameters $\eta(k)=0$ for $k< k_0-1$ and $\eta(k)=\hat{\eta}(k)$ for $k\geq k_0-1$, the recursion reproduces the data exactly, due to the  definition of the  coefficients. Then and only then it becomes {\em tautological}.   
Already if we use  coefficients  $\hat{\eta}_7(k)$ for $k\geq k_0$  from time averaged  numbers of daily newly reported according to eq. (\ref{eq hat(eta)}), the model ceases to be tautological. In this case  the parameter $\eta_0=\eta (k_0-1)$ may be used  for optimizing  (root mean square error)  the   result for the total number of reported infected $A_{tot}$ in comparison with the empirical data (\ref{eq hat-A-tot}).
 The model acquires conditional predictive ability, if longer intervals of constant coefficients are chosen.

 One may prefer to replace $A_{new}$ by the 7-day averages $\hat{A}_{new,7}$ by introducing 
 \[\mathit{Conf}_7(k)=\sum_{j=k_{-1}}^k \hat{A}_{new,7}(j) \; \; (+const)
 \]
 and $\hat{H}_7(k)= (1+\delta)\,Conf_7(k+e+ p_c)$. For constant $\alpha(k)=\alpha$ the replacement of the $\hat{A}_{new}$ in the denominator of (\ref{eq hat(eta)}) then boils down to defining
 \[ \hat{\eta}_7(k)= \frac{\hat{H}_7(k+1)- \hat{H}_7(k)}{\hat{s}(k) 
 \Big(\alpha \big( \hat{H}_7(k-e) -\hat{H}_7(k-(e+p_c))\big) + \xi (1-\alpha) \, \big(\hat{H}_7(k-e) -\hat{H}_7(k-(e+p_d))  \big)     \Big) }
 \]
 Then the model becomes tautological also for the use of daily varying $\hat{\eta}_7(k)$ and ceases to be so only after introducing constancy intervals as described in the next subsection. 
 
 \subsubsection*{Main intervals}
The number of empirically determined parameters can be drastically reduced by approximating  the daily changing empirical infection coefficients $\hat{\eta}_7(k)$ ($k= k_0, k_0+1, \ldots$)  by constant model values $\eta_j$ ($1 \leq j \leq l$)  in appropriately chosen intervals  $J_1, \ldots  J_l$. We call them the 
  {\em constancy, or main, intervals} of the model.  Their choice is crucial for arriving at a full-fledged   non-tautological model of the epidemic.

We thus choose  time markers $k_j$ (``change points'') for the beginning of such intervals and  durations $\Delta_j$ for the transition between  two successive constancy intervals, such that: 
\[ J_j =[ k_j, \, k_{j+1}-\Delta_{j+1}]\; ,  \qquad j= 1, \ldots , l
\]
In the main interval $J_j$ the {\em model strength of infection} $\eta_j$  (this is the constant daily strength of infection during this time) are  generically chosen as the arithmetical mean  of the empirical values   $\mathit{ mean} \{  \hat{\eta}(k)\;  | \; k \in J_j\}$. Small deviations of the mean, inside the 1 $\sigma$ range of the  $\hat{\eta}$-fluctuations in the interval, are admitted if in this way the mean square error of the model $A_{tot}$ can be reduced  noticeably. The  dates $t_j$ of  change points $k_j$  can  be read off heuristically from the graph of the $\hat{\eta}_7$ and may be improved by an  optimization procedure.
$k_1$ is  chosen as the first day of a period in which the daily strength of infection can reasonably be approximated by a constant. In the {\em initial interval} $J_0=[k_0, \, k_1 -1]$ the model uses the empirical daily strength of infection: $\eta(k)=\hat{\eta}_7(k)$ for $k\in J_0$. In the transition intervals $[ k_j - \Delta_j, \, k_j ]$ the model strength of infection  is gradually, e.g. linearly, lowered from $\eta_{j-1}$ to $\eta_j$.\footnote{Here we use the  smoothing function described in \citep{K-S}. Also an elementary optimization procedure for determining the main intervals is described in this paper.}

For the labelling of the days $k$ there are two natural choices; the {\em JHU day count} starting with $k=1$ at 01/22/2020, or a {\em country adapted count} such that  $k_0=1$, where $k_0$ labels the first day for which the reported new infections become non-sporadic (see above). Both choices have their pros and contras; in the following we make use of both in different contexts, declaring of course which one is being used. 

\subsubsection*{Influence of the dark sector} 
Proceeding in this way involves an indirect observance of the dark sector's contribution to the infection dynamics of the visible sector. An unknown part of the counted new infections $\hat{E}_{new}(k)$  is causally due to contacts with infectious persons in $P_d$ of the dark sector, eqs. (\ref{eq E-newdark}) or (\ref{eq recursion for H(k)})). 
 According to   estimates of epidemiologists there is a wide spectrum of possibilities worldwide,   $0 \leq \delta \leq 100$,  for Covid-19, while we  have only rough guesses for the different countries. In the following country reports we work with the simplified dark approach, $p_d=p_c=p=7$ and $\xi=1$ ($\gamma_c =\gamma_d$) and use estimates for the dark factor $\delta$  explained in the respective country section. \\
 

\section{Selected countries/territories \label{section countries}}
In our collection  we include four small or medium sized  European countries (Switzerland, Germany, France, Sweden), three large countries from three different continents (USA, Brazil, India), and a  model for the aggregated data of all world countries and territories. For the first country analysed  here, Switzerland,   relatively reliable data on the dark sector are accessible. We take it as an example for a discussion of the effects of different assumptions on the  branching ration $\alpha$, respectively the factor $\delta$, of the dark sector. For the other countries we lay open on which considerations our  choice of the model the dark factor is based.

\subsection{Four  European countries; Switzerland Germany, France, Sweden\label{subsection 4 European countries}} 
The four European countries discussed here show  different features with regard to the epidemic: Switzerland and Germany have a relatively well organized health  and data reporting system; the overall course of the epidemic with wave peaks in early April and in early November 2020 and a moderately controlled phase in between is is typical for most other European countries.  France, in contrast, shows surprising features in the documentation of statistically recorded new infections (negative entries in the first half of the year); and Sweden has been chosen because of a containment strategy of its own. In the case of Switzerland and Germany first  results of representative serological studies are available. They allow a more reliable  estimate of the size of the dark sector than in most other cases. We therefore start our discussion with these countries.


\subsubsection*{Switzerland}
\begin{figure}[h]
 \includegraphics[scale=0.7]{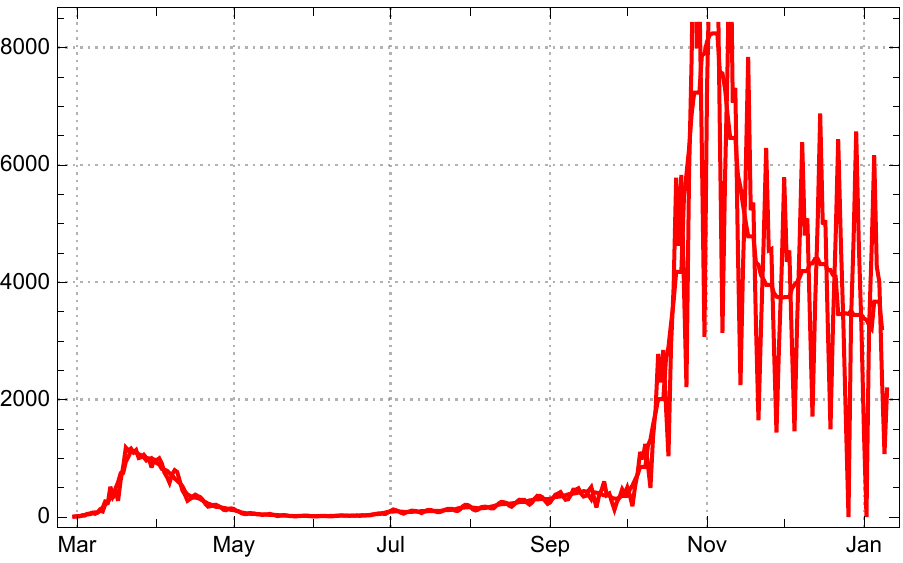} \\
\caption{Daily new reported cases  for Switzerland,  3-day sliding  averages $\hat{A}_{new,3}(k)$ and  7-day  averages $\hat{A}_{new,7}(k)$. \label{fig A-neu CH}}
\end{figure}

\begin{figure}[h]
\includegraphics[scale=0.7]{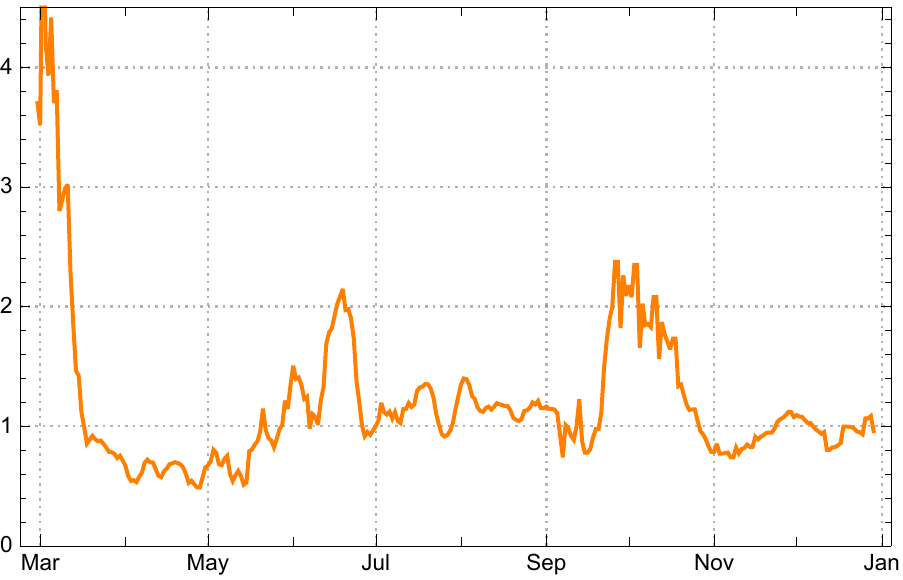} 
\includegraphics[scale=0.7]{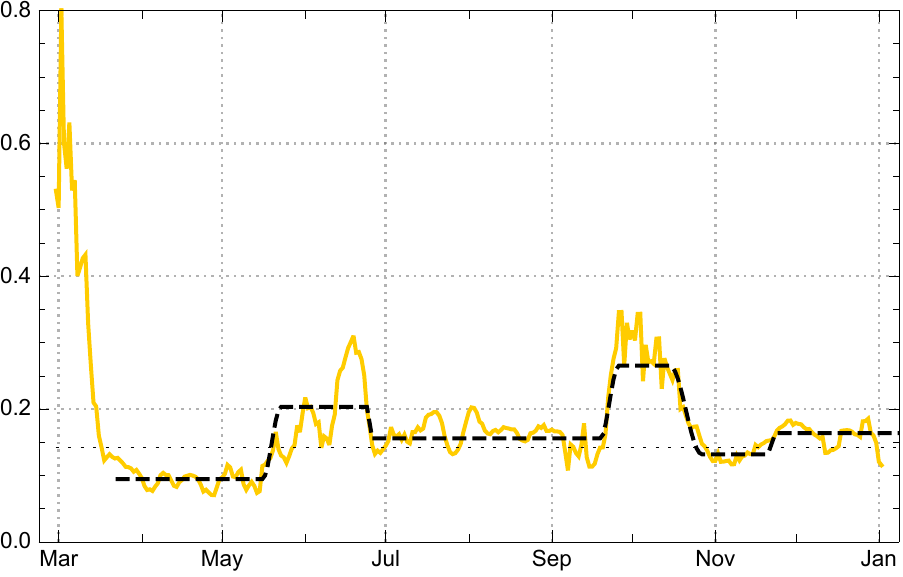} 
\caption{Left: Empirical reproduction rates $\hat{\rho}(k)$ for Switzerland. Right: Daily strength of infection $\hat{\eta}_7(k)$   (yellow) with model parameters $\eta_j$ (black dashed) in the main intervals  $J_j$  for $p=7$ and $\delta=2$. \label{fig rho und eta CH}}
\end{figure}

The numbers of  reported new infected ceased to be sporadic in Switzerland at February 29, 2020; we take  this as our  day $t_0= 1$.  For the reported daily new infections (3-day and 7-day sliding averages) see figure \ref{fig A-neu CH}.
At Feb 28  recommendations for hygiene etc. were issued by the Swiss government and large events   prohibited, including  Basel {\em Fasnacht} (carnival). These regulations were already active at  our day $t_0=1$ (Feb 29) and explain the rapid fall of the empirical strength of infection (7-day averages) $\hat{\eta}_7(k)$  at the very beginning of our period.    During the next 15 days a series of additional general regulations were taken: Mar 13, ban of assemblies of more than 100 persons, lockdown of schools; Mar 16 ($t=12$) lockdown of shops, restaurants, cinemas; Mar 20 ($t=16$),  gatherings of more than 5 persons prohibited. This sufficed for lowering the   growth rate quite effectively. 
 The SEPAR reproduction number  started from peak values close to $5$ and fell down   to below the critical value at March 23, the day $t_1=24$  in  the country count (fig. \ref{fig rho und eta CH}, left). Here it remained with small oscillations until mid May, after which it rose (we choose $t_2=85$, May 23, as the next time mark),  with strong oscillations until late June, before it was brought down to close to 1 in late June ($t_3= 158$, June 23). A long phase of slow growth ($\rho \approx 1.1$) followed until mid September. An extremely swift rise of the reproduction number  to values above 2 in late September ($t_4=242$, Sep. 26)  brought the number of new infections to heights formerly unseen in Switzerland.  In spite of great differences among the differently affected regions ({\em Kantone}) the epidemic was brought under control at the turn to November ($t_5=272$, Oct. 27).

The empirical values of the infection strengths $\hat{\eta}_7(k)$ determined on the basis of the 7-day sliding averages $\hat{A}_{new,7}$ depend on estimates of the dark factor $\delta$ (cf. eq. \ref{eq hat(eta)}). Recent serological studies summarized in  \citep{Kuster:2021} 
 conclude   $\delta \approx 2$. We   choose this value  as generic for the  simplified SEPAR$_d$ model.  The values of 
 $\hat{\eta}_7(k)$, assuming a dark factor $\delta=2$,  are shown in fig. \ref{fig rho und eta CH}, right. How its values are affected by  different assumptions for the dark sector can be inspected by comparing with the results for $\delta =0$ and $\delta=4$   (fig.  \ref{fig eta delta = 0, 4 CH}). The influence of the dark sector on  $\hat{\eta}_7$ becomes visible only late in the  year 2020.
 

\begin{figure}[h]
 \includegraphics[scale=0.7]{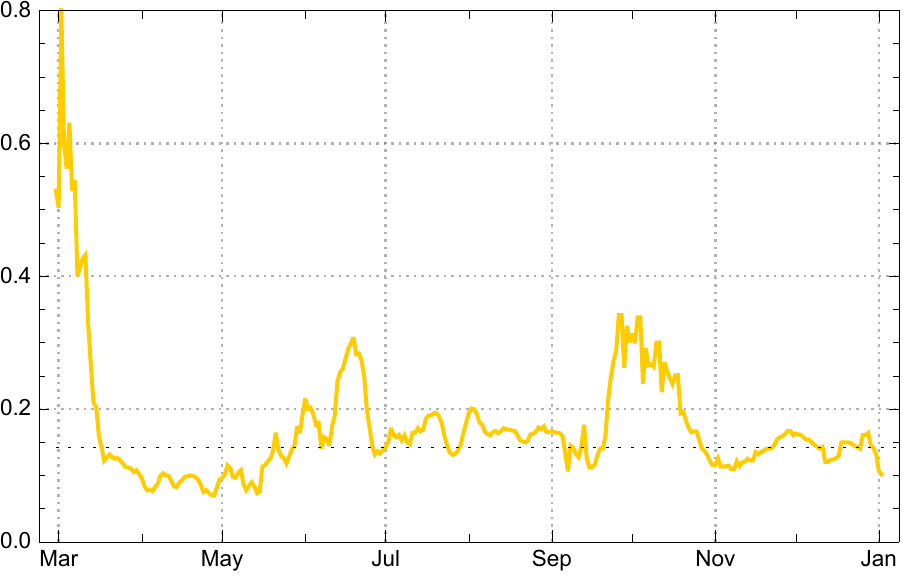}  \hspace{1em}
  \includegraphics[scale=0.7]{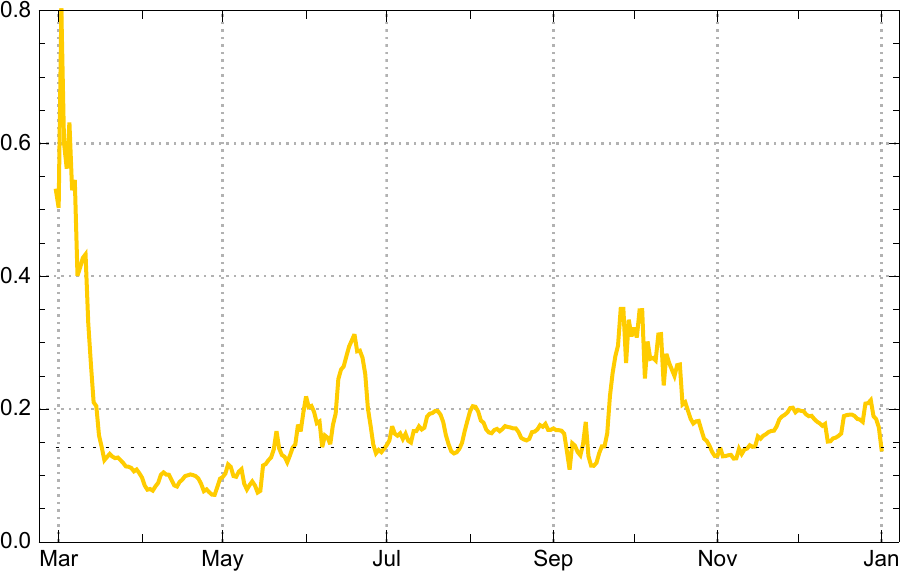}
\caption{Comparison of empirical values $\hat{\eta}_7(k)$ assuming dark factor $\delta=0$ (left) and $\delta=4$ (right); for  comparison with generic choice $\delta=2$ see last figure. \label{fig eta delta = 0, 4 CH}}
\end{figure}

With the time markers between different growth phases of the epidemic indicated above we choose the following main intervals for our model:
$J_0= [1, 22], \; J_1=[24, 78],\; J_2= [85, 117], \; J_3=[119, 204], J_4= [211, 230 ], \; J_5=[242, 264],\; J_6= [270, 321] $, end of data ($t_{eod}=321$) at  Jan 14, 2021. The model values $\eta_j$ in the Intervals $J_j$ are essentially the mean values of $\hat{\eta}_7(k)$ in the respective interval, where small deviations inside the 1 sigma domain are admitted if the (root mean square) fit to the empirical data $\hat{A}_{tot}$ can  be  improved. They are given in the table below and indicated in fig. \ref{fig rho und eta CH} (black dashed lines). (Note that $\eta_0$ has no realistic meaning; it is a free parameter of the start condition for modelling on the basis of the 7-day averages $\hat{\eta}_7(k)$,  see sec. \ref{subsection adaptation to the data}.) The reproduction rates $\rho_j$ in the table refer to the beginning of the intervals; in later times the decrease of $s(k)$ can lower the reproduction rates until the end of the intervals considerably. In the case of Switzerland the latter crosses the critical threshold 1 inside the last constancy interval (see below).

\vspace{0.2cm}
\begin{center}
\begin{tabular}{|l||c|c|c|c|c|c|c|}
\hline 
\multicolumn{8}{|c|}{Model $\eta_j$ and $\rho_j$ in intervals $J_j$   for Switzerland}\\
\hline
 & $J_0$ & $J_1$ & $J_2$ & $J_3$ & $J_4$ & $J_5$& $J_6$\\
\hline 
 $\eta_j$ & -0.127 & 0.095  & 0.203 & 0.156 &  0.265  & 0.132  & 0.165    \\
 $\rho_j$  &  ---  & 0.66   & 1.41 & 1.08  & 1.82 & 0.86 & 1.01  \\
\hline
\end{tabular}
\end{center}
\vspace{0.5em}

The course of the new infections is well modelled with these values (fig. \ref{fig Aneu Mod CH}); the same holds for the total number of counted infected (fig. \ref{fig 3 curves CH} below). 

\begin{figure}[h]
\includegraphics[scale=0.7]{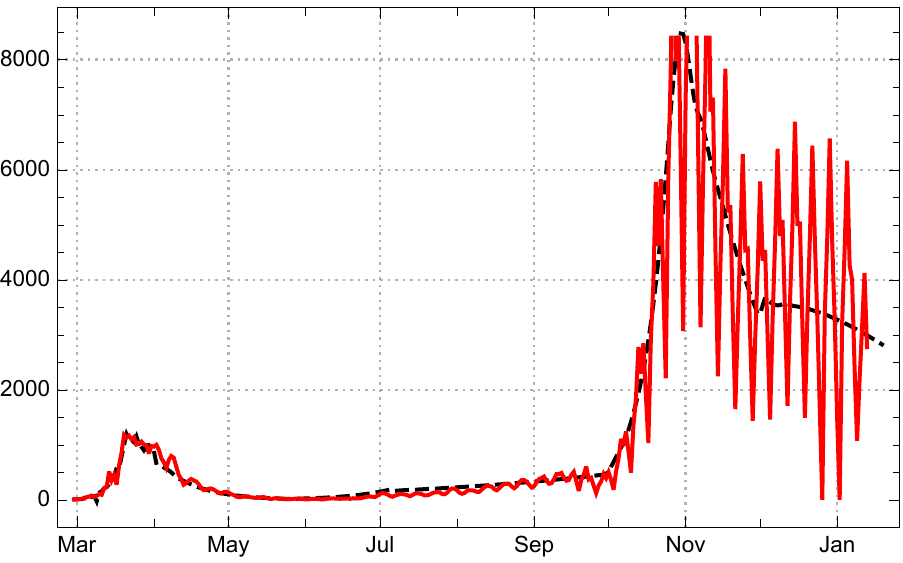} 
\caption{Model reconstruction for the 7-day averages of the number of new infected $A_{new}(k)$ (black dashed) in comparison with the empirical  data, 3-day averages,   $\hat{A}_{new,3}(k)$  (solid red) for  Switzerland. \label{fig Aneu Mod CH} }
\end{figure}

\begin{figure}[h]
\includegraphics[scale=0.6]{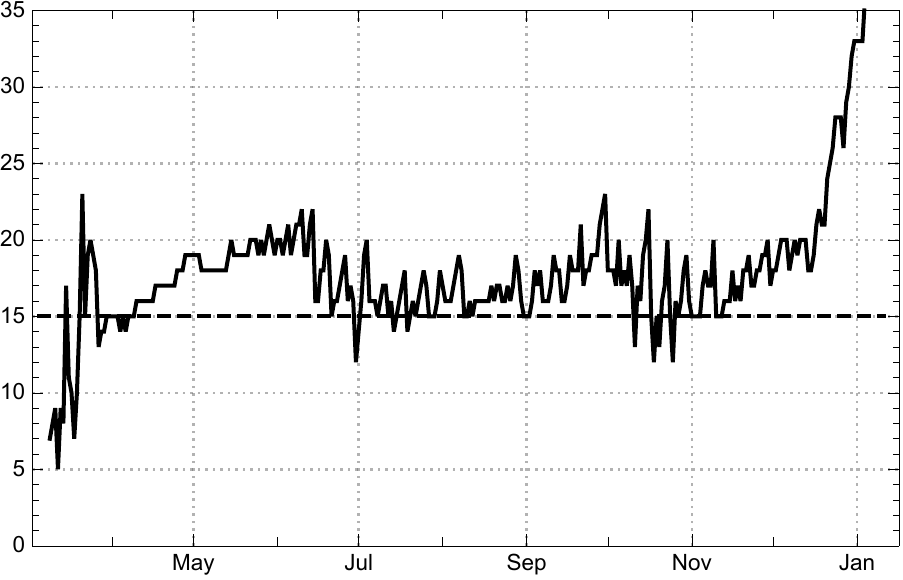} 
\includegraphics[scale=0.7]{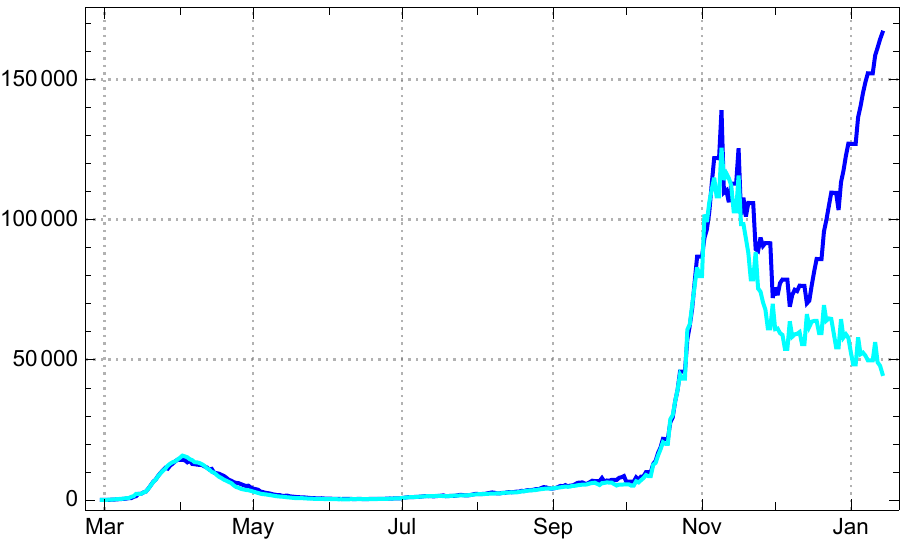}
\caption{Left: Daily values of the mean time of statistically actual   infection $\hat{q}(k)$ for  Switzerland.  Right: Comparison of reported infected $\hat{A}$ (dark blue) and $q$-corrected number ($q=15$) of recorded actual infected $\hat{A}_q$ (bright blue) from the JHU data in Switzerland. \label{fig q(k) CH} }
\end{figure}

The count of days $\hat{q}(k)$, necessary for filling up the numbers of reported actual infected by sums of newly infected during  the directly preceding days shows a relatively stable value,  $\hat{q}(k) \approx 15$, until early November. Since then  the mean time of reported sojourn in the department $A$ increases steeply  (fig. \ref{fig q(k) CH}, left). Initially  this may have been due to a rapid increase of severely ill people during  the second wave of the epidemic; but the number does not fall again with the stabilization  in December 2021. A comparison of the statistically recorded actual infected $\hat{A}(k)$ with the $q$-corrected one $\hat{A}_q(k)$ (eq. \ref{eq hat q(k)}) shows an ongoing increase of the first while the second one falls again after a sharp peak in early November (same figure, right). We  conclude from this that from December 2020 onward the numbers of recovered are no longer reliably reported even in Switzerland.

\begin{figure}[h]
\includegraphics[scale=0.7]{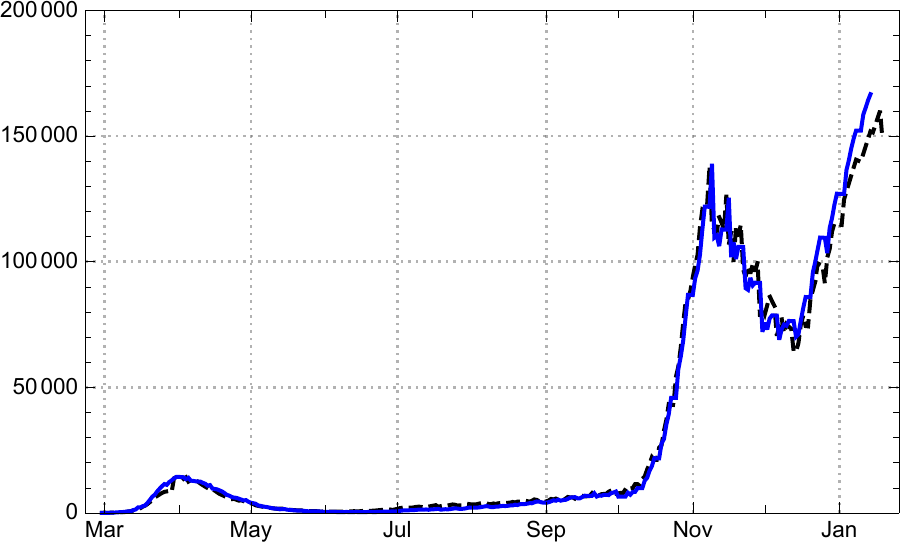} \includegraphics[scale=0.7]{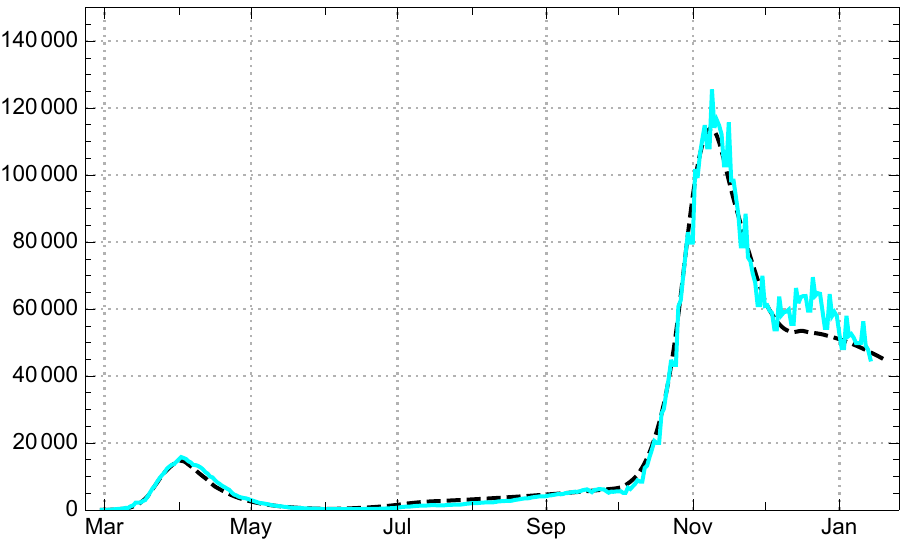}
\caption{Left: Number of actual infected $\hat{A}(k)$ (blue)   for  Switzerland, recorded by the JHU data, and model values calculated with time dependent $q(k)=\hat{q}(k)$  (black dashed). 
Right: Empirical  values,  $q$-corrected with  $q=15$,  for statistically actual cases   $\hat{A}_q(k)$  and the corresponding model values $A(k)$ (black dashed).  \label{fig A Aq CH} }
\end{figure}

Both quantities cam be  well modelled in our approach (fig. \ref{fig A Aq CH}), although we  consider $\hat{A}_q(k)$ as a  more reliable estimate for actually ill persons.  

On this basis, 	the SEPAR$_d$ model expresses the development of the epidemic in Switzerland on the basis of only 6 constancy intervals for the parameters $\eta$ for the whole year 2020. The three main curves of the total number of recorded infected, $A_{tot}(k)$, the number of redrawn $R(k)$ and the number of diseased counted as statistical ``actual'' cases $A(k)$ are shown in figure  \ref{fig 3 curves CH}. 

\begin{figure}[h]
\includegraphics[scale=0.7]{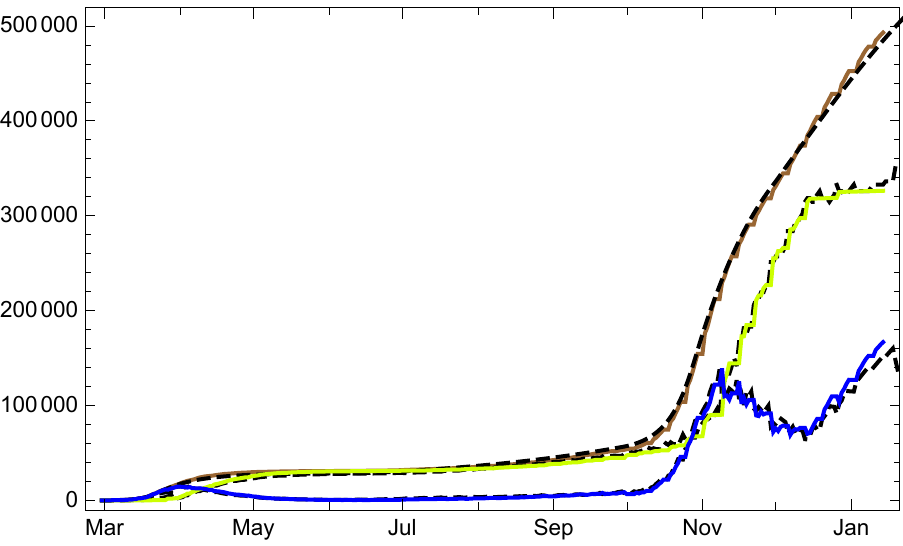}
\caption{Empirical data (solid coloured lines) and  model values (black dashed)  in Switzerland for  numbers of totally infected $\hat{A}_{tot}$ (brown), redrawn $\hat{R}$ (bright green), and  actual  numbers $\hat{A}$, according to the statistic (blue). \label{fig 3 curves CH} }
\end{figure}

\begin{figure}[h]
\includegraphics[scale=0.7]{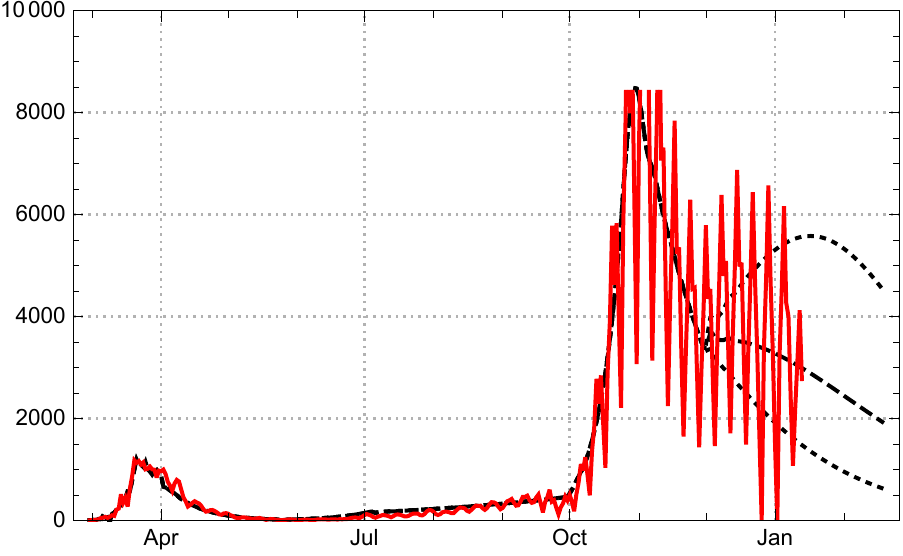} \hspace{2em} \includegraphics[scale=0.7]{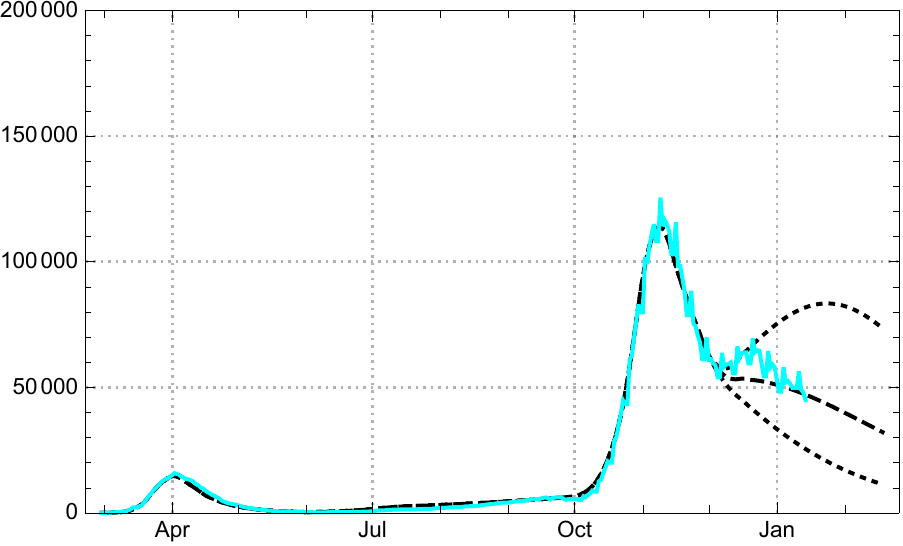}
\hspace{1em}\\
\includegraphics[scale=0.7]{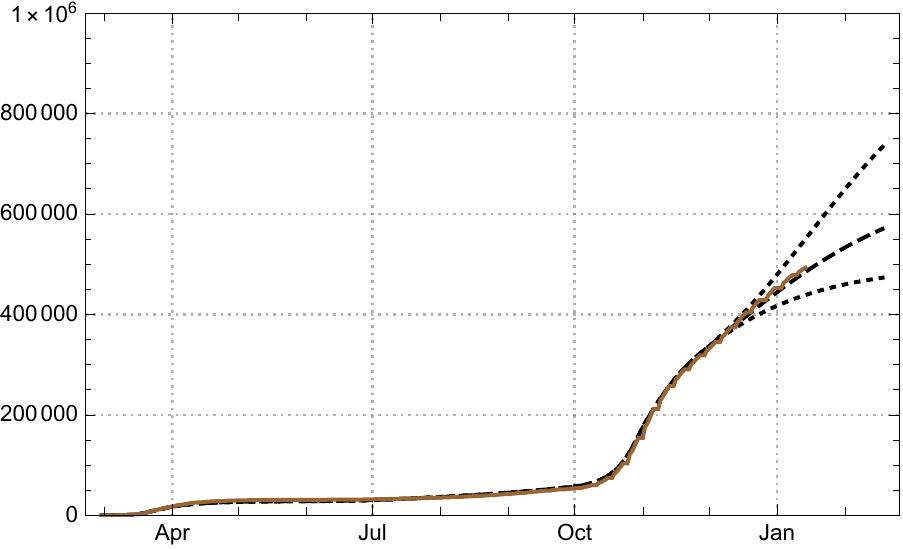} 
\caption{30-day prediction for $A_{new}, \, A_q$ (top) and $A_{tot}$  (bottom) for Switzerland, assuming dark factor $\delta=2$; empirical values  coloured solid lines,  model black dashed (boundaries of 1-sigma region prediction dotted). \label{fig prae-30 CH delta=2}  }
\end{figure}

This may encourage to look at a conditional 30-day prediction for Switzerland given by the SEPAR$_d$ model, the condition being the hypothesis of no considerable change in the contact behaviour of the population and no increasing influence of new virus mutations, i.e., a continuation of the recursion with $\eta_6$, the   strength of infection in the last constancy interval (fig. \ref{fig prae-30 CH delta=2}). 

\begin{figure}[h]
\vspace{2em}
\includegraphics[scale=0.6]{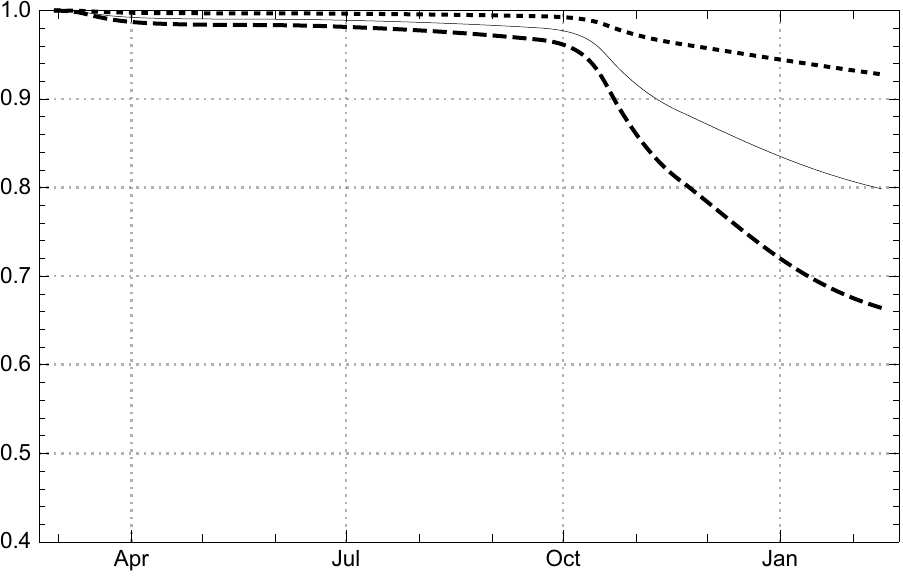}
\caption{Development of ratio of susceptibles $s(k)$ for Switzerland (model values),  assuming dark factor $\delta=0$ (dotted),  $\delta=2$ (solid line) and   $\delta=4$ (dashed). \label{fig srek CH} }
\end{figure}

The figures show clearly that, under the generic assumption $\delta=2$  for Switzerland, the ratio of infected $s(k)$ starts to suppress the rise of new and actual infections already in January 2021 even for the upper bound of the 1-sigma estimate for the parameter  $\eta_7$ (fig. \ref{fig prae-30 CH delta=2}, top). Of course the question arises what would be changed assuming  different values for the dark factor. Figure \ref{fig srek CH} shows how strongly the ratio of susceptibles is influenced by the choice of  $\delta$ already at the end of 2020.

The results  for $\delta=0$ and  $\delta=4$ of the model values  of reported new infected, $A_{new}(k)$ (black dotted or dashed as above) in comparison with  the 3-day averages of the JHU data, $\hat{A}_{new,3}(k)$, is shown in figure \ref{fig prae-30 CH delta=0, 4}. Remember that the  model expresses the dynamics of 7-day averages of the newly infected. 

\begin{figure}[h]
\vspace{2em}
\includegraphics[scale=0.7]{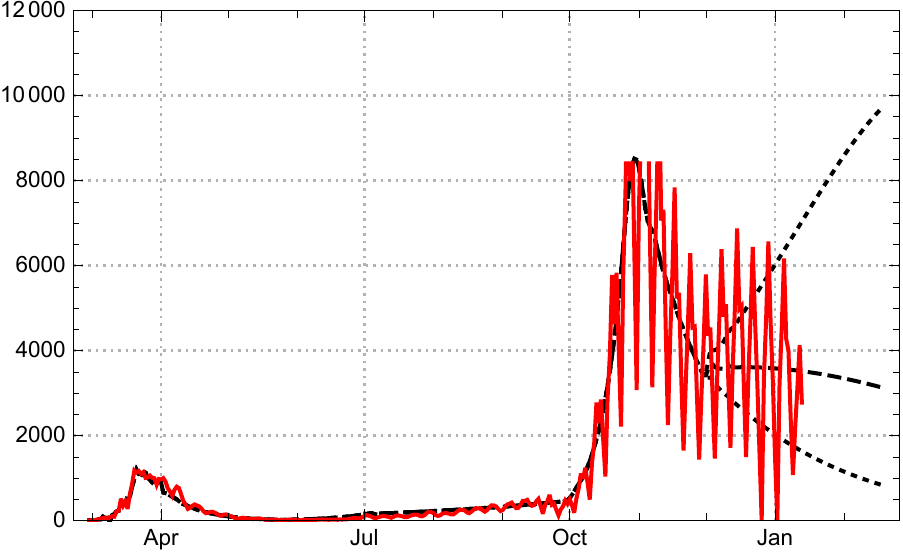} \includegraphics[scale=0.7]{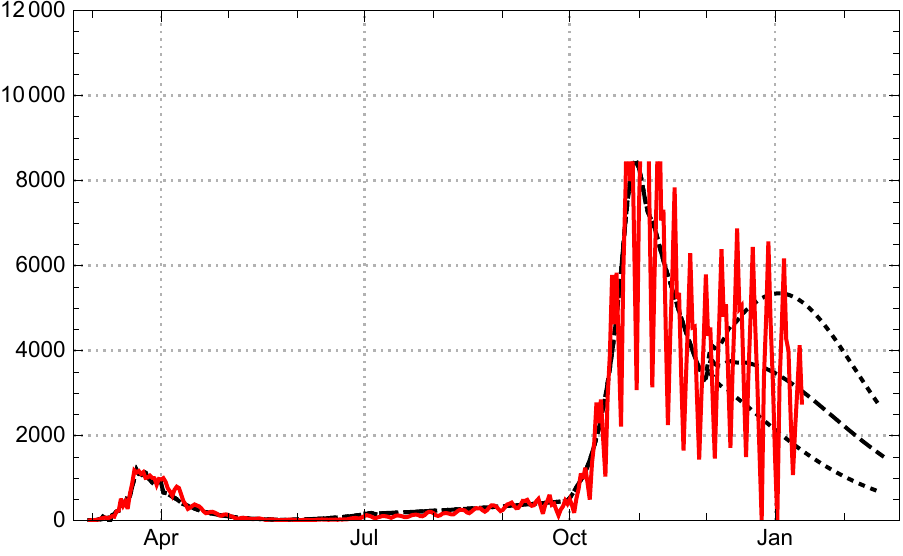}
\caption{30-day prediction for $A_{new}$  for Switzerland, assuming dark factor $\delta=0$ (left), respectively $\delta=4$ (right); empirical values  $\hat{A}_{new,3}$  coloured solid lines,  model values black dashed (boundaries of 1-sigma region  dotted). \label{fig prae-30 CH delta=0, 4}  }
\end{figure}
\vspace*{1em}

Assuming a negligible  dark sector ($\delta=0$) and  the upper boundary of the 1-sigma interval of $\eta_7$, the number of new infected would   continue to rise  deeply into the first quarter of 2021. In all other cases the effective reproduction number is suppressed below the critical value  by the ratio  of susceptibles $s(k)$ already at the turn to the new year. We conclude that the role of the dark sector starts  to have qualitative impact on the development of the epidemic in Switzerland already at this time. \\[3em]

 \subsubsection*{Germany}
\begin{figure}[h]
\includegraphics[scale=0.6]{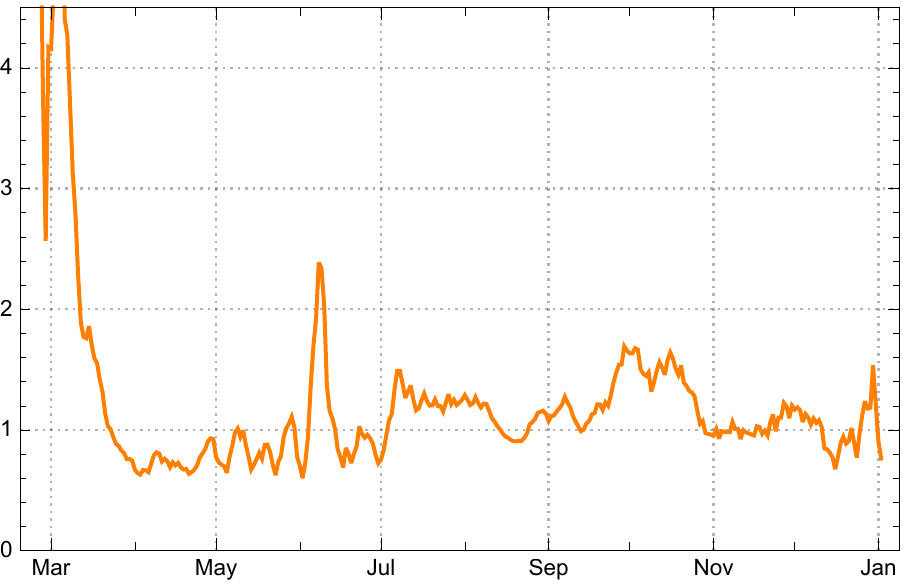} \includegraphics[scale=0.6]{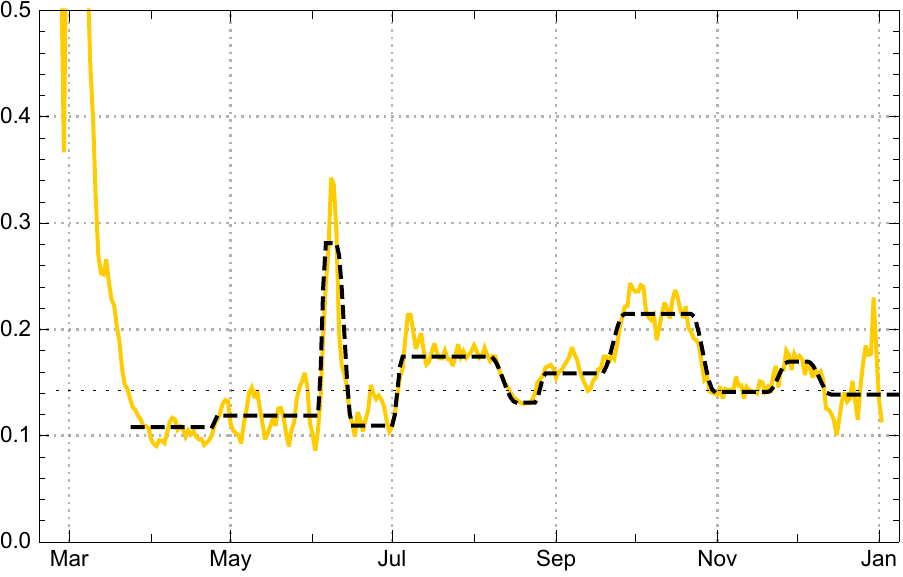} 
\caption{Left: Empirical reproduction rates $\hat{\rho}(k)$ for Germany (yellow). Right: Daily strength of infection $\hat{\eta}(k)$  for Germany (yellow) with model parameters $\eta_j$ in the main intervals  $J_j$ (black dashed); critical value $1/p$ of $\eta$ dotted. \label{fig I-neu rho und ak D}}
\end{figure}

The epidemic entered Germany (population 83 M) in the second half of February 2020;  the recorded new infections seized to be sporadic at $t_0=$ Feb. 25, the day  $k_0=1$ in our country count. With the health institutions being set in a first alarm state and  public advertising of protective behaviour,  the initial reproduction rate (as  determined in our model approach)  fell swiftly from roughly $\rho \approx 4$  to about 1.5 until mid March. 
At May 25 ($t_1=30$) it dropped below 1 and stayed there, with an exceptional week (dominated by a huge infection cluster in the meet factory T\"onnies) until late June 2020 (fig. \ref{fig I-neu rho und ak D} left).  The peak of the first wave was reached  by the 7-day averages of new infections $\hat{A}_{new,7}$ at March 30;  6 days later, i.e. April 6, the local maximum  for the actual numbers of reported infected $\hat{A}$ followed.

Serological studies in the region Munich indicate that during the 
first half of 2020 
the ratio of counted people was about $\alpha =0.25 $  in Germany, i.e. for each counted person there were $\delta \approx \frac{1-\alpha}{\alpha}=3$ persons entering the dark sector  \cite{Radon_ea:2020}. Of course  this ratio  varies in space and time, for example if the number of available tests increases or, the other way round, it is too small for a rapidly increasing number of infected people. The rapid expansion of testing,in Germany during early summer seems to have increased the branching ratio to about  $\alpha \approx 0.5$. In follow up  investigations the authors of the  Munich study    come to the conclusion that during the next months the ratio of unreported infected  has decreased considerably; this  brought the factor   $\delta$  down to $\approx 1$ in early November 2020.\footnote{Short report in \citep{Helmholtz-Zentrum-Muenchen:2020-12}.   This is consistent with the result in \citep{RKI:2020SerologischeStudie}.}
Since the dark segment influences our model mainly through its contribution to lowering of the ratio  of susceptibles $s(k)$, it is nearly negligible in the first six months of the epidemic. We therefore simplify the empirical findings by setting  $\delta=1$ for the  SEPAR$_d$ model of Germany. 

\begin{figure}[h]
 \includegraphics[scale=0.9]{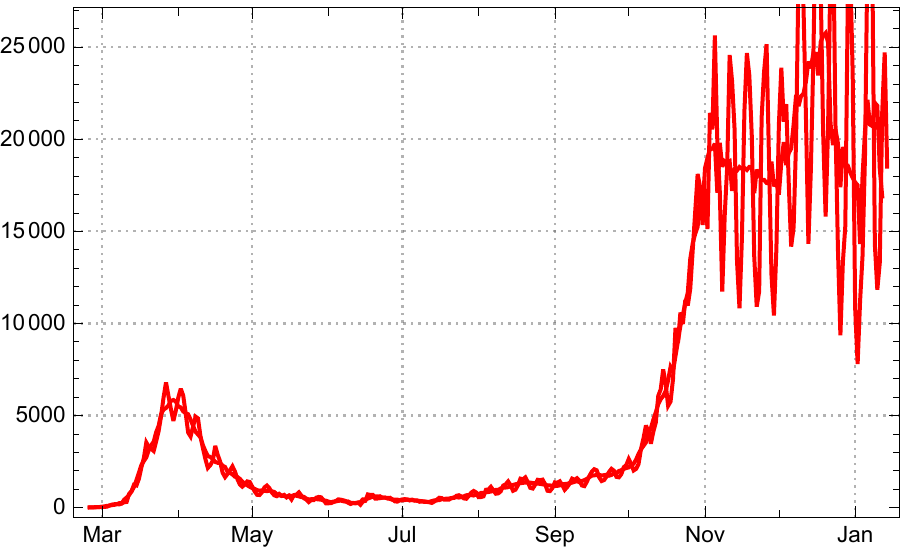} \\
\caption{Daily new reported cases for Germany, 3-day sliding averages  $\hat{A}_{new}(k)$  and  7-day  averages $\hat{A}_{new,7}(k)$.  \label{fig I-neu}}
\end{figure}

Contrary to a widely pronounced assessment (including by experts) according to which  the epidemic  was well under control until September, the reproduction rate rose to $\rho \approx 1.3$  already in July and the first half of August. Only the low level of daily new infections, reached in late May, covered up the expanding tendency and gave the impression of a negligible increase. After a short lowering interlude about mid August 
the  rise came back in late August with $\rho \approx 1.2$, before it accelerated  in late September, brought the reproduction rate to about 1.5, and  led straight into the second wave. Here the weekly oscillations of  recorded new infections rose to an amplitude not conceived before (fig. 
\ref{fig I-neu}). The levels of the mean daily strength of infection used in the model (proportional to the corresponding reproduction rates) are well discernible in the next figure \ref{fig I-neu rho und ak D},  right.

\pagebreak
The dates $t_j$  of the time markers $k_j$  used for our model in the German case are $t_0 =$ 02/25, 2020, $t_1= $ 03/24, $t_2= $ 04/26, $t_3= $ 06/06, $t_4= $ 06/16, $t_5= $ 07/05, $t_6= $ 08/17, $t_7= $ 08/28, $t_8= $ 09/27, $t_9= $ 10/31, $t_{10}= $ 11/28, $t_{11}= $ 12/14, end of data here $t_{eod}= $ 01/15, 2021. In the country day count, where $t_0=1$ ($\sim$ 35 in JHU day count) the main intervals are $J_0= [1, 27], \; J_1=[29, 59],\; J_2= [62, 100], \; J_3=[103, 108]; J_4= [113, 128], \; J_5=[132, 165],\; J_6= [175, 179], \; J_7=[186, 207]; J_8= [216, 241], \; J_9=[250, 270],\; J_{10}= [278, 285], \,  J_{11}= [294,326]  $. 
The  strength of infection and reproduction numbers in the main intervals are given in the following table. 
\vspace{0.2cm}
\begin{center}
\begin{tabular}{|l||c|c|c|c|c|c|c|c|c|c|c|c|}
\hline 
\multicolumn{13}{|c|}{$\eta_0$ and model $\eta_j$, $\rho_j$ in intervals $J_j$   for Germany}\\
\hline
 & $\eta_0$ & $J_1$ & $J_2$ & $J_3$ & $J_4$ & $J_5$& $J_6$ & $J_7$ & $J_{8}$& $J_9$ & $J_{10}$ & $J_{11}$\\
\hline 
 $\eta_j$ & 0 &0.108  & 0.119 & 0.281 &  0.109  & 0.174  & 0.131  & 0.158 &0.215  & 0.141 & 0.170 & 0.139 \\
 $\rho_j$  &  ---  & 0.75   & 0.83 & 1.96  & 0.76 & 1.21 & 0.91 & 1.10 &1.49 &  0.97& 1.15 &0.93 \\
\hline
\end{tabular}
\end{center}
\vspace{0.5em}

\begin{figure}[h]
\includegraphics[scale=0.8]{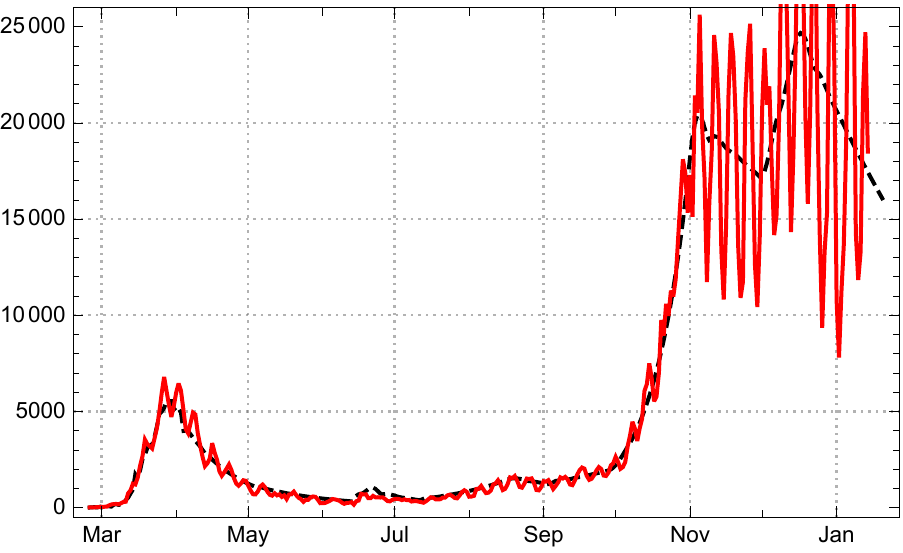}  \includegraphics[scale=0.8]{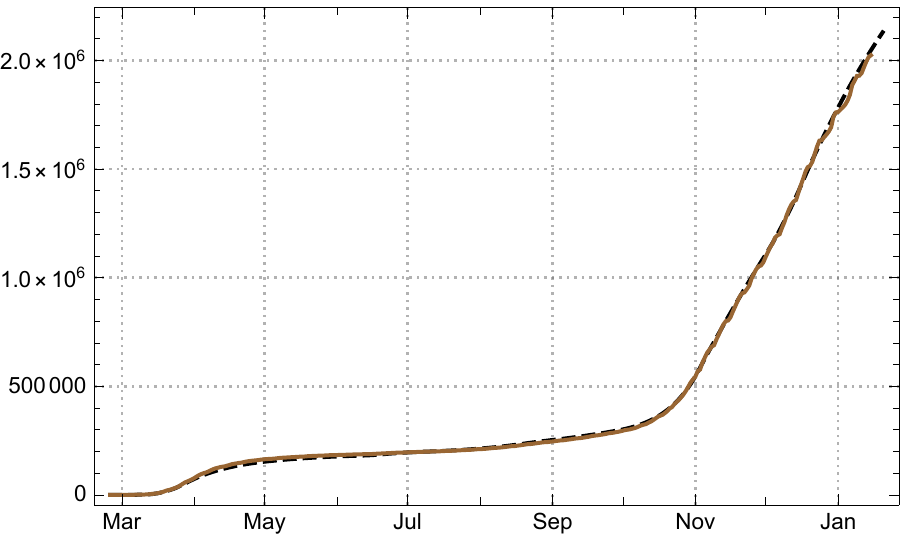} 
\caption{Left: Daily new reported infected  (3-day averages) for  Germany; empirical $\hat{A}_{new}$  solid red, model $A_{new}$ black dashed.\newline Right: Total number of reported infected;  empirical $\hat{A}_{tot}$ solid brown, model  $A_{tot}$ black dashed. \label{fig Ineu and Itot D}}
\end{figure}

With these parameters the SEPAR model reproduces (pre- or better ``post''-dicts) the averaged daily new infections and the total number of reported infected well (fig. \ref{fig Ineu and Itot D}), while one has to be more careful for treating the  reported number of actual infected  $\hat{A}$.

\begin{figure}[h]
\includegraphics[scale=0.6]{graph-D-q} 
\includegraphics[scale=0.7]{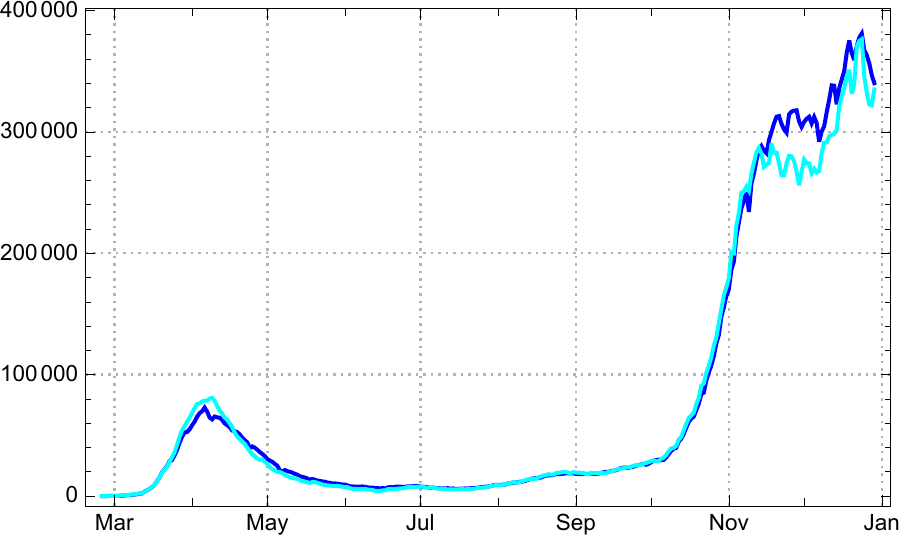}
\caption{Left: Daily values of the mean time of statistically actual   infection $\hat{q}(k)$ for  Germany.  Right: Comparison of reported infected $\hat{A}$ (dark blue) and $q$-corrected number ($q=15$) of recorded actual infected $\hat{A}_q$ (bright blue) from the JHU data in Germany. \label{fig q(k) D} }
\end{figure}

If one checks  the mean duration of being recorded as actual case in the JHU  statistics for Germany by eq. (\ref{eq hat-A_q}) one finds a  good approximation  $q \approx 15$ after the  early phase;  but the result also  indicates that in May/June, and again in the second half of November, the reported duration of the infected state surpassed this value (fig. \ref{fig q(k) D} left). Accordingly the empirical data $\hat{A}$ and the corrected ones $\hat{A}_q$  ($q=15$) drop apart in late November (same figure, right).

\begin{figure}[h]
\includegraphics[scale=0.7]{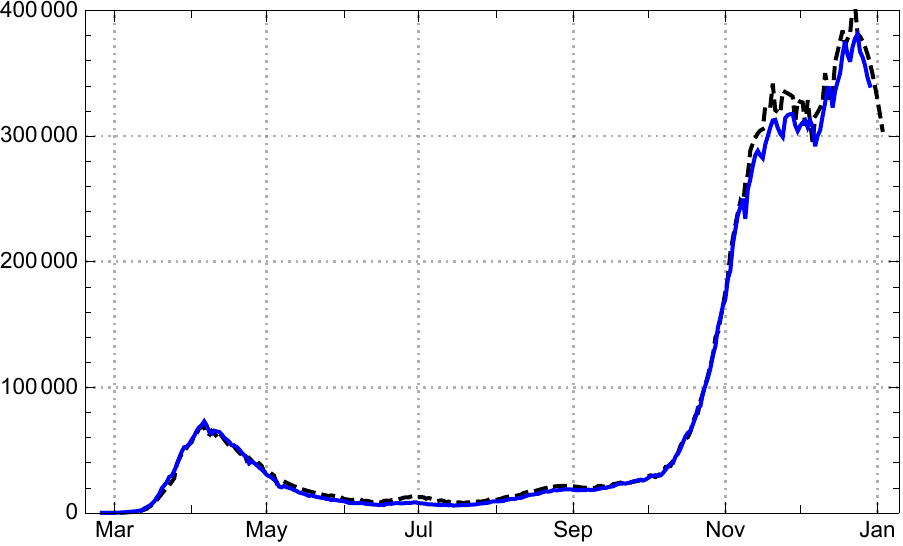} \includegraphics[scale=0.7]{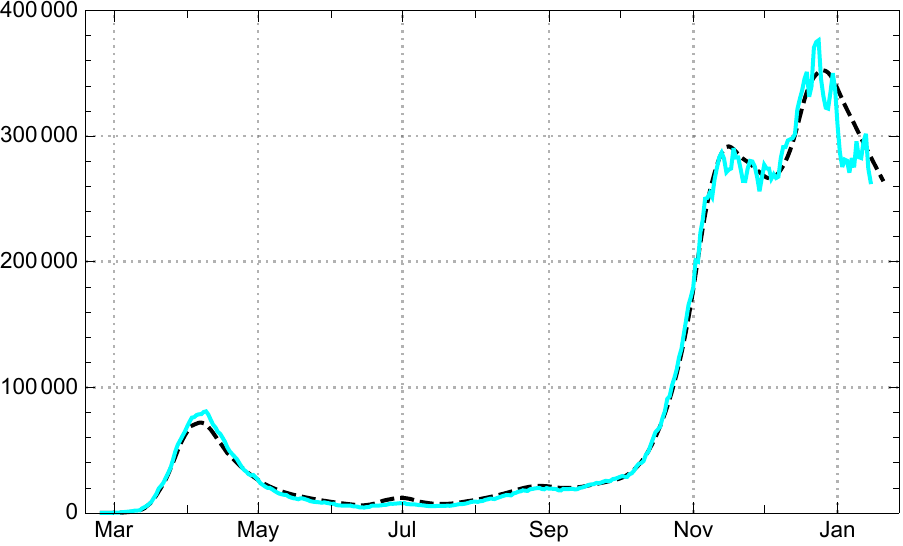}
\caption{Left: Statistically actual cases   $\hat{A}(k)$  for  Germany  and the corresponding model values $A(k)$ (black dashed) calculated with time dependent model values for $q$  (see text).
Right: Empirical  values,  $q$-corrected,   for  actual cases   $\hat{A}_q(k)$  and the corresponding model values $A(k)$ (black dashed), $q=15$. \label{fig I Iq D} }
\end{figure}

In consequence the model value for the actual infected $A(k)$ agree with the JHU data  $\hat{A}(k)$ only if the model uses time varying values $\hat{q}(k)$ (fig. \ref{fig I Iq D} left), while the $q$-corrected numbers $\hat{A}_q(k)$  are well reproduced by the model with constant $q=15$ (same figure, right).

\begin{figure}[h]
\includegraphics[scale=0.7]{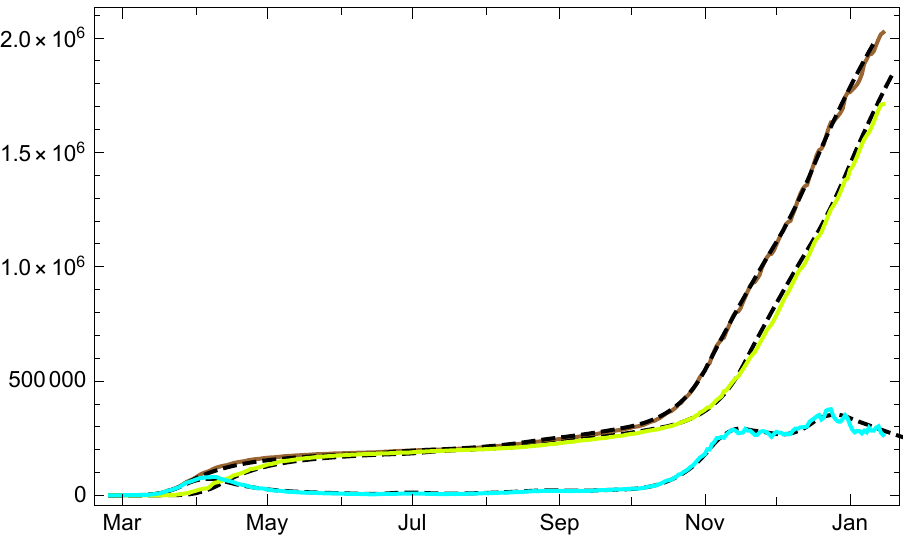}
\caption{Empirical data (solid coloured lines) and  model values (black dashed) for Germany:  numbers of totally infected $\hat{A}_{tot}$ (brown), redrawn $\hat{Rq}$ (bright green), and $q$-corrected actual  numbers $\hat{A}_q$  (bright blue) . \label{fig 3 curves D} }
\end{figure}

As a result, the 3 model curves representing the total number of (reported) infected $A_{tot}$, the redrawn $R$ and the actual infected   fit  the German data well, if the last two  are compared with the $q$-corrected empirical numbers $\hat{A}_q$ (fig. \ref{fig 3 curves D}).

\begin{figure}[h]
\includegraphics[scale=0.7]{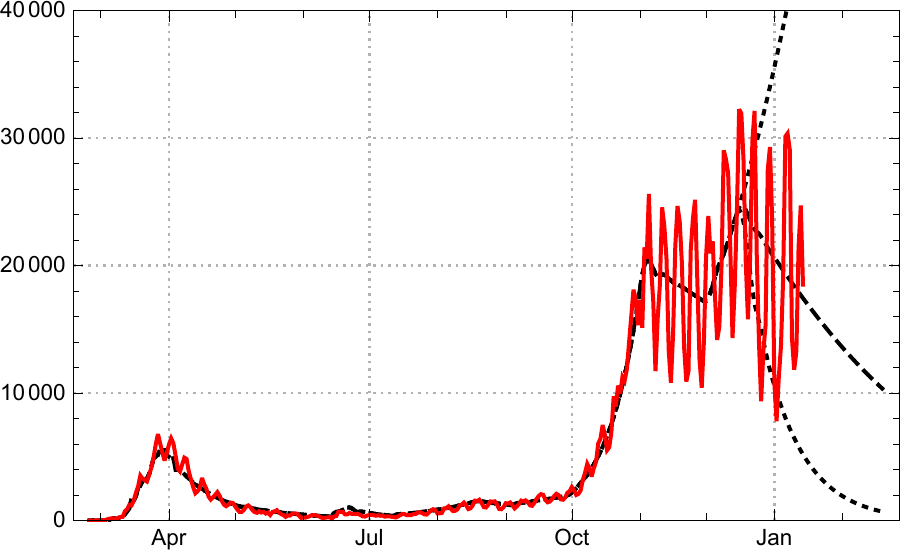} \includegraphics[scale=0.7]{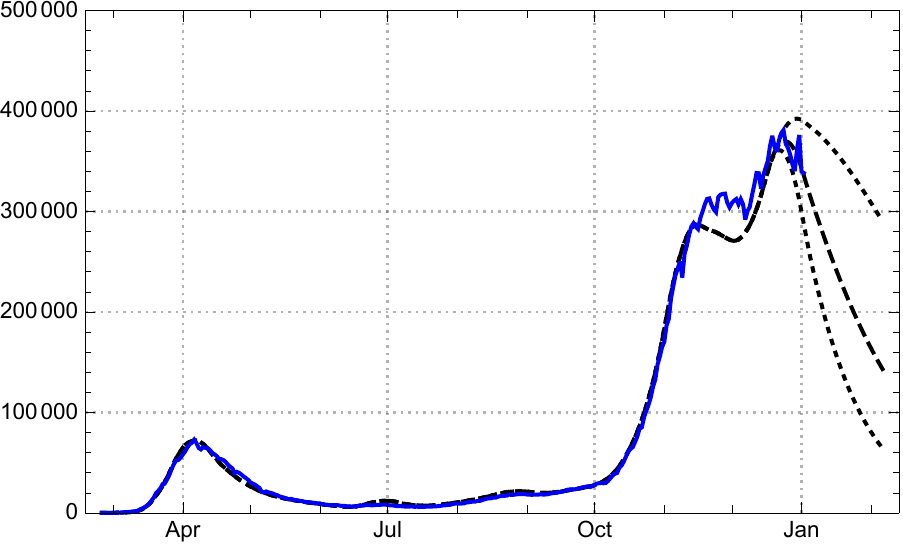} 
\hspace{0.5cm}\\
\includegraphics[scale=0.7]{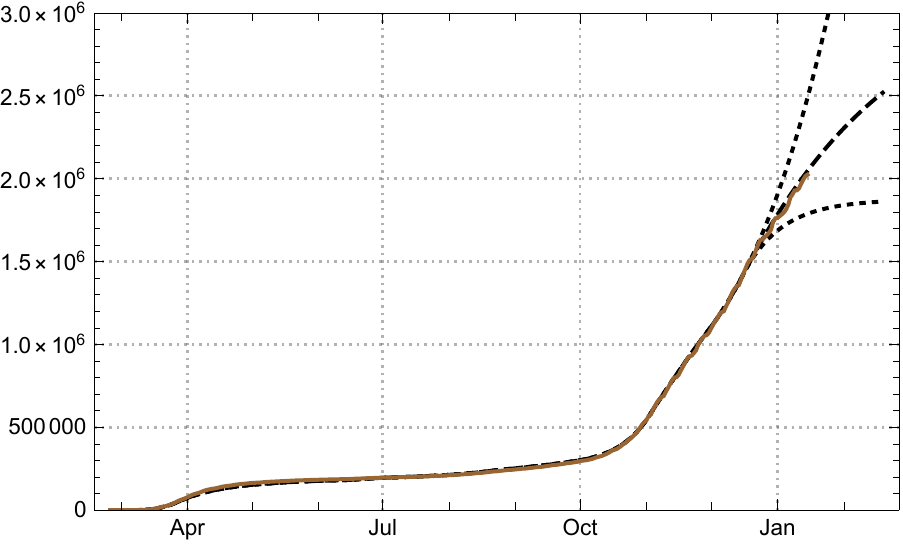}
\caption{30-day prediction for $A_{new}, \, A$ (top) and $A_{tot}$  (bottom) for Germany; empirical values  coloured solid lines,  model black dashed (boundaries of 1-sigma region prediction dotted). \label{fig prae-30 D}  }
\end{figure}

\newpage
Conditional predictions for $A_{new}, \, A_q$  and $A_{tot}$, assuming no essential change of the behaviour, contact rates and the reproduction number  from the last main interval $J_{10}$ are given in fig. \ref{fig prae-30 D}. The dotted lines indicate the  boundaries of the prediction for the 1-$\sigma$ domain for the variations of the values of $\hat{\eta}$ in the last main interval $J_{10}$.

\newpage

\subsubsection*{France}
\begin{figure}[h]
 \includegraphics[scale=0.7]{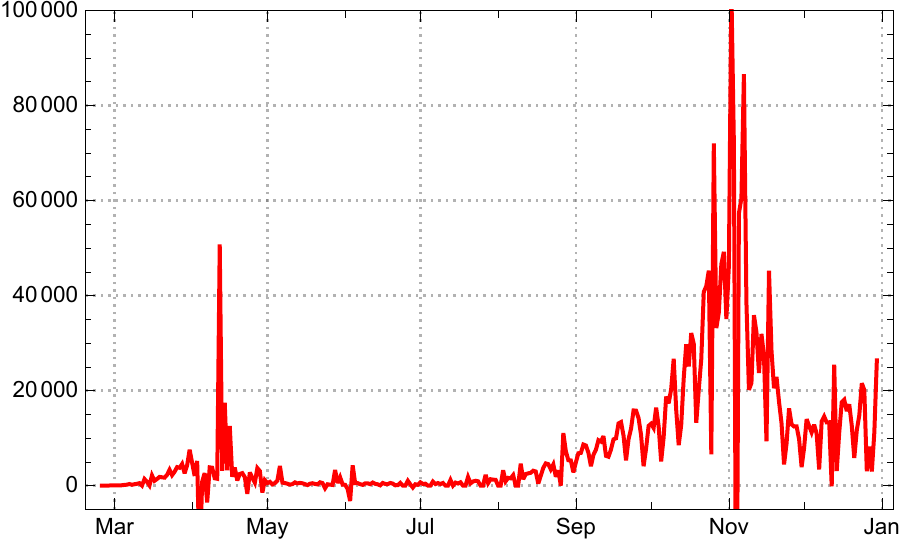}
 \includegraphics[scale=0.7]{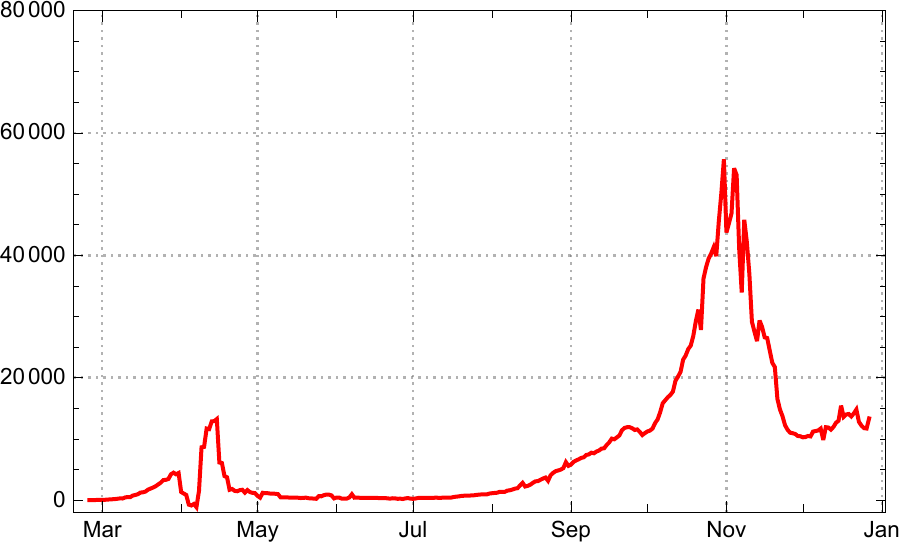} 
\caption{Left: Number of  daily newly reported in France $\hat{A}_{new}(k)$. Right: 7-day sliding averages of new infections $\hat{A}_{new,7}(k)$ for France. \label{fig I-neu Ineu7 JHU F}}
\end{figure}

The overall picture of the epidemic in France (population 66 M) is similar to  other European countries. But the 
French JHU data show anomalies  which are not found elsewhere: The differences of two consecutive values of the confirmed cases,  which ought to represent the number of newly reported, is sometime {\em negative}!
This happens in particular in the early phase of the epidemic (until June 2020)  where, e.g.,   $\hat{A}_{new}(58)=-2206$, and there are other days  with negative entries. Presumably this is due to ex-post data corrections   necessary in the first few months of the epidemic. Later on 
 the control over the documented data seems to have been improved; negative values are avoided, although  null entries in  $\hat{A}_{new}$ still appear. 
So the French data are a particular challenge to any  modelling approach. 
Even  in this extreme case the smoothing by 7-day sliding averages works  well, as shown in fig.  \ref{fig I-neu Ineu7 JHU F}.  

\begin{figure}[h]
\includegraphics[scale=0.6]{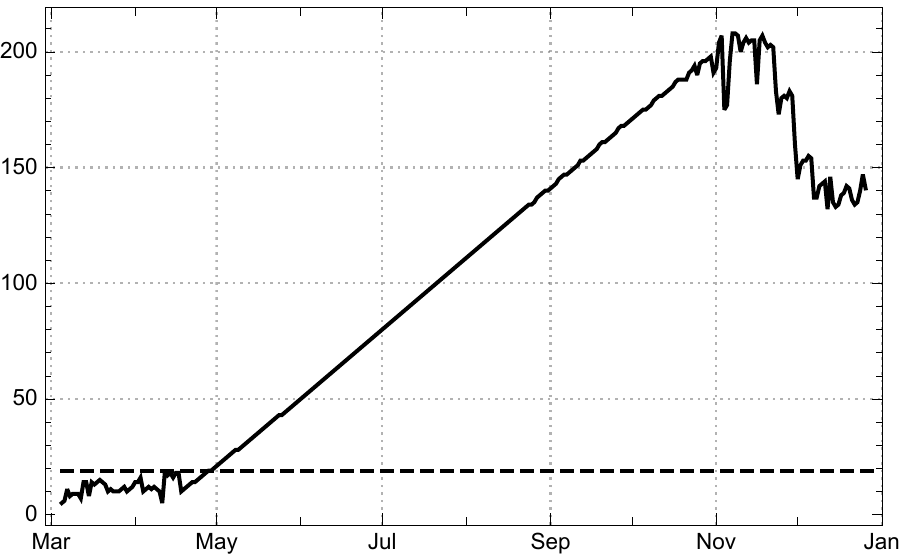} 
\includegraphics[scale=0.6]{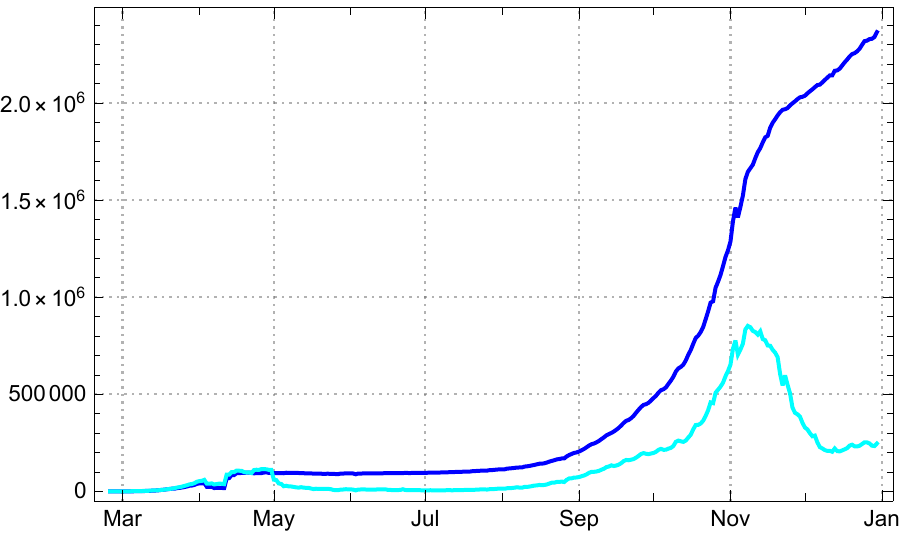}
\caption{Left: Daily values of the mean time of statistically actual   infection $\hat{q}(k)$ for  France.  Right: Comparison of reported infected $\hat{A}$ (dark blue) and $q$-corrected number ($q=19$) of recorded actual infected $\hat{A}_q$ (bright blue) from the JHU data in France. \label{fig q(k) F} }
\end{figure}

Another surprising feature of the French (JHU) statistics is an amazing increase in the number of days which  infected persons are  being counted as ``actual cases''. It starts close to 15, but shows a  monotonous increase until late October where a few downward outliers appear, before the curve turns moderately down in early November 2020 (fig. \ref{fig q(k) F}, left). 
 
As a consequence  the peak of the first wave in early April (clearly visible in the number of newly reported)  is suppressed in the curve of the actual infected; $\hat{A}(k)$ has no local maximum in the whole period of our report. It  even continues to rise, although with a reduced slope,  after  the second peak of the daily newly reported, $\hat{A}_{new}$, in early November. 
The  decrease of the slope of $\hat{A}$ starts  shortly after this peak,  accompanied by a  local maximum of the 
$q$-corrected values for the actual infected reaches $\hat{
A}_q$ (fig.  \ref{fig q(k) F}). Both effects seem to be  due to a  downturn  of $\hat{q}(k)$ . This extreme behaviour of the  data cannot be ascribed to  medical reasons;  quite  obviously it results from a high degree of uncertainty in   data taking and recording in the French health system.

\begin{figure}[h]
\includegraphics[scale=0.7]{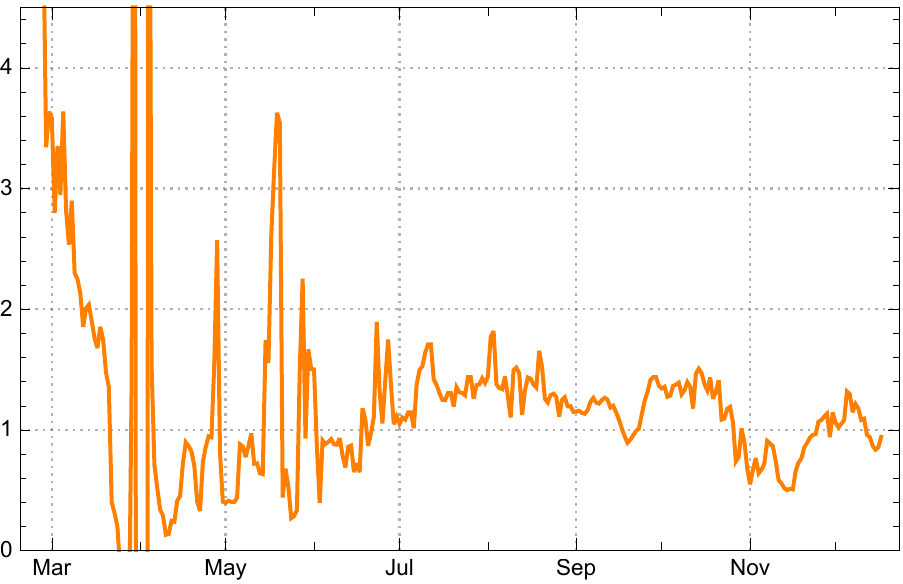} \includegraphics[scale=0.7]{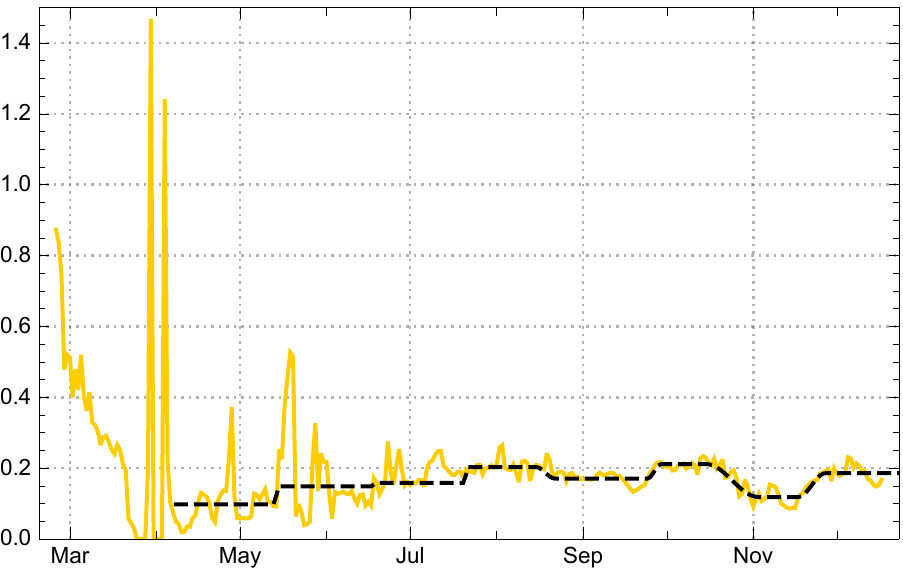} 
\caption{Left: Empirical reproduction rates $\hat{\rho}(k)$ for France. Right: Daily strength of infection are shown in figure \ref{fig F rho eta} for France (yellow) with model parameters $\eta_j$ in the main intervals  $J_j$ (black dashed). \label{fig F rho eta}}
\end{figure}

The  reproduction numbers   are shown in figure \ref{fig F rho eta}, left. For the determination of the daily strength of infection we have to fix a value for the dark factor. Lacking data from representative serological studies in France  we assume that it is  larger than in Switzerland and 
 choose as a reference value for the model $\delta=4$. With this value the determination of the     $\hat{\eta}_7(k)$ are given in figure \ref{fig F rho eta}, right, here again with dashed black markers for the  periods modelled by constancy intervals in our approach. 

The  markers of change times are here 
 $t_0 =$ 02/25, 2020, $t_1= $ 05/17, $t_2= $ 06/15, $t_3= $ 07/21, $t_4= $ 08/22, $t_5= $ 09/29, $t_6= $11/03, $t_7= $11/27, end of data  $t_{eod}= $ 12/30, 2020. In the country day count, $t_0 =1$ ($\sim 35$ in JHU day count), the main intervals are $J_0= [1, 82], \; J_1=[83, 110],\; J_2= [112, 146], \; J_3=[148, 173]; J_4= [180, 213], \; J_5=[218, 234],\; J_6= [253, 267],\; J_7= [277, 310] $.
The model strength of infection  $\eta_j$ and corresponding reproduction numbers $\rho_j$ for the main intervals are given in the following table. 
\vspace{0.2cm}
\begin{center}
\begin{tabular}{|l||c|c|c|c|c|c|c|c|}
\hline 
\multicolumn{9}{|c|}{$\eta_0$ and model $\eta_j$,  $\rho_j$ in intervals $J_j$   for France}\\
\hline
 & $a_0$ & $J_1$ & $J_2$ & $J_3$ & $J_4$ & $J_5$& $J_6$ & $J_7$ \\ 
\hline 
 $\eta_j$ & 2.008 &0.149 & 0.158 & 0.203 &  0.170  & 0.211  & 0.118 & 0.186 \\
 $\rho_j$  &  ---  & 1.03  & 1.09 & 1.39  &1.16 & 1.40& 0.71 & 1.07\\
\hline
\end{tabular}
\end{center}
\vspace{0.5em}

\begin{figure}[h]
\includegraphics[scale=0.8]{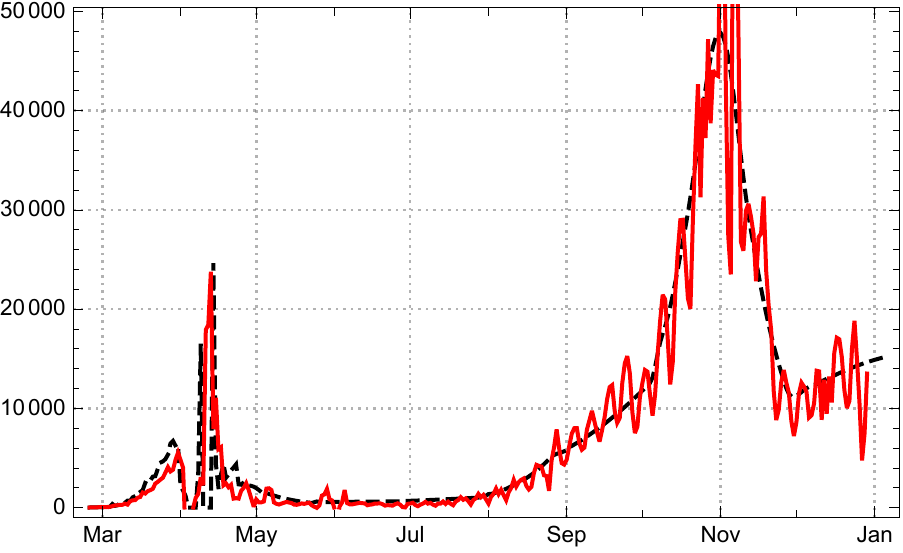}  \includegraphics[scale=0.8]{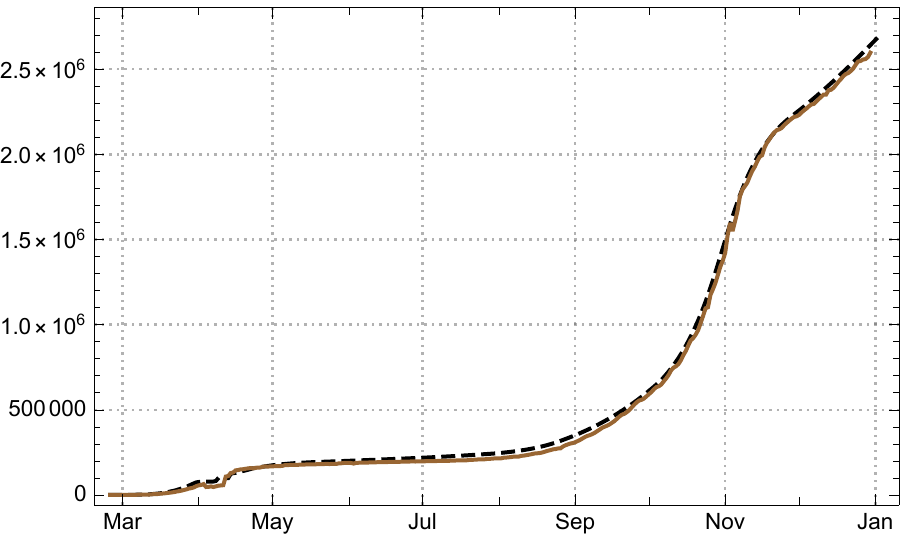} 
\caption{Left: Daily new reported number of infected for  France (7-day averages); empirical $\hat{A}_{new}$  solid red, model $A_{new}$ black dashed. Right: Total number of reported infected (brown);  empirical $\hat{A}_{tot}$ solid, model  $A_{tot}$ black dashed. \label{fig Ineu and Itot F}}
\end{figure}

\begin{figure}[h]
\includegraphics[scale=0.7]{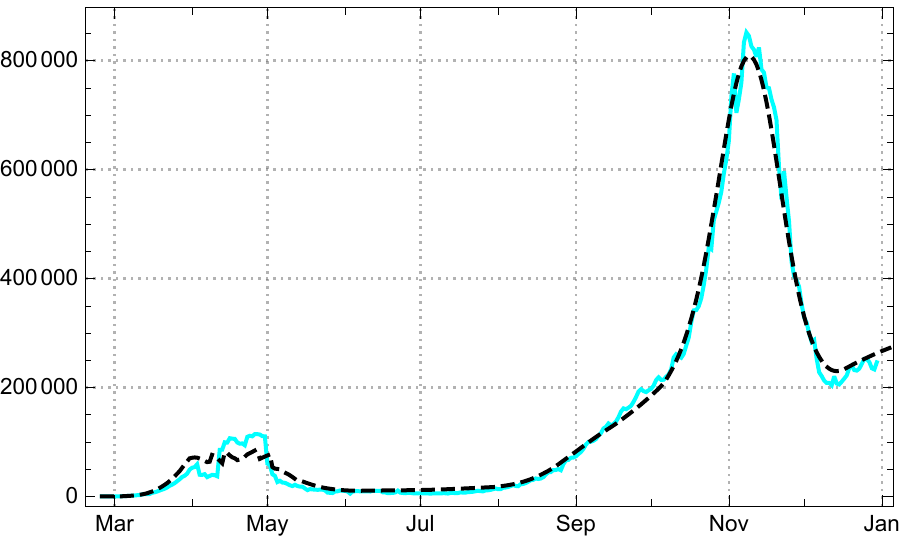} 
\caption{Empirical  values for statistically actual $q$-corrected cases   $\hat{A}_q(k)$  and the corresponding model values $A(k)$ (black dashed)  for  France. \label{fig Aq F} }
\end{figure}

\begin{figure}[h]
\includegraphics[scale=0.7]{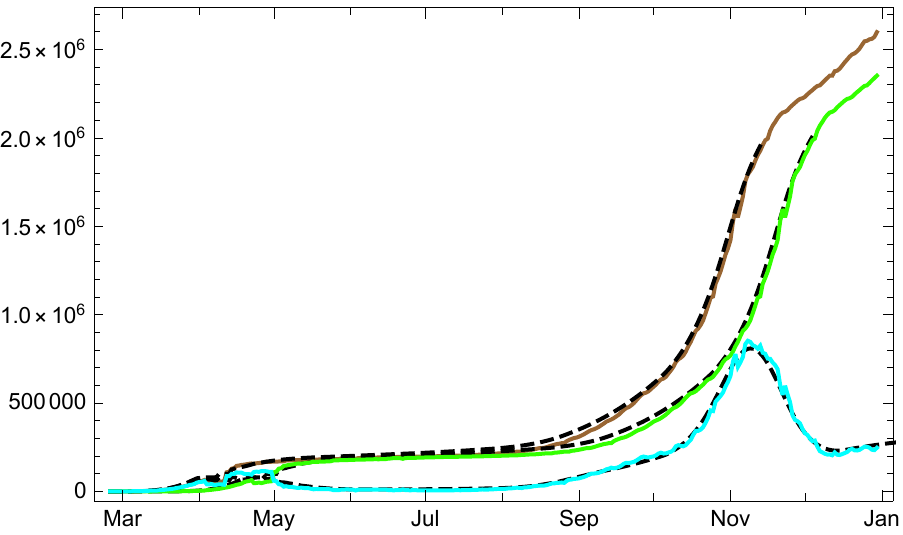}
\caption{Empirical data (solid coloured lines) and  model values (black dashed)  in France for  numbers of totally infected $\hat{A}_{tot}$ (brown), redrawn $\hat{R}$ (bright green), and  $q$-corrected actual  numbers $\hat{A}_q$ (bright blue). \label{fig 3 curves F} }
\end{figure}

An overall picture of the French development of new infections and total number of recorded cases is given in fig. \ref{fig Ineu and Itot F}. The so-called ``actual'' cases are  well modelled in our approach (fig.\ref{fig Aq F}), if the reference are the $q$-corrected numbers  $\hat{A}_q(k)$  of actual infected   (or if time dependent durations $q(k)$ (read off from the JHU data, $q(k)=\hat{q}(k)$) are used). For a combined graph of the 3 curves see fig.  \ref{fig 3 curves F}.

\pagebreak \vspace*{1em}
\subsubsection*{Sweden}
Sweden (population 10 M) has chosen a path of its own for containing  Covid-10, significantly  different from  most other European countries. In the first half year of the epidemic no general lockdown measures were taken; the  general strategy consisted in advising the population to reduce personal contacts and to go into self-quarantine, if somebody showed symptoms which  indicate an infection with the SARS-CoV-2 virus. One might assume that the number of undetected infected, the dark sector, could be larger than in other European countries. As we see below such a hypothesis is not supported by the analysis of the data  in the framework of our model.

\begin{figure}[h]
 \includegraphics[scale=0.7]{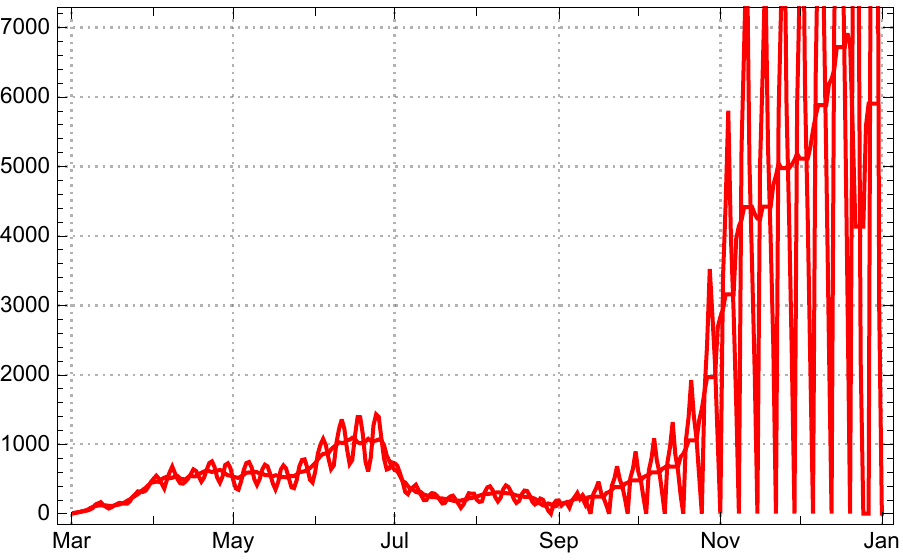} \quad 
\includegraphics[scale=0.7]{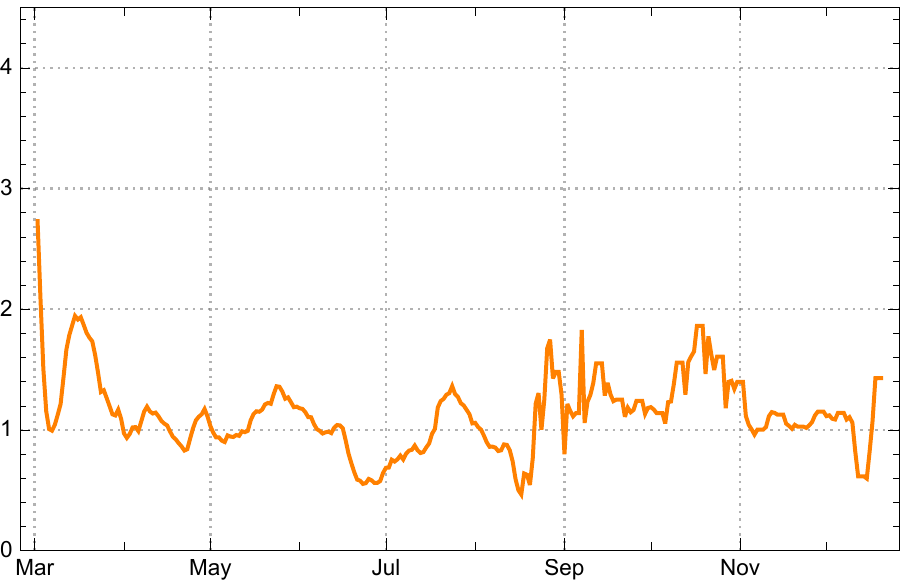}
\caption{Left: Daily new reported cases $\hat{A}_{new}(k)$ (JHU data)  for Sweden and  7-day sliding  averages $\hat{A}_{new,7}(k)$. Right: empirical reproduction rates $\hat{\rho}(k)$ for Sweden. 
\label{fig I-neu rho und ak S}}
\end{figure}

Under the conditions of the country (in particular the relative low population density in Sweden) the first wave of the epidemic was fairly well kept under control, if we abstain from discussing death rates like  in the rest of this paper. Once the initial phase was over (with  reproduction numbers already lower than in comparable countries, but still up to about $\rho \approx 2$), the reproduction rate was  close to 1   for about two weeks in late June,  and even lower in early July  (fig. \ref{fig I-neu rho und ak S}, right). In the second half of October 2020, however,  Sweden was hit by a second  wave  like all other European countries. After a sharp rise of the daily new infections in early October (fig.  \ref{fig I-neu rho und ak S}, left), the Swedish government decided to decree a (partial) lockdown. By this the reproduction number, which already in late August had risen to above 1.2 and went up to 1.45 in mid October, was brought down to  the former range ($\rho \approx 1$), although now on a much higher level of actual infected than during the first wave.

The higher the assumptions for the  dark sector,  the  larger  the  calculated values for $\hat{\eta}(k)$ on the basis of the same data, and vice versa. Figure \ref{fig S eta delta=0, 15, 25} shows the differences of $\hat{\eta}_7(k)$  in the case of Sweden under the   hypotheses  $\delta = 0, \; 4, \; 15, \; 25$. Until July/August 2020 the four curves show minor differences and indicate relative stable values for the infection strengths leading to reproduction rates close to 1. In September/October the values rose considerably; they went down after the October lockdown  only under the assumption of a small dark sector, $\delta=0$ or $4$, while for the  larger dark factors $\delta=15, \; 25$ the values of the daily strength of infection continues to increase. This seems  implausible.\footnote{Such a hypothesis for the dark sector could be explained only by a drastic and irresponsible change of contact behaviour of the Swedish population or an increased infectivity of the virus. Neither of these  explanations is  supported by  available empirical evidence.} 
We therefore  choose $\delta =4$ also for Sweden. 

\begin{figure}[h]
 \includegraphics[scale=0.7]{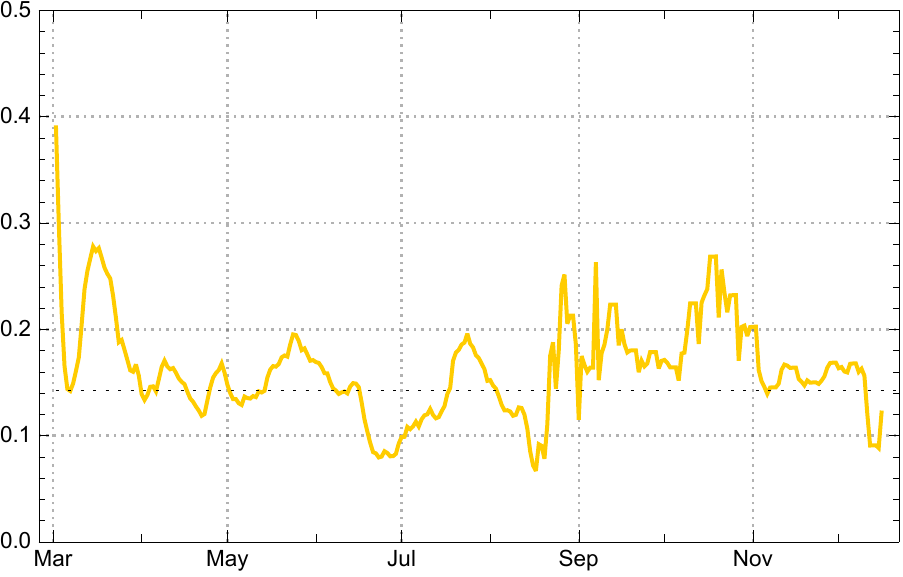}  \includegraphics[scale=0.7]{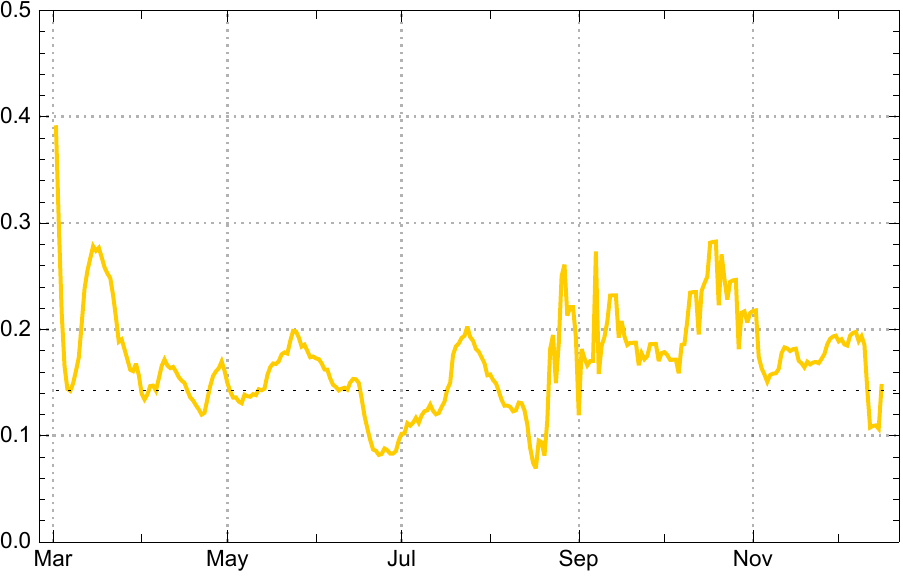} \\
\vspace{1.5em}
\includegraphics[scale=0.7]{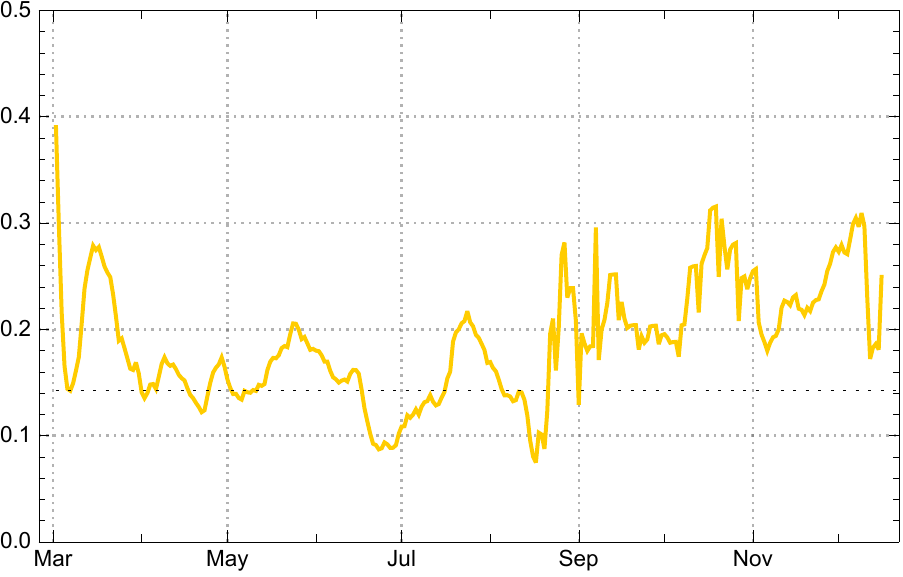} \includegraphics[scale=0.7]{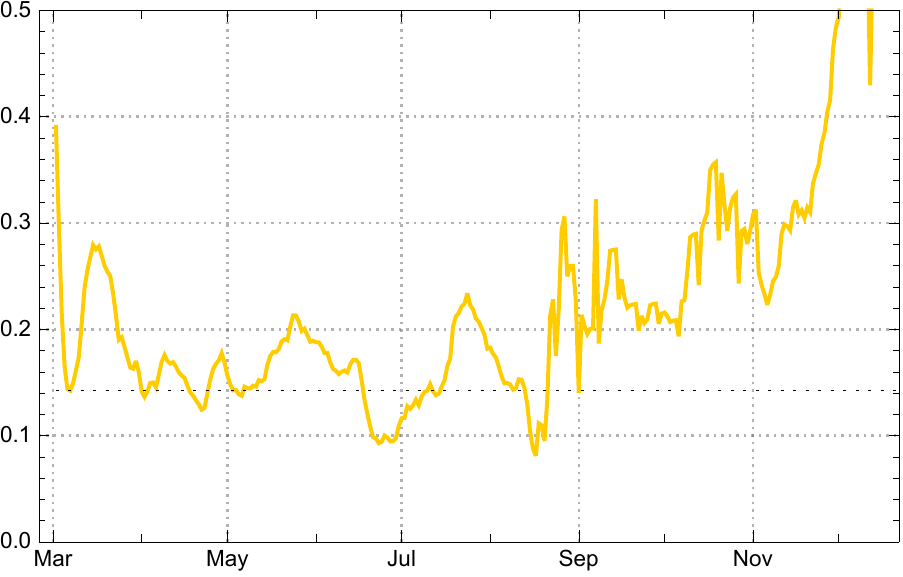} 
\caption{Daily strength of infection  (sliding 7-day averages) $\hat{\eta}_7(k)$ for Sweden, assuming different dark factors $\delta$. 
Top:   $\delta=0$ (left) and  $\delta=4$  (right). Bottom: $\delta=15$ (left) and  $\delta=25$  (right). \label{fig S eta delta=0, 15, 25}}
\end{figure}

 In our definition the new infections in Sweden ceased to be sporadic at $t_0 =$ 03/02, 2020, the day 41 in the JHU day count. After  a month of strong ups and downs of the strength of infection, the  approach of constancy intervals  gains traction;   with time markers of the main intervals  $t_1= $ 03/31, $t_2= $ 05/17, $t_3= $ 06/18, $t_4= $ 07/17, $t_5= $ 08/25, $t_6= $ 10/10, $t_7= $ 11/05, $t_8= $ 12/12, 
  end of data  $t_{eod}= $ 12/30, 2020. In the country  count the main intervals are $J_0= [1, 29], \; J_1=[30, 75],\; J_2= [77, 99], \; J_3=[109, 126]; J_4= [138, 169], \; J_5=[177, 218],\; J_6= [223, 214],\; J_7= [223, 284],\; J_8= [286, 303]    $.

\vspace{0.2cm}
\begin{center}
\begin{tabular}{|l||c|c|c|c|c|c|c|c|c|}
\hline 
\multicolumn{10}{|c|}{Model $\eta_0$ and $\eta_j$,  $\rho_j$ in intervals $J_j$   for Sweden}\\
\hline
 & $\eta_0$ & $J_1$ & $J_2$ & $J_3$ & $J_4$ & $J_5$& $J_6$& $J_7$& $J_8$ \\
\hline 
 $\eta_j$ & 0.659 & 0.146  & 0.174 & 0.095 &  0.149  & 0.177  & 0.232 & 0.178& 0.150  \\
 $\rho_j$  &  ---  & 1.01  & 1.20 & 0.65  & 1.00 & 1.19 & 1.54 & 1.13& 0.84\\
\hline
\end{tabular}
\end{center}
\vspace{0.5em}

\begin{figure}[h]
\includegraphics[scale=0.8]{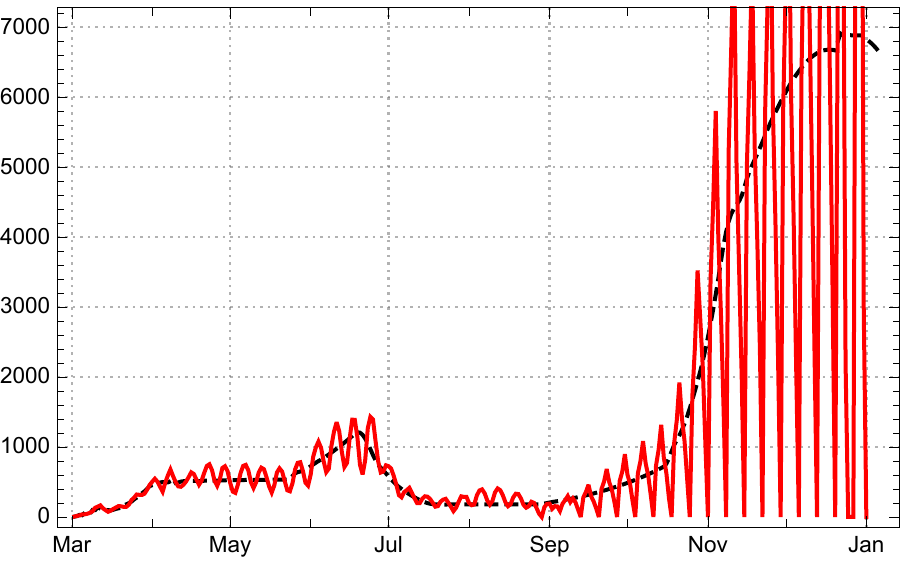}  \includegraphics[scale=0.8]{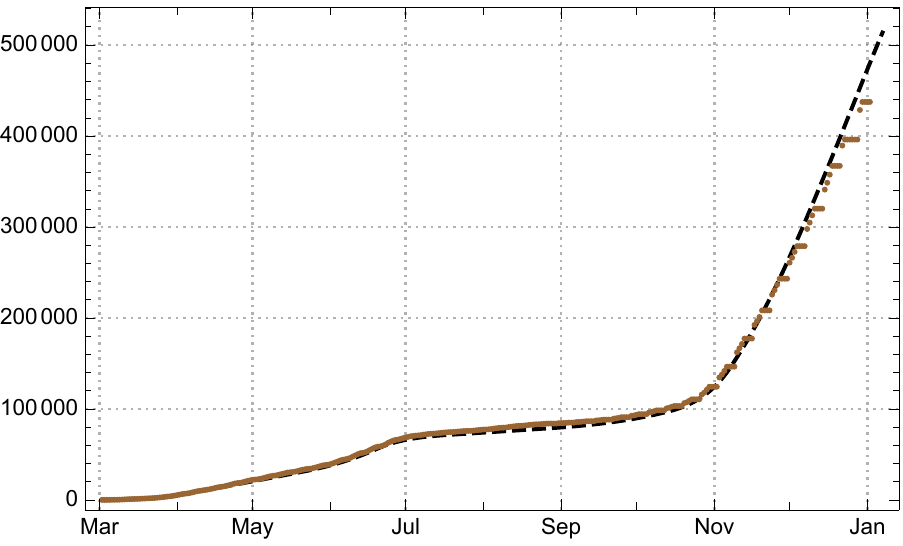} 
\caption{Left: Daily new reported infected  for the Sweden, empirical 3-day averages  $\hat{A}_{new,3}$  solid red; model $A_{new}$ black dashed. Right: Total number of reported infected (brown);  empirical $\hat{A}_{tot}$ solid, model  $A_{tot}$ dashed. \label{fig Ineu and Itot S}}
\end{figure}

Although one might want to refine the constancy intervals, already these  intervals  lead  to a fairly good model reconstruction of the mean motion of new infections and the total number of infected (fig. \ref{fig Ineu and Itot S}). Note that since early September the reported numbers of new infections show strong weekly oscillations between null  at the weekends and high peaks in the middle of the week.

\begin{figure}[h]
\includegraphics[scale=0.6]{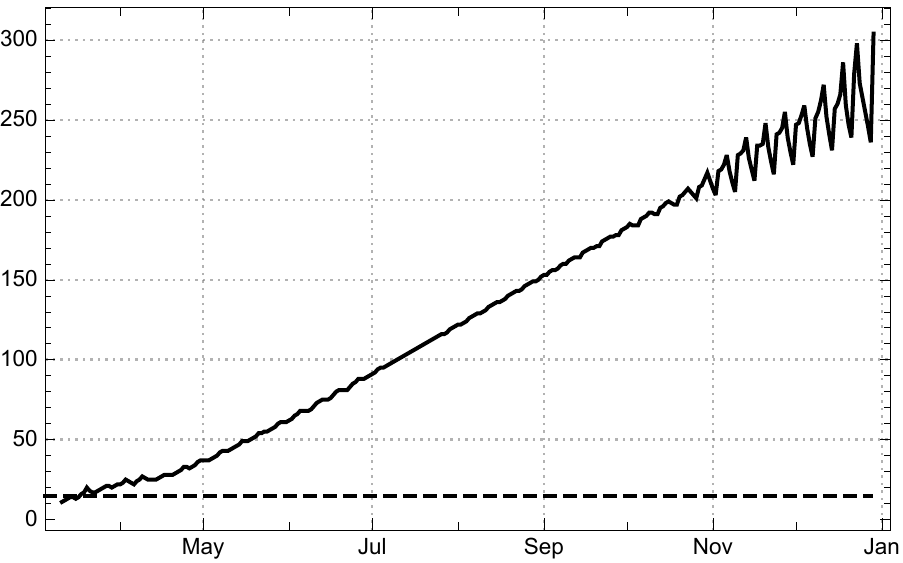} 
\includegraphics[scale=0.63]{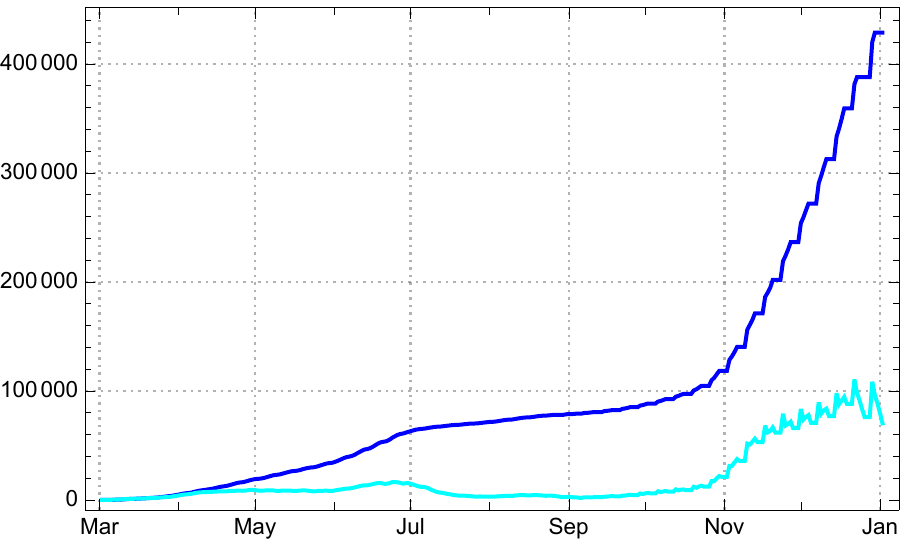}
\caption{Left: Daily values of the mean time of statistically actual   infection $\hat{q}(k)$ for  Sweden.  Right: Comparison of reported infected $\hat{A}$ (dark blue) and $q$-corrected number ($q=15$) of recorded actual infected $\hat{A}_q$ (bright blue) from the JHU data in Sweden. \label{fig q(k) S} }
\end{figure}

The JHU statistics does not register recovered people for Sweden at all; only deaths are reported. In consequence the usual interpretation of  (\ref{eq hat-A}) as  characterizing the  ``actual'' infected   breaks down for Sweden\footnote{The same holds for the UK.} and the estimation (\ref{eq hat q(k)}) for the mean duration of illness becomes meaningless  (fig. \ref{fig q(k) S}, left). An indication of the extent of reported actual diseased is given by the  $q$-corrected number $\hat{A}_q$ (same figure, right).

\begin{figure}[h]
\includegraphics[scale=0.6]{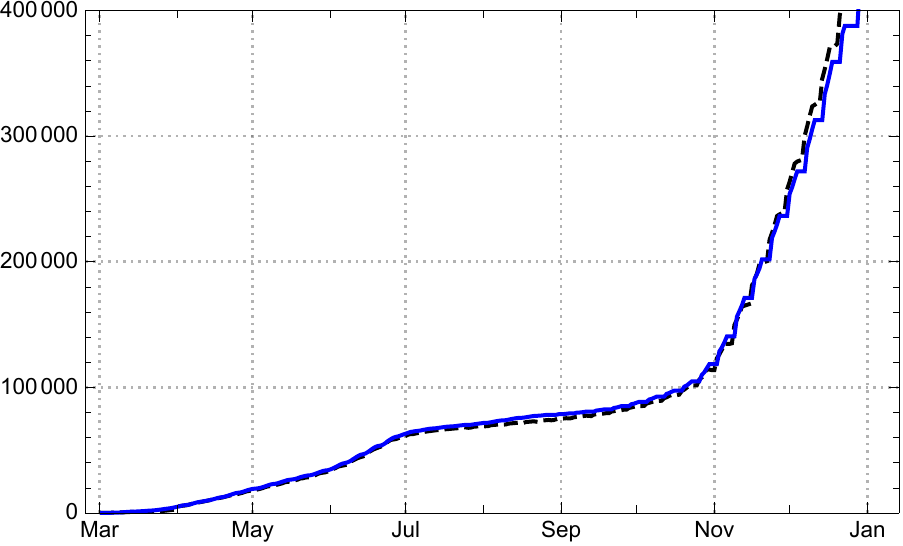} \includegraphics[scale=0.6]{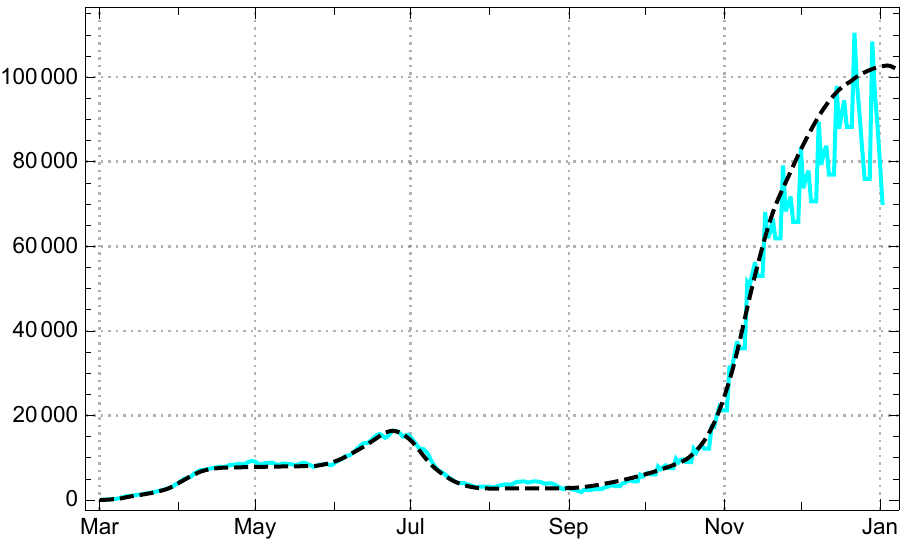}\\
\caption{Left: Empirical data $\hat{A}(k)$  for Sweden (blue) and model values $A(k)$ determined with time varying $q(k)=\hat{q}(k)$ (black dashed).
Right: Empirical  values,  $q$-corrected,   for statistically actual cases   $\hat{A}_q(k)$  and the corresponding model values $A_q(k)$ (black dashed).
 \label{fig I Iq S} }
\end{figure}

\begin{figure}[h]
\includegraphics[scale=0.58]{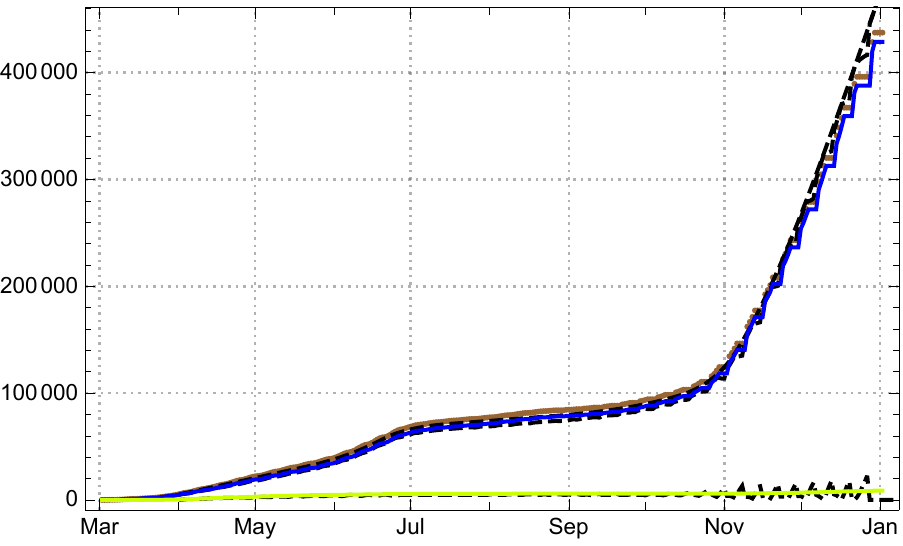} \quad\includegraphics[scale=0.58]{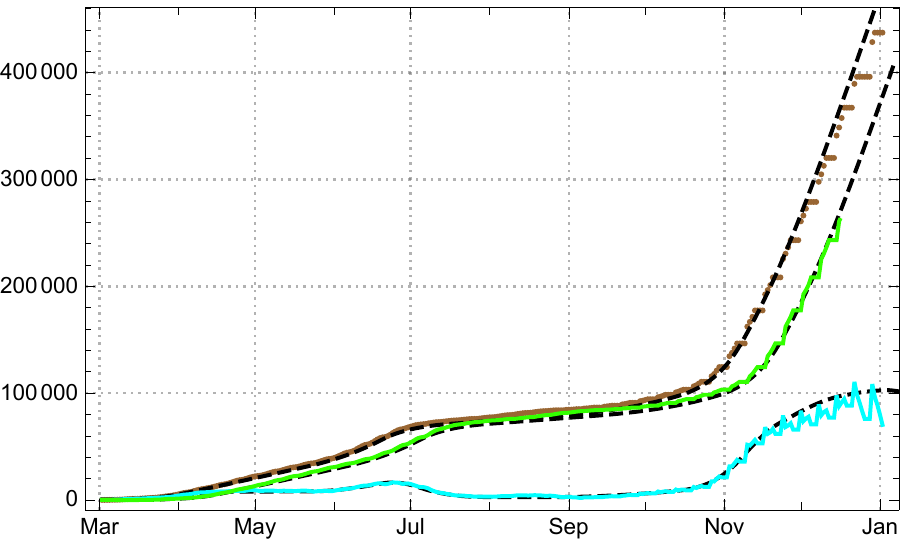}
\caption{Left: Numbers of totally infected $\hat{A}_{tot}$ (brown), reported redrawn $\hat{R}$ (bright green) -- here deaths only --  and the difference $\hat{A}$  (blue) for Sweden;  model values black dashed. Right: the same for $\hat{A}_{tot}$ (brown), $q$-corrected  $\hat{A}_q$ (bright blue)  and the  redrawn $\hat{R}_q$ (green) as the difference (model values black dashed). 
\label{fig 3 curves S} }
\end{figure}

In this sense, the synopsis with a collection of the ``3 curves'' can be given for Sweden like for any other country (fig. \ref{fig 3 curves S}). 
Of course  the model  reproduces the empirical values $\hat{A}(k)$ even in such an extreme case if the time dependent   empirical values of (\ref{eq hat q(k)}) are used for  the  the model calculation, $q(k)=\hat{q}(k)$, while it reconstructs the $q$-corrected numbers for the estimate of actually infected $\hat{A}_q(k)$ if the respective constant is used, here $q=15$ (fig. \ref{fig I Iq S}).



\pagebreak \vspace*{1em}
\subsection{The three most stricken regions:  USA,   Brazil, India}
In this section we give a short analysis of the course of the pandemic during 2020 for the three countries which have to bemoan the largest numbers of deceased and  huge numbers of infected (USA, Brazil, India). We expected higher dark factors $\delta$ than for the European countries discussed above and checked this expectation by the same heuristic approach as used for Sweden, i.e. by  a comparative judgement of the  changes of the empirically determined strength of infection $\hat{\eta}_7(k)$, which result from  different assumptions of the values for $\delta$. To our surprise we found no clear evidence for an overall larger dark factor for the USA than for the European countries and work here with $\delta =4$, while for India there are strong indications of a large dark factor which we estimate as $\delta\approx 35$ (see below). For Brazil we consider  $\delta \approx 8$ a reasonable choice for the overall development of the epidemic.   Don't forget, however, that all these are plausibility considerations which are not based on representative serological studies.  

\subsubsection*{USA}
\begin{figure}[h]
 \includegraphics[scale=0.9]{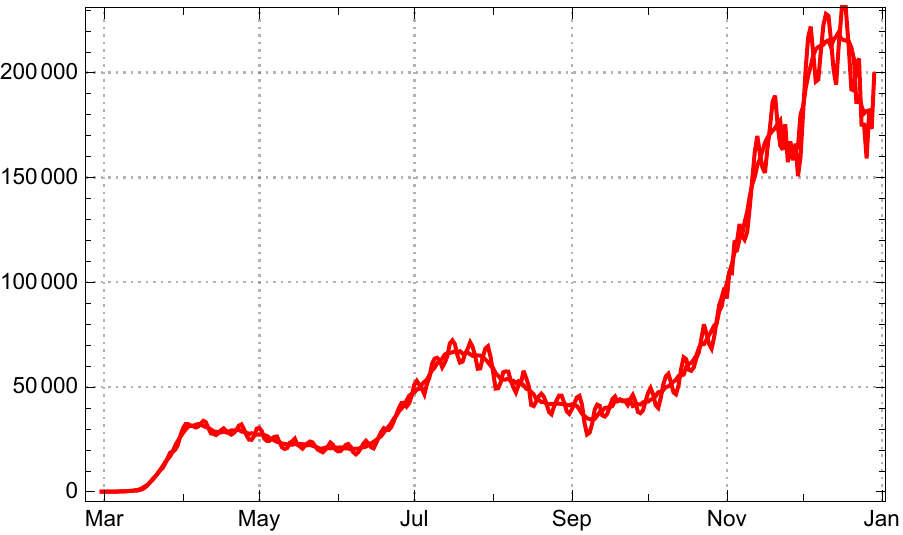}\\
\vspace{1.5em}
\includegraphics[scale=0.7]{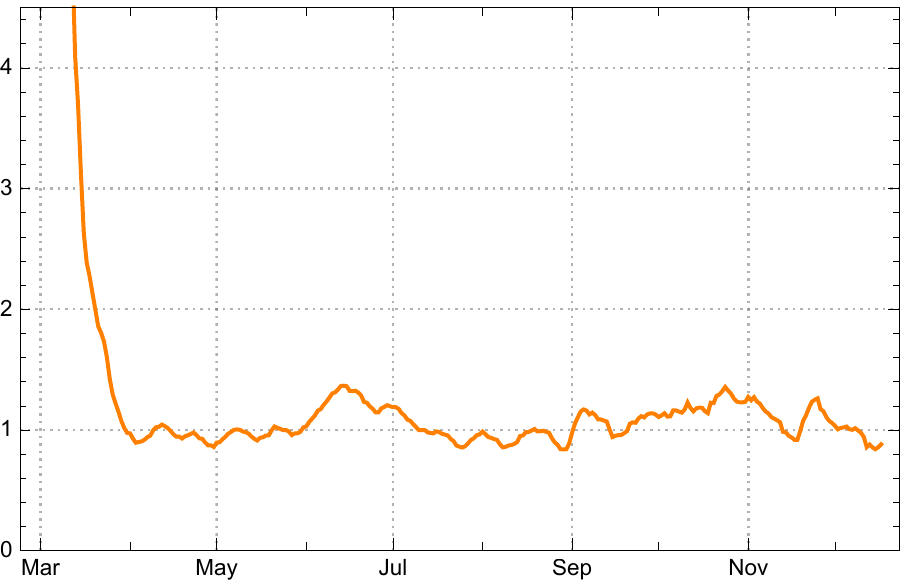} \includegraphics[scale=0.7]{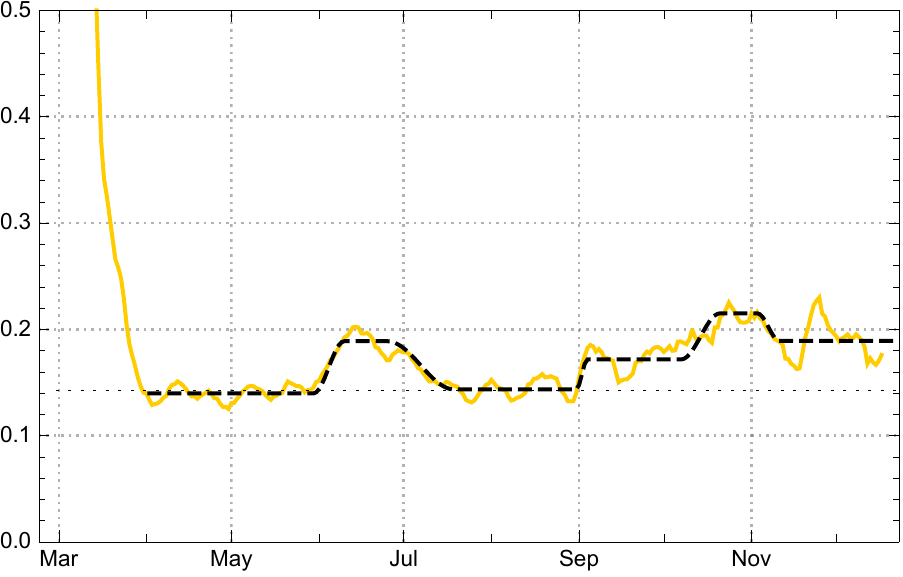} 
\caption{Top:  3-day and 7-day sliding  averages of daily new reported cases, $\hat{A}_{new, 3}(k)$, $\hat{A}_{new, 7}(k)$, for the USA.  Bottom, left: Empirical reproduction rates $\hat{\rho}(k)$ for the USA. Right: Daily strength of infection  $\hat{\eta}_7(k)$  for the USA (yellow), with $\delta=4$;  model parameters $\eta_j$ in the main intervals  $J_j$  black dashed. \label{fig I-neu rho und ak USA}}
\end{figure}

At the beginning of the pandemics the United States of America (population 
333 M) suffered a rapid rise of infections with an initial reproduction rate well above 5.   In early April 2020 this dynamics was broken and a slow decrease started for about 2 months. In mid  June a second wave  with an upswing for about a month and a reproduction rate shortly below   $1.3$ followed.  In late August and early September the subsequent downswing faded out. After  a short phase of indecision the beginnings of a third wave became clearly visible; it lasted until (at least) mid December.  Figure \ref{fig I-neu rho und ak USA} shows  the daily new infections,  the reproduction rate $\hat{\rho}(k)$ determined from the JHU data  and the strength of infection $\hat{\eta}_7(k)$ assuming $\delta=4$ .

\begin{figure}[h]
 \includegraphics[scale=0.5]{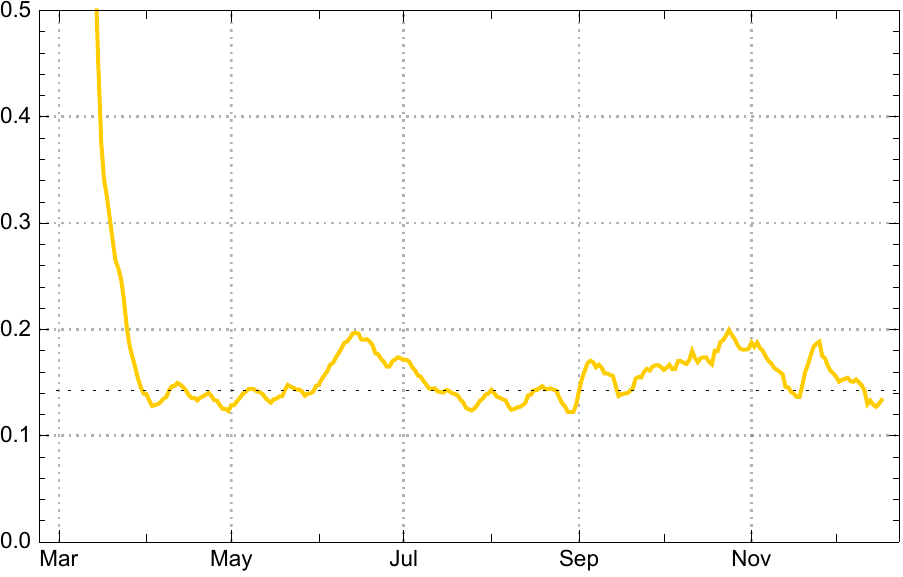}  \includegraphics[scale=0.5]{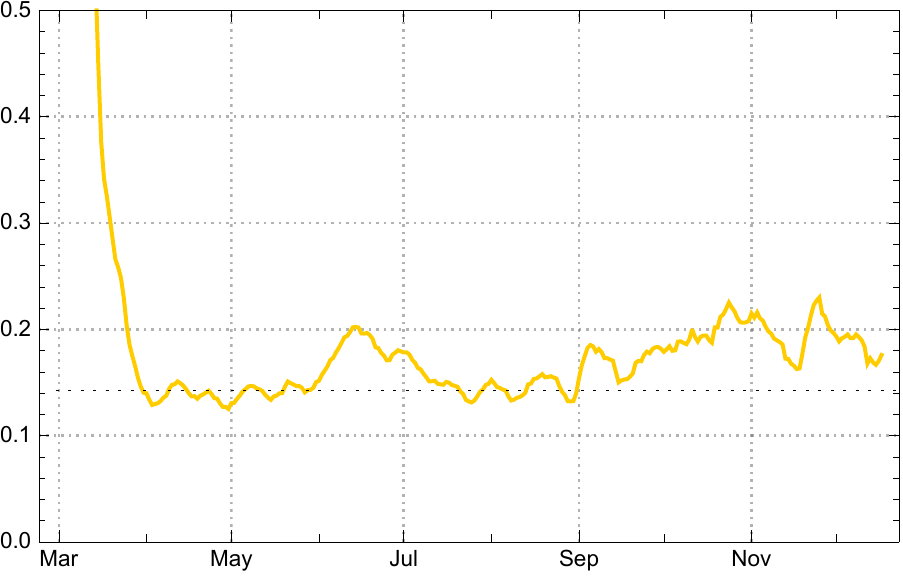} 
\includegraphics[scale=0.5]{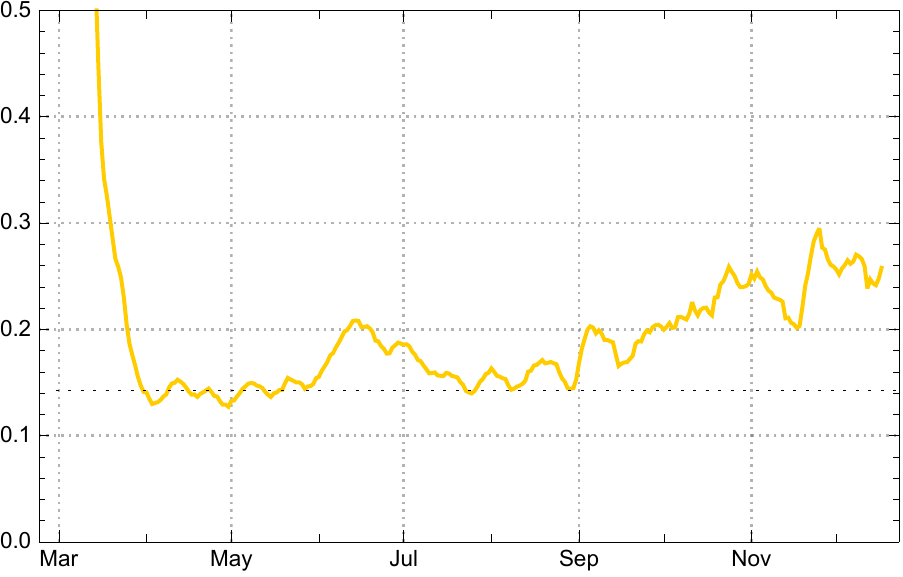} 
\caption{Daily strength of infection  (sliding 7-day averages) $\hat{\eta}_7(k)$ for USA, assuming different dark factors:  
  $\delta=0$ (left),  $\delta=4$  (middle), and $\delta=8$ (right). \label{fig US eta delta=0, 4, 8}}
\end{figure}

Figure \ref{fig US eta delta=0, 4, 8} displays three  variants of $\hat{\eta}_7(k)$ for $\delta= 0, \; 4, \; 8$. 
The third one shows an  implausible increase for the strength  of  infection at the end of the year, which would seem reasonable only if one of the  new,   more aggressive mutants of the virus had started to spread in the USA in September 2020 already. Without further evidence we do not assume such a strong case.  As $\delta=0$  contradicts all evidences collected on unreported infected,  we  choose $\delta=4$  for the SEPAR$_d$ model of the USA. Also here we find a moderate increase of the mean level of the $\hat{\eta}_7$.
To judge whether this may be due to the inconsiderate behaviour of part of the US population (supporters of the outgoing president) or  the first influences of a virus mutation and/or still other factors is beyond the scope if this paper and our competence.

In section \ref{section data} it was already noted that the estimation of the time of sojourn in the ``actual'' state of infectivity,  suggested by the statistics for the USA, leads to surprising effects. It rises from about 15 in March 2020 to above 100 in early November, with a moderate platform in between; then it starts falling,before it makes an abrupt jump  (fig \ref{fig US q(k) and A, Aq}, left). The jump of $\hat{q}(k)$ is an artefact of  a  change in the  record keeping: from December 14, 2020 onward the reporting of data of recovered people  was given up ($Rec(k)=0$ for date of $k$ after  2020/12/14). Of course this jump is also reflected in the numbers of recorded actual infected $\hat{A}(k)$ (same figure, right).

\begin{figure}[h]
\includegraphics[scale=0.6]{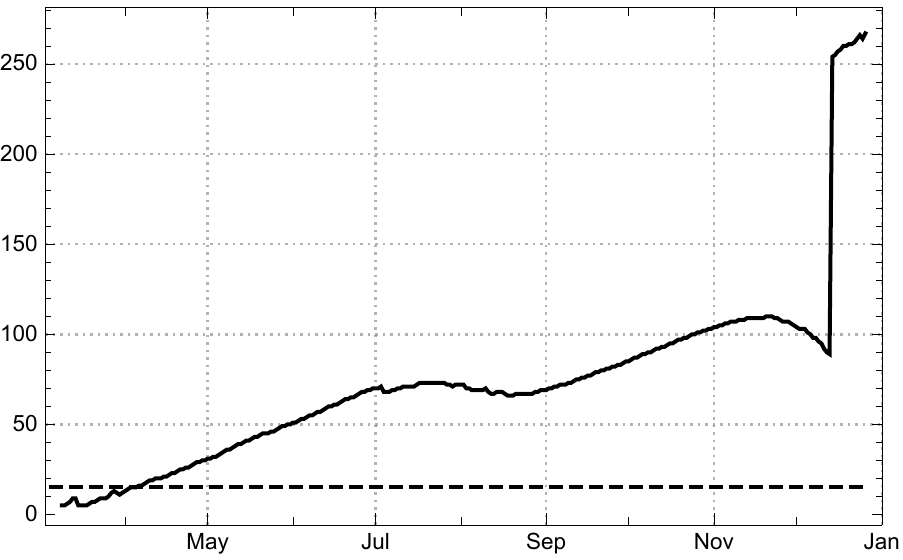} \quad \includegraphics[scale=0.6]{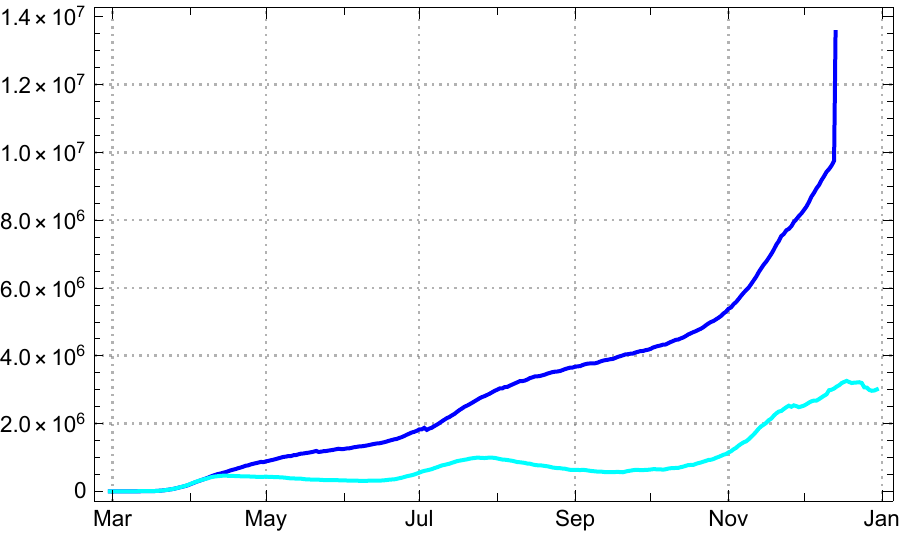}
\caption{Left: Daily values of the mean time of statistically actual (``active'')  infection $\hat{q}(k)$ for  USA.  Right: Empirical data (JHU) for statistically actual cases $\hat{A}(k)$ (dark blue) versus $q$-corrected ones, $q=15$, $\hat{A}_q(k)$ (bright blue)  for  the USA.  \label{fig US q(k) and A, Aq} }
\end{figure}

The main (constancy) intervals of the 
 model are visible  in the graph of  the  daily strength of infection  $\hat{\eta}_7(k)$ in fig. \ref{fig I-neu rho und ak USA}, bottom right. 
The dates of the time marker  between the intervals are: 
$t_0$= 02/29 2020; $t_1=$ 04/02, $t_2=$ 06/10, 
 $t_3=$ 07/19, $t_4=$ 09/04, $t_5=$ 10/21 $t_6=$ 11/11,   end of data $t_{eod}=$ 12/30.
Expressed in terms of the country  day count with $k_0=1$ ($\sim 39$
in the JHU day count) 
the main intervals for the USA are:
$ J_0=[1 , 33], \, J_1=[34, 92], \, J_2 = [103, 118], \,J_3 = [142, 185], \,
J_4 = [189, 222] ,$ $\, J_5 = [236, 249], \, J_6 = [257, 322] 
$.

The start parameter for the strength of infection $\eta_0$ and a slightly adapted choice of parameter values $\eta_j$ inside the 1-sigma domain of  the respective interval $J_j$ are given by the following table.

\vspace{0.2cm}
\begin{center}
\begin{tabular}{|l||c|c|c|c|c|c|c|}
\hline 
\multicolumn{8}{|c|}{Model $\eta_0$ and  $\eta_j$,  $\rho_j$ in intervals $J_j$   for the USA}\\
\hline
 & $\eta_0$  & $J_1$ & $J_2$ & $J_3$ & $J_4$ & $J_5$& $J_6$\\
\hline 
$\eta_j$ &-1.011 &  0.140  & 0.189 &  0.143  & 0.172  & 0.215  & 0.194 \\
$\rho_j$ & ---  &    0.97   & 1.28 & 0.94  & 1.08 &  1.30 & 1.12  \\
\hline
\end{tabular}
\end{center}
\vspace{0.5em}
\noindent

\begin{figure}[h]
 \includegraphics[scale=0.8]{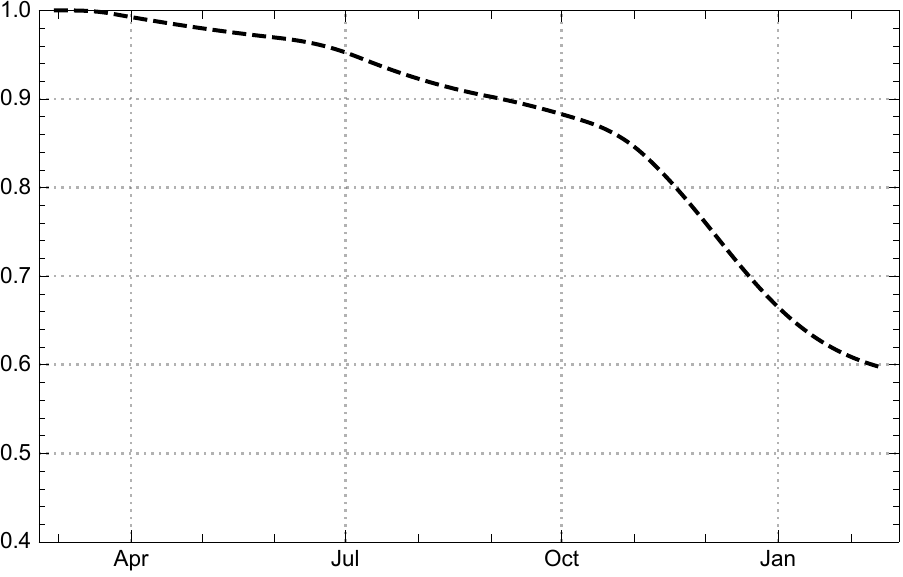} 
  \caption{Ratio of susceptibles $s(k)$ (model values) for the USA, $\delta=4$.  \label{fig srek US}}
\end{figure}
\noindent 
Here, as for the other countries, the $\rho_j$ denote  the model reproduction rates at the beginning of the respective intervals. Assuming the dark factor $\delta=4$, the effective reproduction  rate $\rho(k)$ changes  considerably  from the beginning of $J_6$ until the end of the year,  $s(11/11\, 2020)=0.82$ to $s(12/31)=0.67$, (cf.fig. \ref{fig srek US}).

\begin{figure}[h]
 \includegraphics[scale=0.8]{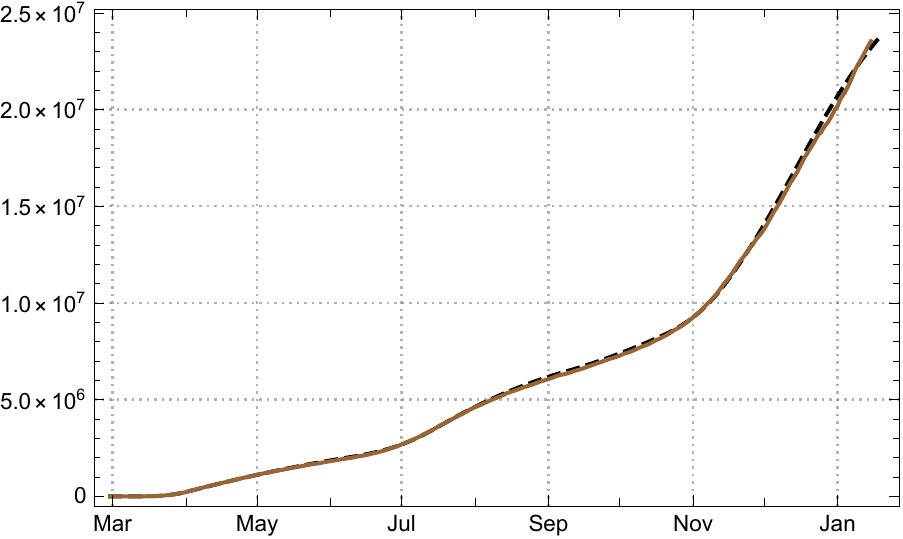} 
 \includegraphics[scale=0.8]{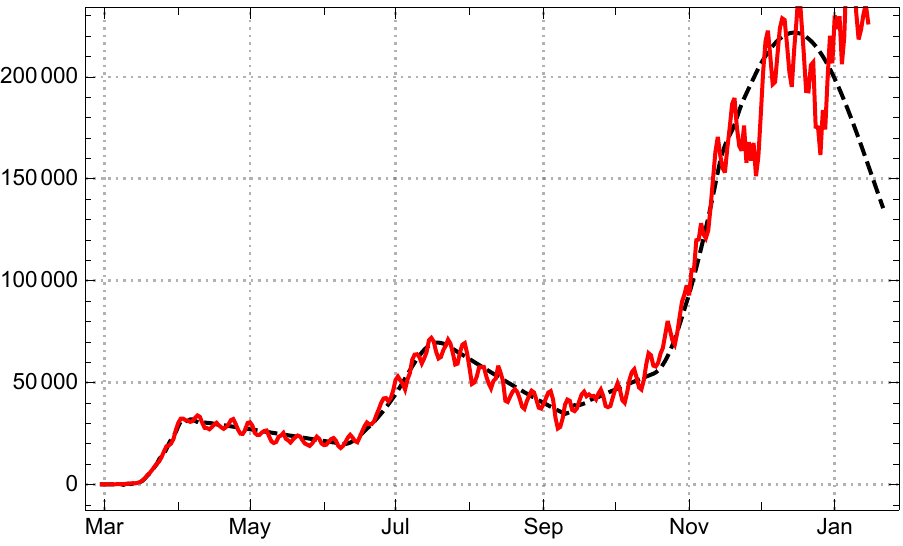}  
\caption{Left: Total number of infected (brown);  empirical $\hat{A}_{tot}$ solid, model  $A_{tot}$ dashed. Right: Daily newly reported for the USA, 7-day averages (red); empirical $\hat{A}_{new}$  solid red, model $_{new}$ black dashed ($\delta=4$).  \label{fig Ineu and Itot US}}
\end{figure}

In consequence, the reproduction rate at the end of the year is down to  $\rho(12/31)=0.86$; and the newly reported infected are expected to reach a peak in late December (fig.  \ref{fig Ineu and Itot US}, right).  Apparently this  does not agree with the data, while  the total number of infections  is well  reproduced by   the SEPAR$_d$ model (same figure, left).  The difference for the new infected may be an {\em indication that our estimate of the dark factor $\delta$ is wrong    or the strength of infection at the beginning of the new year 2021 is drastically boosted}, e.g., by a rapid spread of a new, more aggressive mutant of the virus.


\begin{figure}[h]
 \includegraphics[scale=0.8]{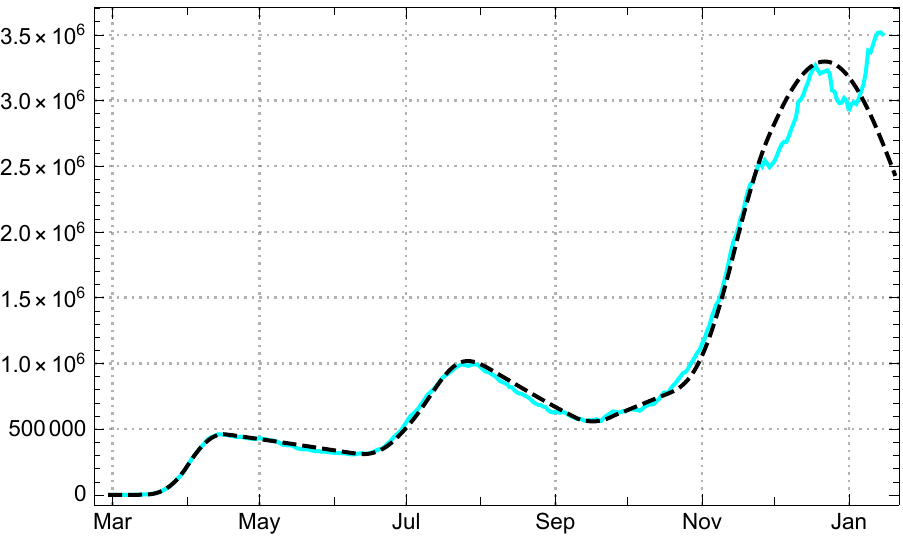} 
  \caption{Number of $q$-corrected actual infected   for the USA;  empirical $\hat{A}_q$ (bright blue) and model values $A_q$ (black dashed).  \label{fig Aq US}}
\end{figure}

As noted above, the recovered people are 
documented  with increasingly large time delays in the records for the USA. Therefore it seems preferable to compare the model value of   actually infected, $A(k)$,  with  $\hat{A}_q(k)$  rather than with  $\hat{A}(k)$ (fig. \ref{fig Aq US}). The empirical determined values of $\hat{A}_q(k)$ are shown in  bright blue in the  figure. They are marked by three local maxima indicating the peak values of  three waves of the epidemics in the USA.   These peaks are blurred in the graph of $\hat{A}$ because too many of the effectively redrawn are dragged along as acute cases in the statistics. Due to under-reporting  at the end of the year, the  local extremum in December  may  be fuzzier than it appears here. But keep in mind that the SEPAR$_d$ model with $\delta=4$  {\em predicts a local maximum} inside the interval $J_6$, if the contact behaviour and the resulting strength of infectivity $\eta_6$ do not change considerably.

\begin{figure}[h]
\includegraphics[scale=0.7]{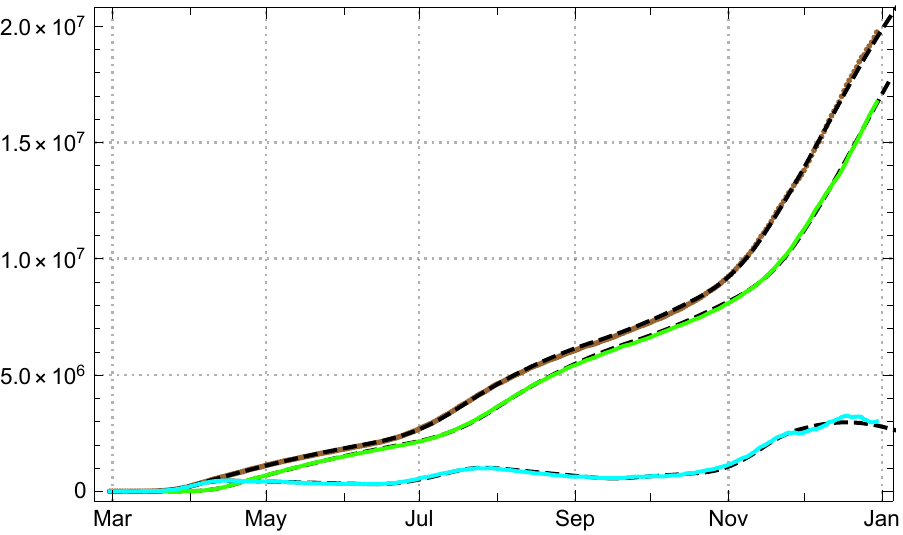}\quad 
\includegraphics[scale=0.7]{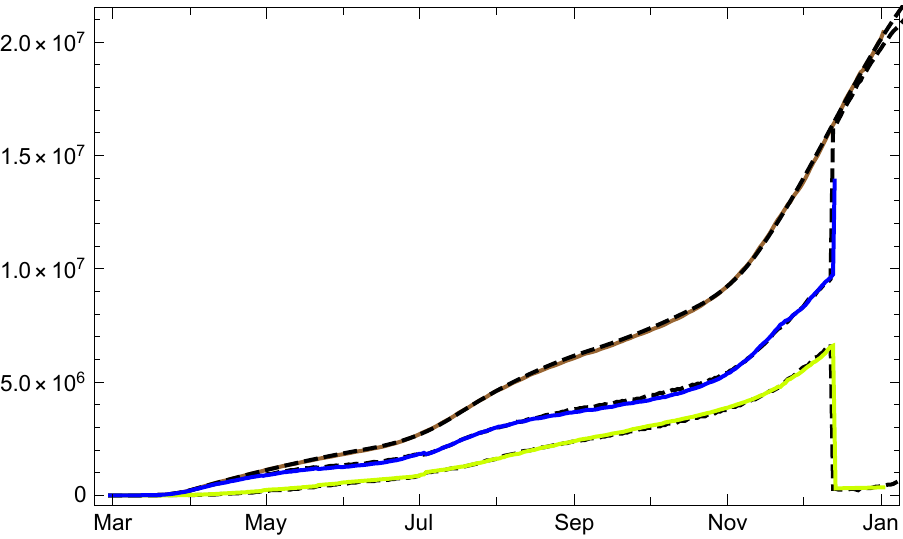} 
\caption{Left:  numbers of totally infected $\hat{A}_{tot}$ (brown), redrawn $\hat{R}_q$ (green),  and $q$-corrected actual  numbers $\hat{A}_q$  (bright blue).   Right:   Numbers of totally infected $\hat{A}_{tot}$ (brown), reported redrawn $\hat{R}$ (brigth green),  and  actual  numbers $\hat{A}$ of the statistic (blue)  for the USA.  Empirical data (solid coloured lines) and  model values (black dashed). \label{fig 3 curves USA} }
\end{figure}

 Figure \ref{fig 3 curves USA}, left, 
shows the 3 curves for the model values (black dashed) of the total number of infected $A_{tot}$, the actually infected  $A$  in terms of estimates with constant   $q=15$, and the redrawn $R$,  all of them compared with the corresponding values $\hat{A}_{tot}(k),\; \hat{A}_q(k), \;\hat{R}_q(k)$ determined from the  JHU data (coloured solid lines).
 By using  time dependent values $q(k)$, like, e.g., in the case of Germany,     SEPAR$_d$  is able to  
model the statistically ``actual'' cases also here  (fig.  \ref{fig 3 curves USA}, right). 
Because of the growing fictitiousness of the numbers $\hat{A}(k)$ in the case of the USA  we prefer, however,  to look at the corrected values $\hat{A}_q(k)$, as  stated already.


\newpage \vspace*{1em}
\subsubsection*{Brazil}
\begin{figure}[h]
 \includegraphics[scale=0.7]{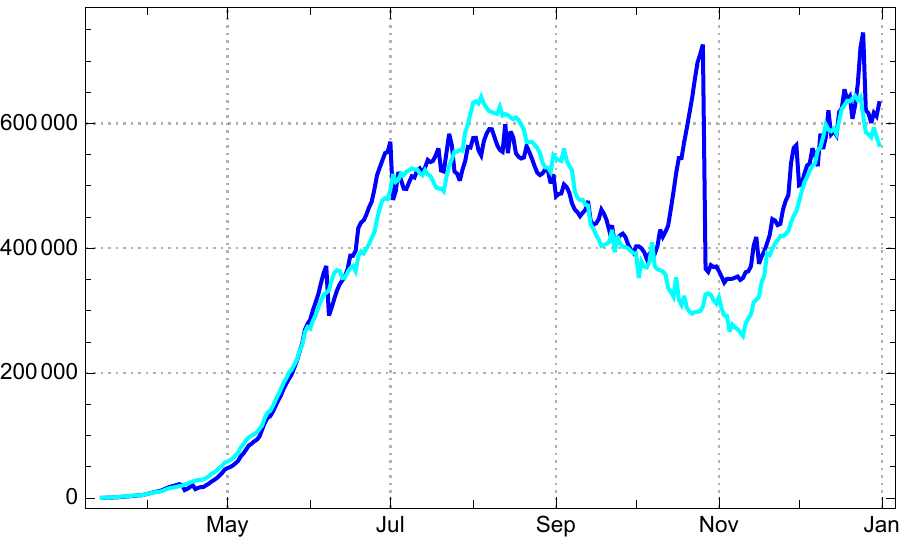} \hspace{1em} \includegraphics[scale=0.7]{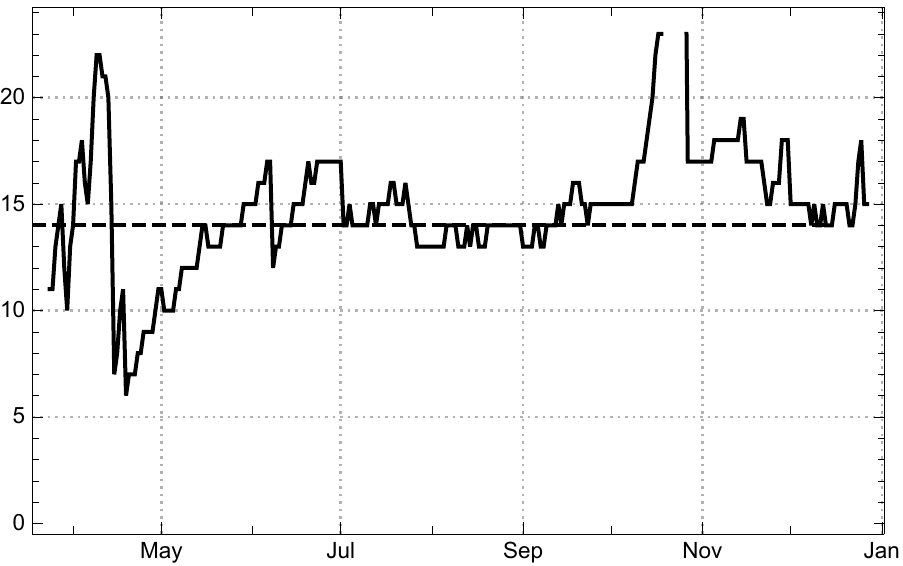} 
\caption{Left: Acute infected $\hat{A}(k)$ recorded by the statistics (dark blue) in comparison with $q$-equalized number $\hat{A}_q(k)$ (bright blue) for Brazil ($q=14$).   Right: Empricial estimate $\hat{q}(k)$ of mean duration of active infective according to the statistics for Brazil.  \label{fig I Iq Br} }
\end{figure}

The documentation of newly reported became non-sporadic in Brazil (population 212 M) at  $t_0=$ March 15, 2020. 
A first peak for the officially recorded  number of actual infected $\hat{A}(k)$ 
was surpassed in early August 2020 with a decreasing phase until late October, after which  a second wave started (fig. \ref{fig I Iq Br} left). In  contrast to the USA  we find here a comparatively stable estimate for the mean time $\hat{q}(k) \approx 14$ (fig. \ref{fig I Iq Br}  right). 

\begin{figure}[h]
\includegraphics[scale=0.7]{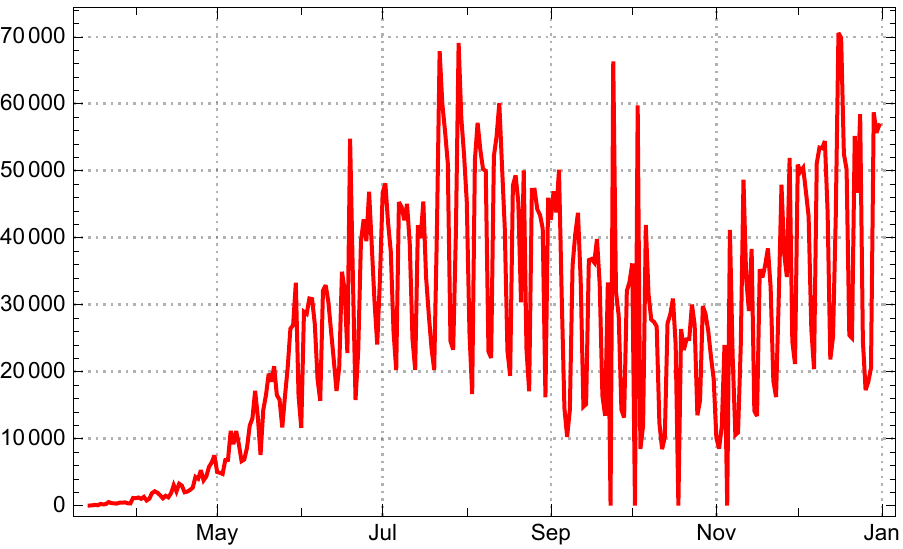} \includegraphics[scale=0.7]{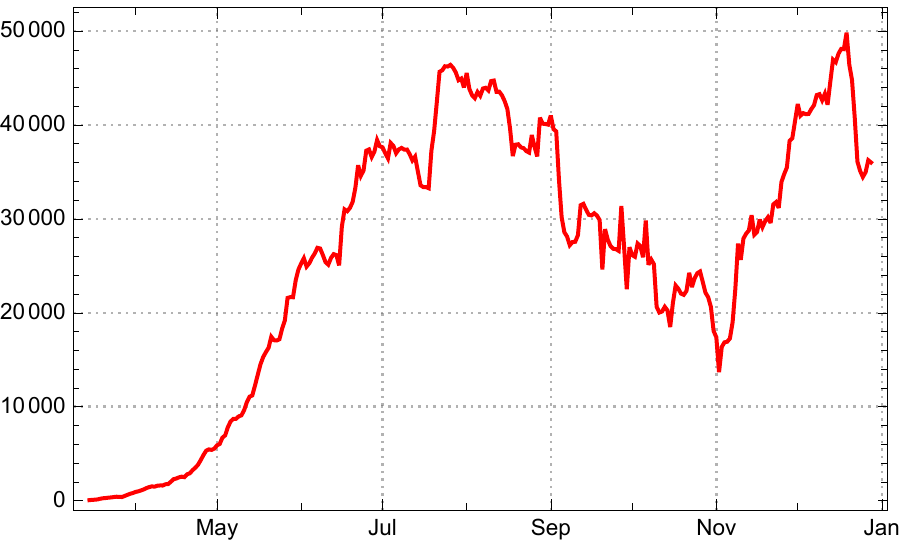}\\84
\caption{Daily varying numbers $\hat{A}_{new}(k)$ (left)versus 7-day sliding averages $\hat{A}_{new,7}(k)$ (right) of new  infections for Brazil. \label{fig I-neu Br}}
\end{figure}

The outlier peak of $\hat{A}(k)$ about October 25  appears  also as  an exceptional peak in the $\hat{q}(k)$. Apparently it is due to an interruption of  writing-off actual infected to the redrawn (compare fig. \ref{fig 3 curves Br}). 
Up to this exceptional phase there  is a close incidence between the $\hat{A}(k)$ and the $q$-corrected number $\hat{A}_q(k)$. The minimum of the mean square difference is  acquired for $q=14$.  The numbers of newly reported $\hat{A}_{new}(k)$ show strong daily fluctuations which are  smoothed by the 7-day sliding average $\hat{A}_{new,7}(k)$ (fig. \ref{fig I-neu Br}).

\vfill


\begin{figure}[h]
 \includegraphics[scale=0.6]{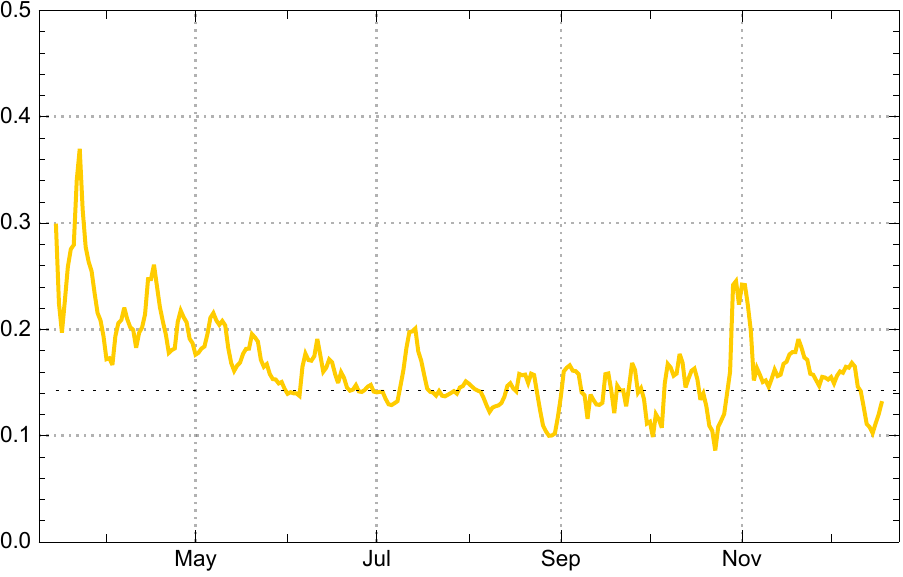}  \includegraphics[scale=0.6]{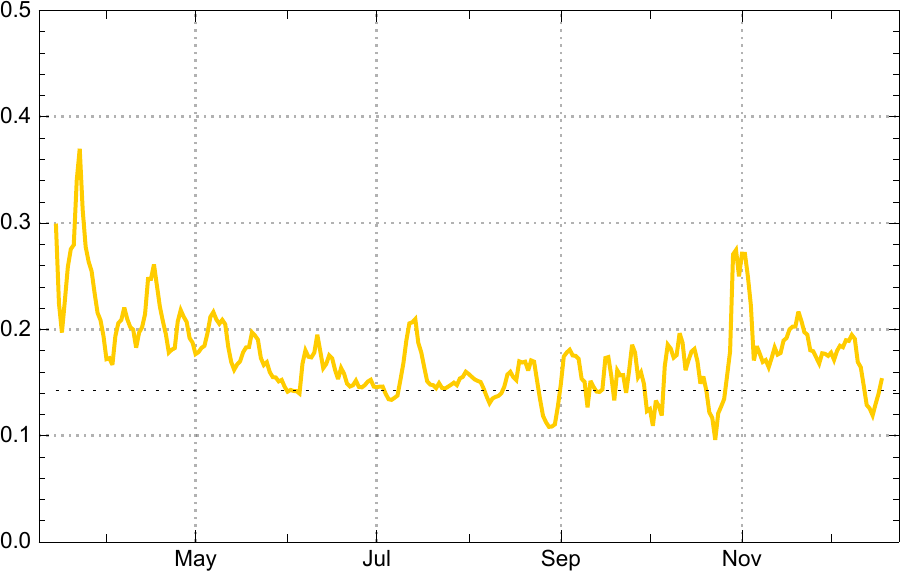} \\
\vspace{1.5em}
\includegraphics[scale=0.6]{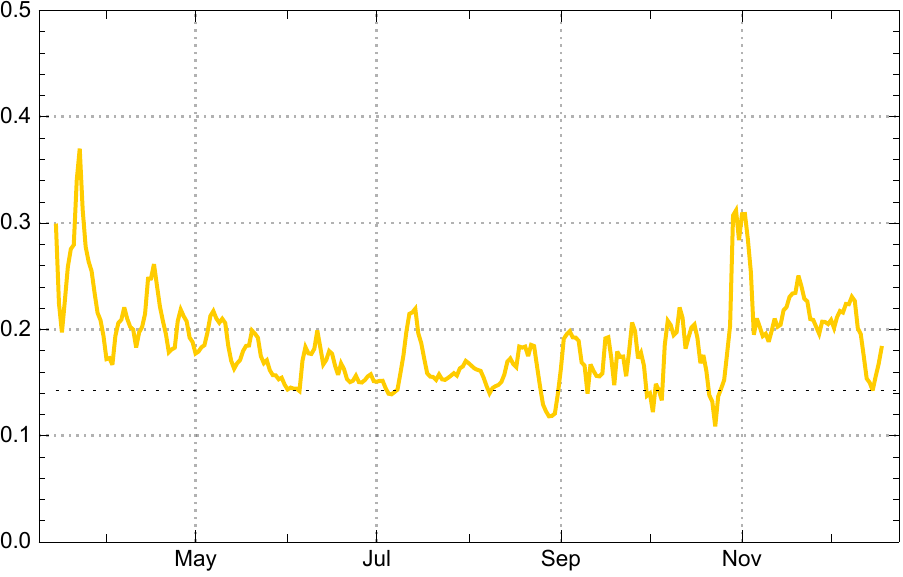} \includegraphics[scale=0.6]{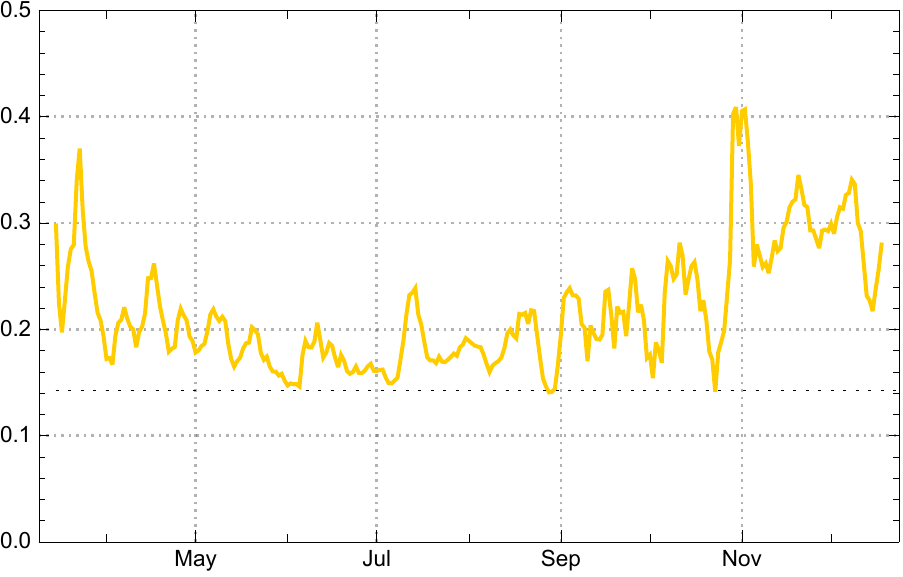} 
\caption{Daily strength of infection  (sliding 7-day averages) $\hat{\eta}_7(k)$ for Brazil, assuming different dark factors. 
Top:   $\delta=0$ (left) and  $\delta=4$  (right). Bottom: $\delta=8$ (left) and  $\delta=15$  (right). \label{fig Br eta delta=0, 4, 8, 15}}
\end{figure}
A comparison of different strengths of infection $\hat{\eta}_7(k)$  indicates    a value between $\delta=0$ and $\delta=8$ as a plausible choice (fig. \ref{fig Br eta delta=0, 4, 8, 15}). For higher values an unnatural increase of the infection strength would appear close to the end of the year (if not due to a new mutant. 
As we assume that in Brazil the dark factor is higher than in European countries, we  use  $\delta=8$ as  a plausible model hypothesis. 

 \begin{figure}[h]
\includegraphics[scale=0.7]{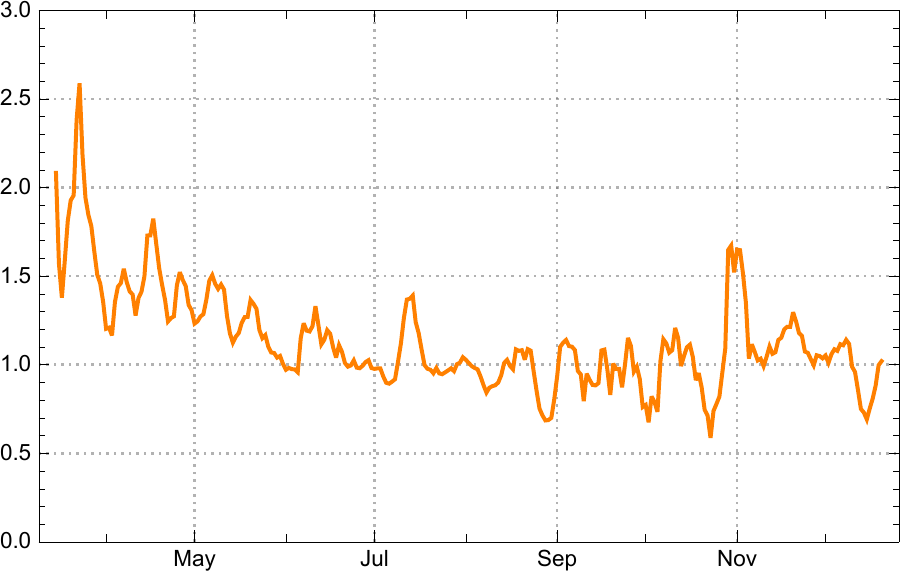} 
\includegraphics[scale=0.7]{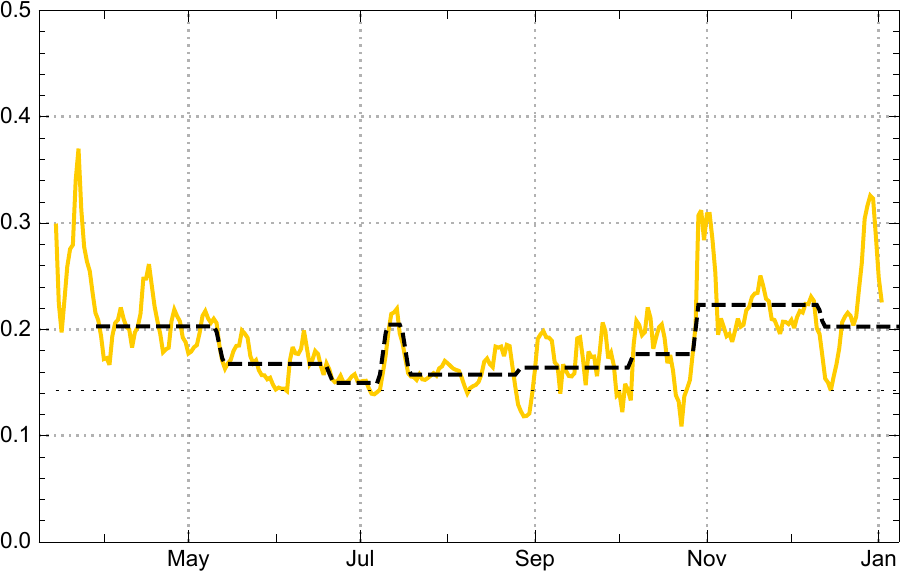} 
\caption{Left: Empirical reproduction rates $\hat{\rho}(k)$ for Brazil (orange). 
Right: Daily strength of infection, empirical $\hat{\eta}_7(k)$ (yellow),  for Brazil  assuming $\delta=8$, and model parameters $\eta_j$ in the main intervals  $J_j$ (black dashed). \label{fig rho eta Br}}
\end{figure}

  The daily reproduction numbers $\hat{\rho}(k)$ (independent of  $\delta$) start from a lower level than for many other countries, slightly above 2. They show a relatively stable downward trend, falling below 1 for a few days in early June and for longer periods after June 22, 2020 (fig. \ref{fig rho eta Br}, left). 
But already at the end of March, when  the reproduction rate was still considerably above 1  ($t_0 = $ March 30),  its downward trend was already slow enough to allow for approximation by constancy intervals. 

In the case of Brazil the initial interval   starts at $t_0=$ 03/15, 2020. The following time can be subdivided into main intervals $J_j$ in which the averaged daily strength of infection  $\hat{\eta}(k)$ can be replaced by their mean values $\eta_j$, starting with   $t_1=$ 03/30. The main intervals are separated by the days 
  $t_2=$ 05/14,  $t_3=$ 06/22, $t_4=$ 07/11,  $t_5=$ 07/19 $t_6=$ 08/27,  $t_7=$ 08/10,  $t_8=$ 10/29,  $t_9=$ 12/13, 2020. They 
can well be discerned in fig. \ref{fig rho eta Br}, right,   showing the daily strength of  infection  $\hat{\eta}_7(k)$ (yellow) and their mean values (black dashed) in these intervals.
In  the country day count $k_0=1$ ($\sim 70$ in the JHU count) the main  intervals are 
$J_0=[1, 15], \,J_1=[16, 59], \; \;  J_2 = [61,  97], \, J_3 = [100, 115],  \, J_4 = [119, 122], \, 
J_5 =[127, 164],
J_6= [166, 204],\, J_7= [206, 220], \,  J_8= [222, 271], \,  J_9= [274, t_{eod}] $, 
 here with the end of data $t_{eod}=292$.

The start parameter  $\eta_0$   and the  model reproduction numbers  $\rho_j$ in the respective interval $J_j$ are  

\vspace{0.2cm}
\begin{center}
\begin{tabular}{|l||c|c|c|c|c|c|c|c|c|c|}
\hline 
\multicolumn{11}{|c|}{Model $\eta_0$ and $\eta_j$ and $\rho_j$ in $J_j$   for  Brazil}\\
\hline
 & $\eta_0$ & $J_1$ & $J_2$ & $J_3$ & $J_4$ & $J_5$ &  $J_6$ &  $J_6$ &  $J_8$ &  $J_9$ \\
 \hline 
$\eta_j$ &  0.031 & 0.202 & 0.164 & 0.142 & 0.191 & 0.143 & 0.141 & 0.148 &0.180  & 0.172 \\
$\rho_j$ & ---  &  1.42   & 1.14 &  0.98  &1.30  & 0.96 & 0.94 & 0.97 & 1.17& 1.00\\
\hline
\end{tabular}
\end{center}
\vspace{0.5em}

\begin{figure}[h]
 \includegraphics[scale=0.6]{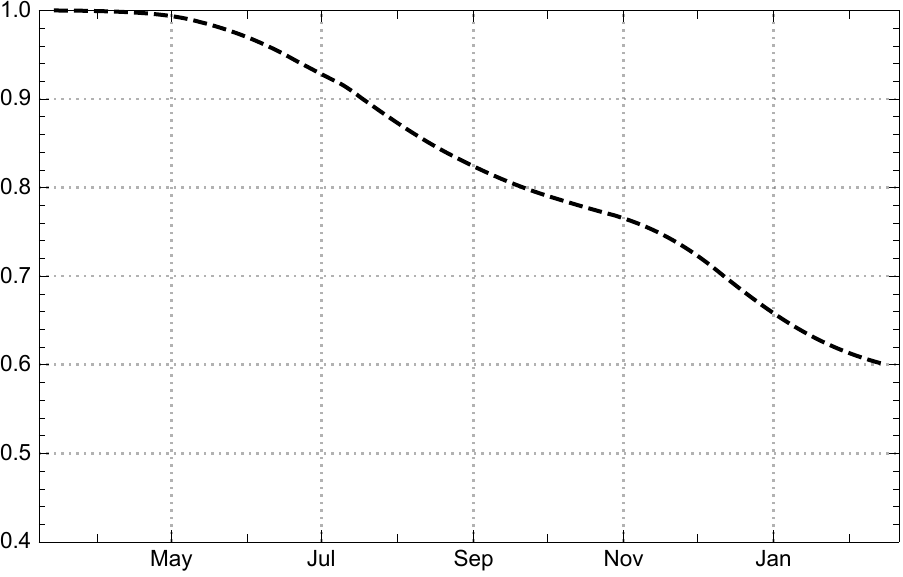} 
  \caption{Model values of the ratio of susceptibles $s(k)$  for Brazil ($\delta=8$).  \label{fig srek Br}}
\end{figure}

Also here the $\rho_j$  designate reproduction numbers at the beginning of the  $j$-the interval and the fall of $s(k)$ makes the reproduction number cross the critical value in the last main interval (cf. \ref{fig srek Br}). This does not mean that it will stay there.

\begin{figure}[h]
\includegraphics[scale=0.8]{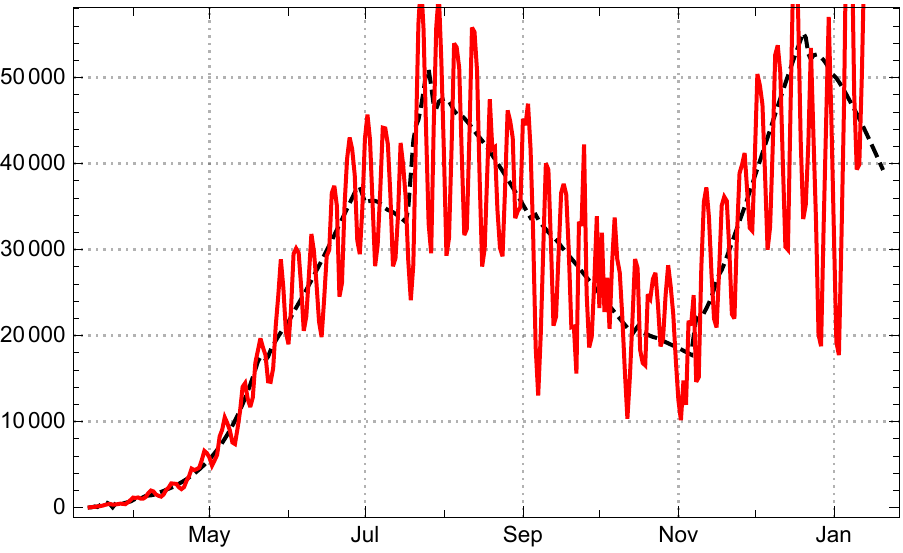}  
\includegraphics[scale=0.8]{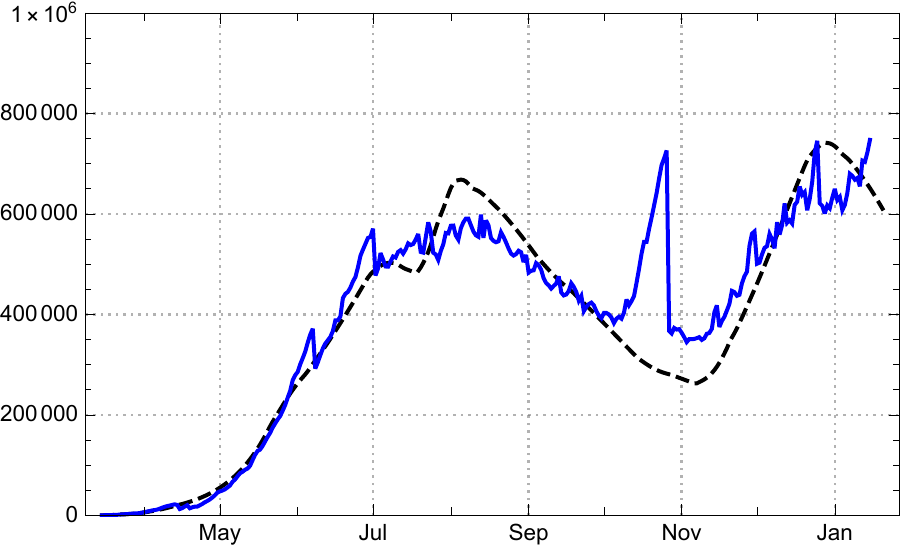}
\caption{Left: Empirical values  (3-day average) for daily newly reported for Brazil $\hat{A}_{new,3}$  (solid red line)) and  model values  black dashed). Right: Actual cases $\hat{A}$ (blue), model values (black dashed). 
 \label{fig Ineu and I Br}}
\end{figure}

\begin{figure}[h]
\includegraphics[scale=0.7]{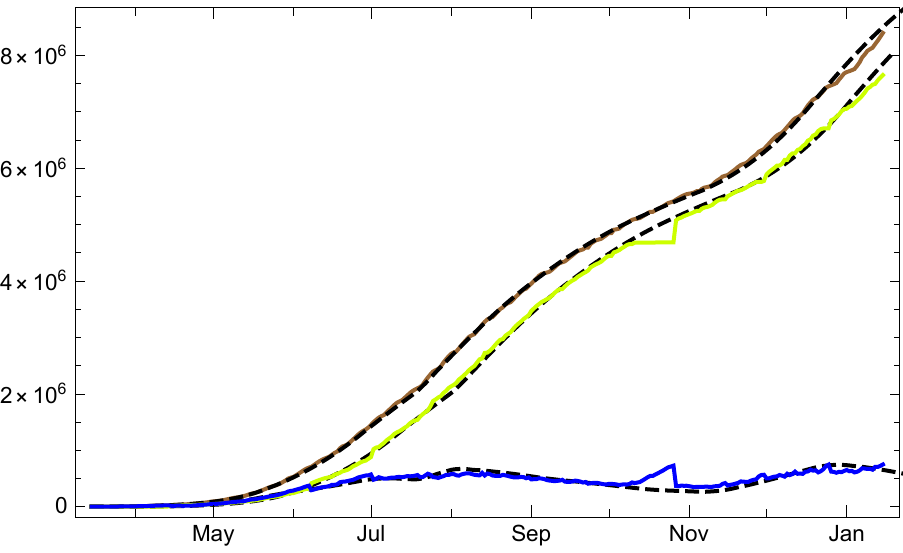}
\caption{Empirical data (coloured solid lines) for the  numbers of totally infected $\hat{A}_{tot}$ (brown), redrawn $\hat{R}$ (bright green), actual infected $\hat{A}$  (blue)  and the respective model values  (black dashed) for Brazil ($\delta=8$). \label{fig 3 curves Br} }
\end{figure}

The resulting model curves and their relationship to the empirical data for new infections and acual infections are shown in fig. \ref{fig Ineu and I Br}.   
A panel of the {\em three curves} $A_{tot}, \, A, R$ is shown  in  fig. \ref{fig 3 curves Br}. Here one sees clearly that the outlier bump of $\hat{A}(k)$ is accompanied by an inverse outlier in $\hat{R}(k)$.

\vspace{5em}

\vspace*{7em}

\pagebreak 

\subsubsection*{India}
\begin{figure}[h]
 \includegraphics[scale=0.7]{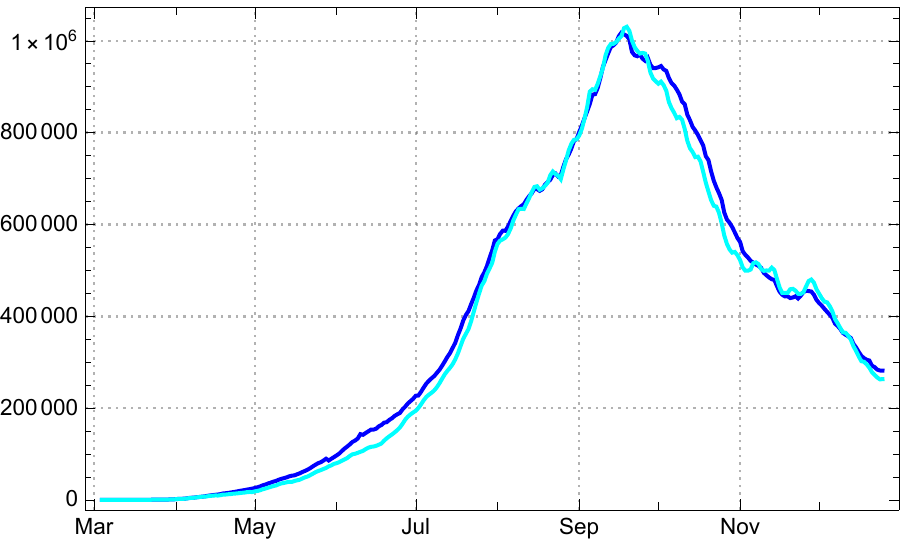} \hspace{1em} \includegraphics[scale=0.7]{graph-Ind-q} 
\caption{Left: Actual infected $\hat{A}(k)$ recorded by the statistics (dark blue) in comparison with $q$-equalized number $\hat{A}_q(k)$ (bright blue) for India ($q=11$).   Right: Empiricial estimate $\hat{q}(k)$ of mean duration of  actual infectived according to the statistics, i.e. in $\hat{A}$,  for India.  \label{fig I Iq Ind} }
\end{figure}

The recorded data on $\hat{A}_{new}(k)$ for India (population 1387 M) start to be non-sporadic at $t_0=$  March  4, 2020. From this time on we find a steady growth of the number of  reported actual infected $\hat{A}(k)$ until early September. Because the size of the country and the life conditions  in large parts of it a comparatively high number of unrecorded infected   may  be assumed, with a dark factor at the order of magnitude $\delta \sim 10$ probably $20 \leq  \delta \leq 50$.\footnote{A serological investigation of over 4000 inhabitants found 24 \% infected (from which over 90 \% were asymptomatic). With about 140 $k$ reported infected in a population of roughly 31 $M$ this amounts to a dark factor $\delta \approx 50$ (source ANI  retrieved 12/21 2020 {\tiny  \url{https://www.aninews.in/news/national/general-news/second-sero-survey-finds-2419-pc-of-punjab-population-infected-by-covid-1920201211181032/}}).}     
In early September the tide changed and  a nearly monotonous decline of actual infected started. With the exception of a short intermediate dodge the decline continues at the end of 2020 (fig. \ref{fig I Iq Ind}, left). Although in late December 2020 there were only about 0.7 \% recorded infected in India, the high quota of unreported infected  poses the question whether  the downturn  in late summer  may already be due to a the decrease of the fraction of susceptibles  $s(k)$.  

 Before we discuss this point let us remark that the time of being statistically recorded as actual case is relatively stable in the Indian data, with a good approximative constant value $q\approx 11$ (fig. \ref{fig I Iq Ind}, right).   In consequence $\hat{A}(k)$  does not differ much from $\hat{A}_q(k)$ (same fig. left). This allows to use  the recorded data  $\hat{A}$  in the following without the  proviso to be made in the case of the USA.

\begin{figure}[h]
\includegraphics[scale=0.7]{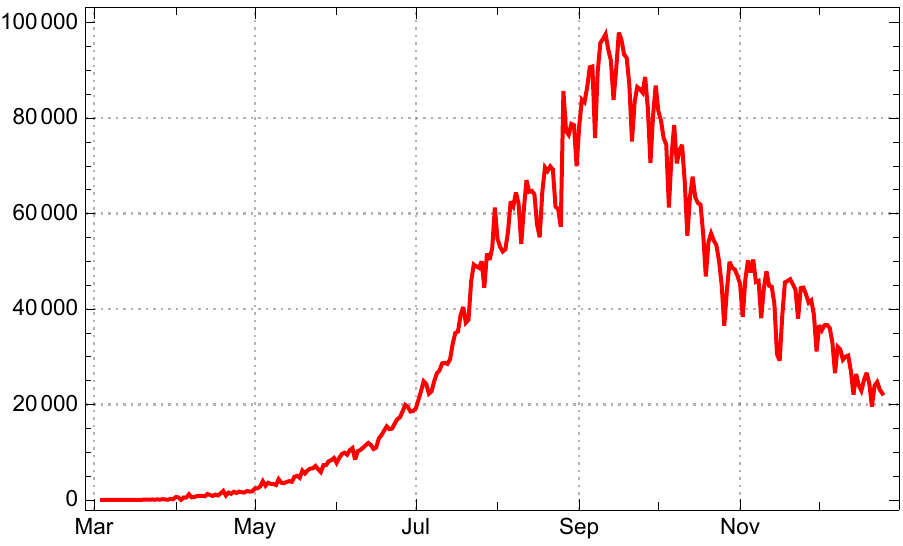} \includegraphics[scale=0.7]{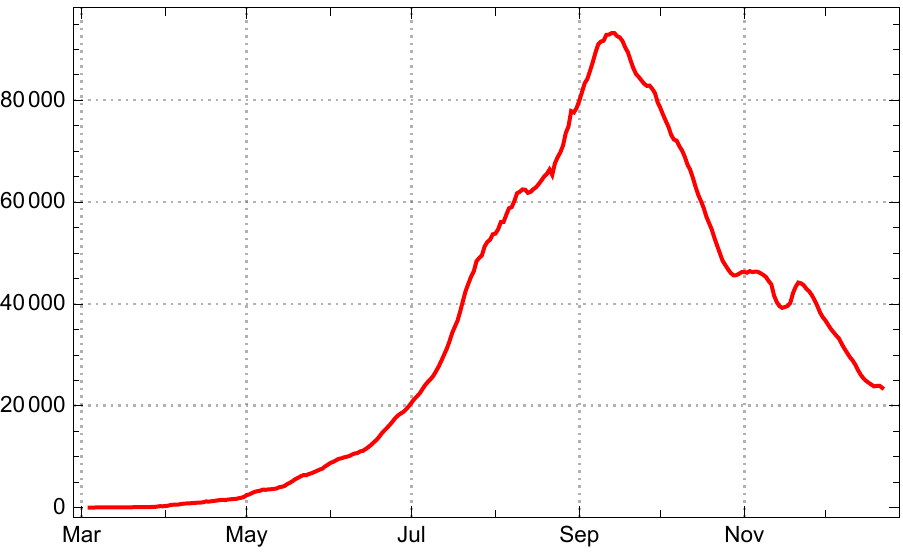}\\
\caption{Top: Empirical number of daily new infections $\hat{A}_{new}$ and 7-day sliding average $\hat{A}_{new,7}$ for India. 
 \label{fig I-neu Ind}}
\end{figure}

\pagebreak
The reproduction numbers and the corresponding daily strength of infection of the model are derived from the 7-day sliding averages of newly reported.  Figure \ref{fig I-neu Ind} shows both the daily varying $\hat{A}_{new}$ and the averaged $\hat{A}_{new,7}$.

\begin{figure}[h]
\includegraphics[scale=0.7]{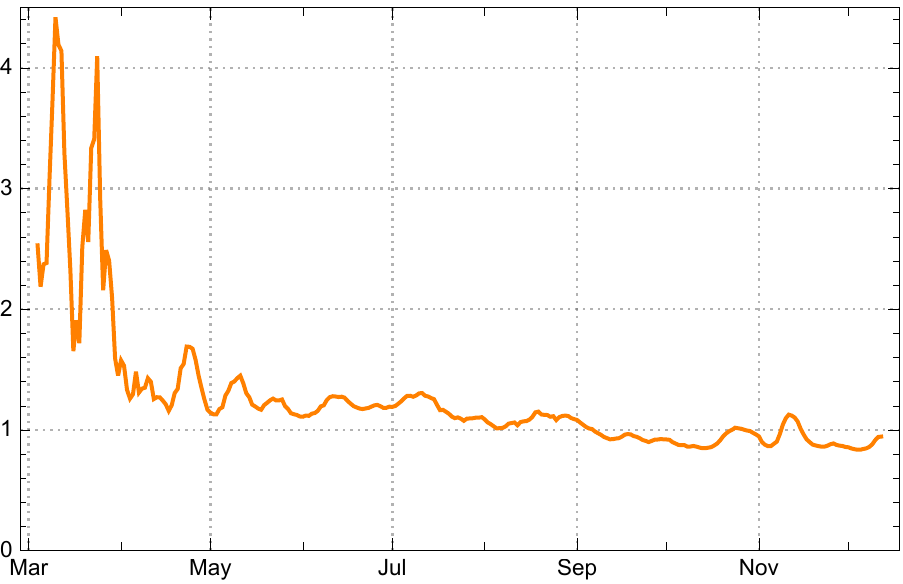} \quad \includegraphics[scale=0.7]{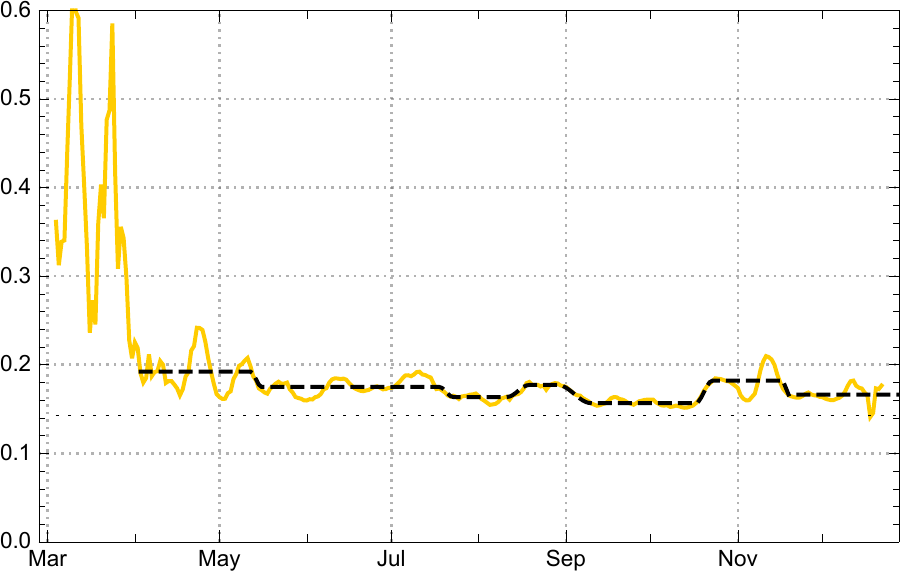}
\caption{Left: Empirically determined reproduction rates $\hat{\rho}(k)$ for  India  (iorange). Right: Empirical infections strength $\hat{\eta}_7$ (yellow) assuming  a dark sector with $\delta=35$;  model values for $\eta$ black dotted. \label{fig rho Ind} }
\end{figure}

\begin{figure}[h]
\includegraphics[scale=0.8]{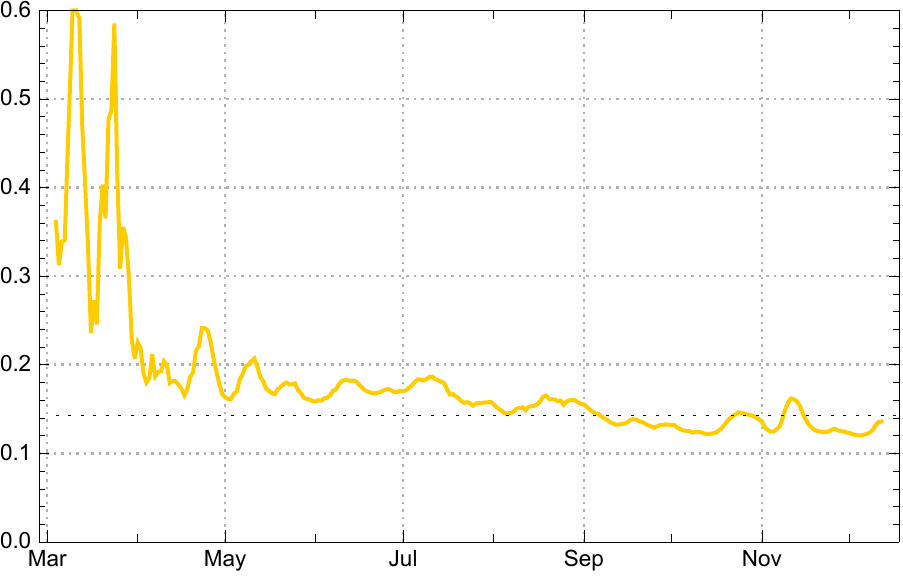} \includegraphics[scale=0.8]{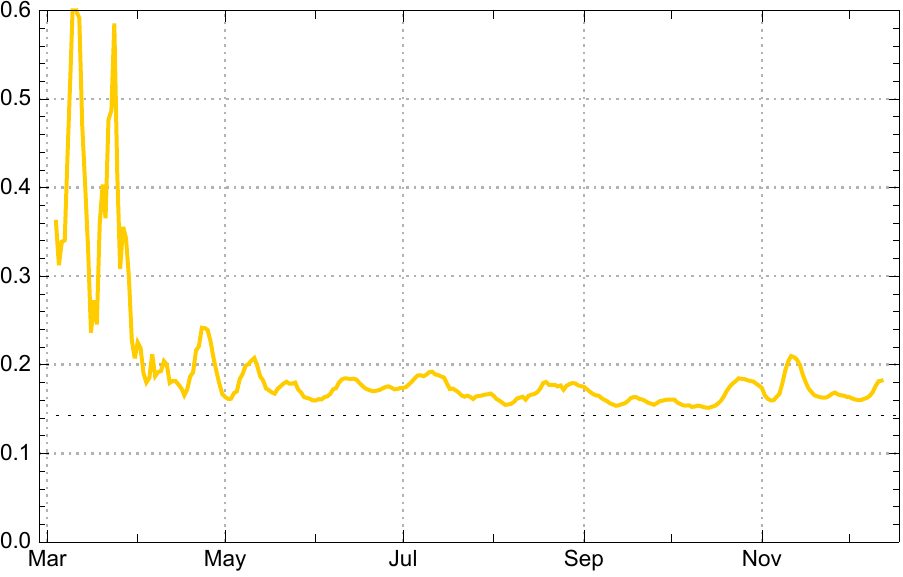} 
\caption{Left: Empirical daily strength of infection  $\hat{\eta}_7(k)$ for  India (yellow) with no dark sector, i.e. assuming $\delta =0$.
Right:   Empirical strength of infection  $\hat{\eta}_7(k)$  for India (yellow), assuming a dark sector with factor $\delta = 35$ \label{fig eta Ind} }
\end{figure}

The reproduction number 
 fell rapidly from roughly 3.5 at the beginning to below 1.5  in early April, and 1.2 in late May,  after which  it  continued to decrease with minor fluctuations. In early September it dropped  below the critical value 1, where it stayed with few exceptional fluctuations until December  (fig. \ref{fig rho Ind}).  It runs, of course, parallel to the daily strength of infection  $\hat{\eta}(k)$ calculated from the 7-day averages if abstraction is made from the dark sector,  $\delta=0$ (fig. \ref{fig eta Ind}, left). Here we confront it with the more realistic graph of $\hat{\eta}_7$ calculated under the assumption of a dark sector.

 Such an idealized scenario with $\delta =0$ is shown in fig \ref{fig eta Ind}, left. 
If, on the other hand,  the  empirical daily strength of infection
are determined under the more realistic assumption of a non-negligible dark sector,  e.g.  $\delta = 35$,  the picture is different (same figure, right). Here one finds a  daily strength of infection  moderately fluctuating in a narrow band between 10 and  20 \% above the critical value $\eta_{crit} = \frac{1}{7} \approx 0.147 $, rather than  dropping  below it  in  early September like in the first case. We do not know of any indications for a changing contact behaviour of the population in India; neither can we assume a decreasing  aggressiveness of the virus.  Therefore the first scenario ($\delta \approx 0$) looks  highly unrealistic. In both cases the empirically determined reproduction rates  $\hat{\rho}(k)$ are the same (fig. \ref{fig rho Ind}). In the second case the fall of $\hat{\rho}(k)$ below 1 in early September is due to the lowering of $\hat{s}(k)$, i.e.  as an effect of an incipient herd immunization.

But how can that be with a  herd immunization quota of  $(1+\delta) \frac{\hat{A}_{tot}(k)}{N}  \approx 12\, \%$, even with $\delta = 35$, in September 2020?\footnote{In 12/2020 it was already  twice as much.}
  The reason lies in the comparatively low overall daily strength of infection. Between May and December 2020 it fluctuated  between 10 and 20 \% above the critical level (fig. \ref{fig eta Ind}, right). Even if part of the low level had to be ascribed to an intentional under-reporting of the $\hat{A}_{new}(k)$, this would mean an increasing size of the dark sector; the   overall effect would be the same.\footnote{Only a permanently increasing amount of under-reporting could emulate a fake picture of a non-existing  downswing of the epidemic for several months. We exclude such a hypothesis.}


 \noindent
 For  modelling the epidemic in India we can do with 7  constancy intervals. 
After the initial interval $ J_0=[1 , 30]$  in the day count of the  country ($k_0 \sim 43$ in the JHU count) the main intervals  are 
$J_1=[31, 70], \, J_2 = [74,  137], \, J_3 =[142, 161], \, J_4= [169, 179], \, J_5= [191, 228], \, J_6= [234, 256], \, J_7= [260, k_{eod}]$, 
with end of data $k_{eod}=297$. The date of the interval separators are $t_0=$ 03/04, $t_1= $ 04/03, $t_2=$ 05/16, $t_3= $ 07/23,  $t_4=$ 08/19,  $t_5=$ 09/10,  $t_6=$ 10/23,   $t_7=$ 11/18, end of data  $t_{eod}= $ 12/26 2020. 

Similar to the Brazilian case, the reproduction rate surpasses 1 in the first four  main intervals;  only in mid September a downswing of the epidemic started, interrupted by an intermediate dodge at the beginning of November. In mid September the total  number of acknowledged infected was roughly $A_{tot}(240) \approx 5 \; M$, about 3.8 per mill of the total population; but with a dark quota of $\delta=35$ the total number of infected had probably already risen above the 10 \% margin (see above).

The start parameter of the model $\eta_0$ is chosen according to the best adaptation  to the 7-day averaged  data (without a claim for a directly  realistic interpretation) and the  parameter values $\eta_j$  essentially as the mean values of the $\hat{\eta}(k)$ in the respective interval $J_j$. They are given in the table.

\vspace{0.2cm}
\begin{center}
\begin{tabular}{|l||c|c|c|c|c|c|c|c|}
\hline 
\multicolumn{9}{|c|}{Model $\eta_0$ and  $\eta_j$, $\rho_j$ in $J_j$   for India}\\
\hline
 & $\eta_0$ & $J_1$ & $J_2$ & $J_3$ & $J_4$  & $J_5$ & $J_6$ & $J_7$\\
\hline 
$\eta_j$ &  -0.138 & 0.192  & 0.175 &  0.164  & 0.177 & 0.157  & 0.182 & 0.166 \\
$\rho_j$ &  ---    & 1.34 & 1.2 & 1.07 & 1.12 & 0.90 & 0.99 & 0.87 \\
\hline
\end{tabular}
\end{center}
\vspace{1.5em}
\noindent 

\begin{figure}[h!]
\includegraphics[scale=0.8]{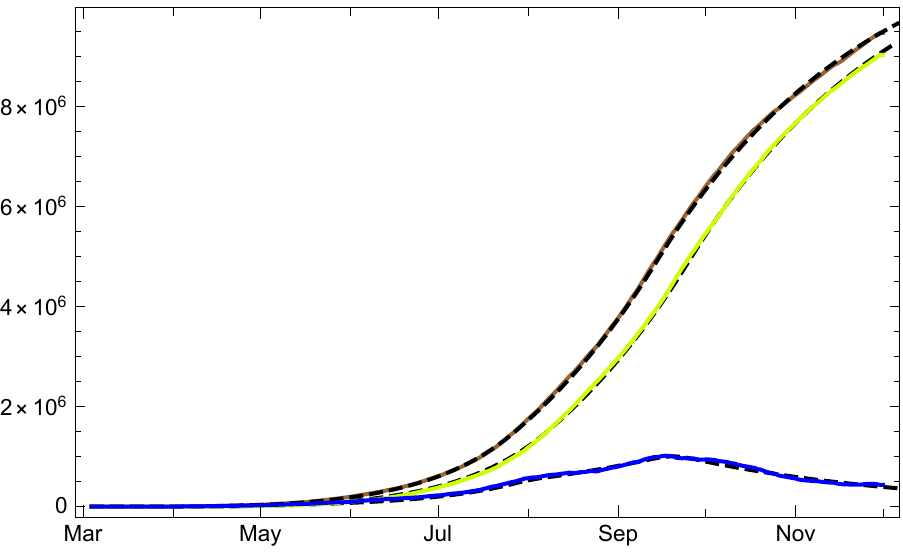}
\caption{Empirical data (colored solid lines) for the  numbers of totally infected $\hat{A}_{tot}$ (brown), redrawn $\hat{R}$ (bright green), actual infected $\hat{A}$  (blue),  and the respective model values  (black dashed) for India. All with dark sector, $\delta=35$ and constancy intevals (see main text). \label{fig 3 curves Ind}}
\end{figure}

With these parameters the SEPAR model leads to a convincing reconstruction  of the epidemic in India. This is shown by the graph showing the 
 three curves $A_{tot}, \, A, R$ (fig. \ref{fig 3 curves Ind}). 

\begin{figure}[h]
\includegraphics[scale=0.7]{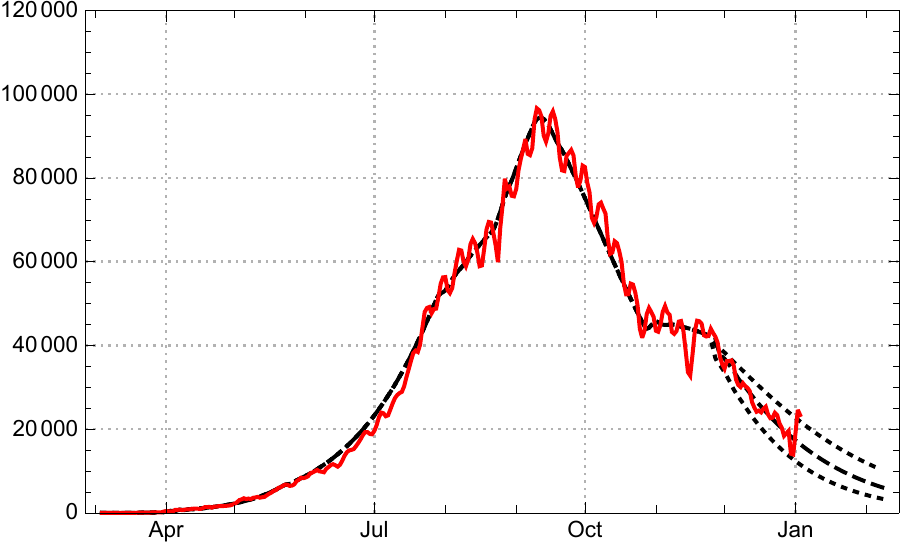} 
\hspace{0.5cm} \includegraphics[scale=0.7]{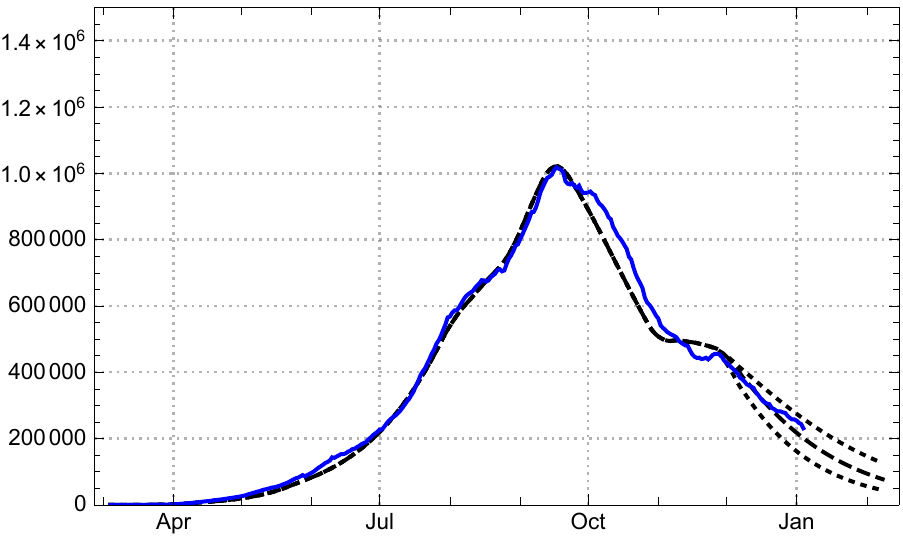}
\caption{Left: Daily newly reported for India, empirical $\hat{A}_{new}$ (solid red line) and model $A_{new}$ (black dashed). Right: Reported actual cases for India, empirical $\hat{A}$ (blue) versus  model  $A$ (black dashed). Both with dark sector, $\delta=35$ and 30-day conditional prediction  assuming no large vchange of the infection strenght in $J_6$, the last main interval. \label{fig Aneu and A Ind}}
\end{figure}

\newpage
 The numbers of newly reported $\hat{A}_{new}(k)$ and the number of actual  cases $\hat{A}(k)$, including a conditional prediction for the next 30 days on the basis of the last infection strength $\eta_6$  (fig.  \ref{fig Aneu and A Ind}).  Black dotted the boundaries of the 1 $\sigma$ domain for the $\hat{\eta}$ -variation in $J_6$. In the case of India the  data show  exceptional low variability inside the constancy intervals. Thus the width  of the  1 $\sigma$ domain is smaller than in any of the other countries considered.

\vspace{1.5em}


With a  dark sector  roughly as large as assumed in the model ($\delta = 35$) the ratio of susceptibles went down in late 2020 to  $s(k) \approx 0.7$ (fig \ref{fig  Ind srek}). This is the background for the reassuring prognosis for the development in India at the beginning of 2021 (fig. \ref{fig Aneu and A Ind}).

\begin{figure}[h]

\includegraphics[scale=0.7]{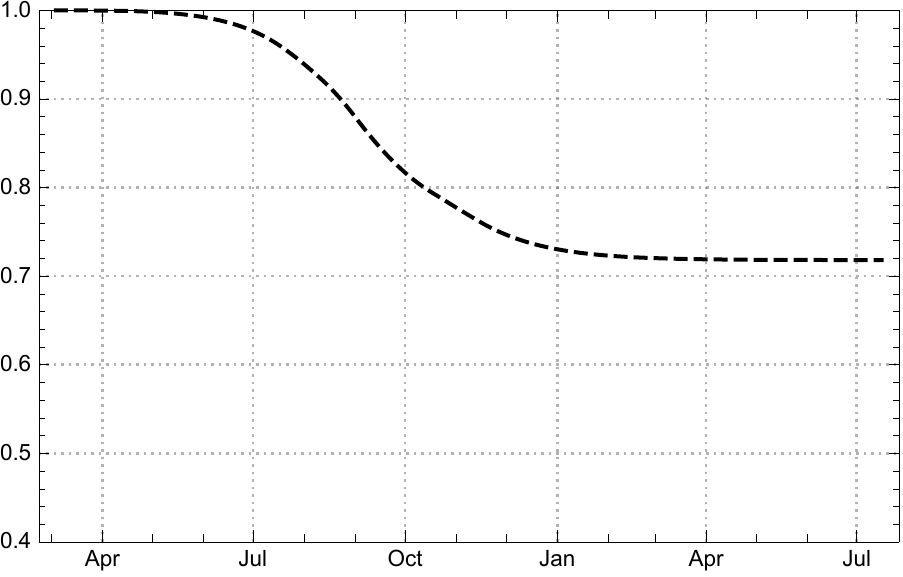}
\caption{
Ratio of susceptibles $s(k)$ for India. \label{fig Ind srek} }
\end{figure}

.


\pagebreak

\subsection{Aggregated data of the World\label{subsection World}}
\begin{figure}[h]
 \includegraphics[scale=0.7]{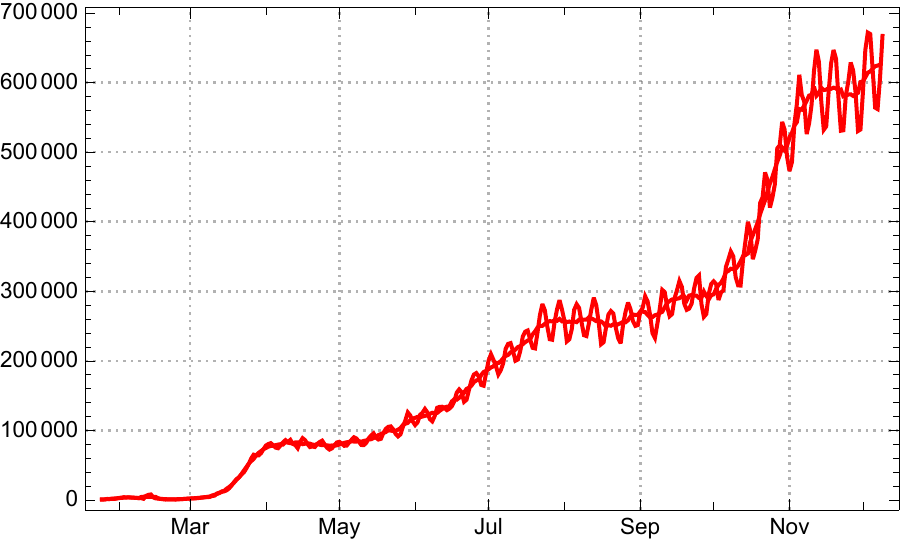} 
 \caption{Daily new reported cases $\hat{A}_{new}(k)$ for the World and  7-day sliding  averages $\hat{A}_{new,7}(k)$. \label{fig I-neu rho W}}
\end{figure}

\vspace{5em}
Let us now see how the aggregated data of all countries and territories documented in the JHU data resource can be analysed in our  framework, and how they are reproduced by the SEPAR$_d$ model. For the sake of simplicity we speak simply of  {\em the World}.
The number of daily new infections $\hat{A}_{new}$ shows clearly three or four steps, expressed by phases of accelerated growth of  $\hat{A}_{new}$ between February and December 2020 (fig. \ref{fig I-neu rho W}):  March (European countries), May to July (two waves in the US, bridged by rising numbers in Brazil  and India), October (second wave in Europe and Brazil, third wave in US), and less visible the January/February wave in China and South-Korea. 

\begin{figure}[h]
\includegraphics[scale=0.6]{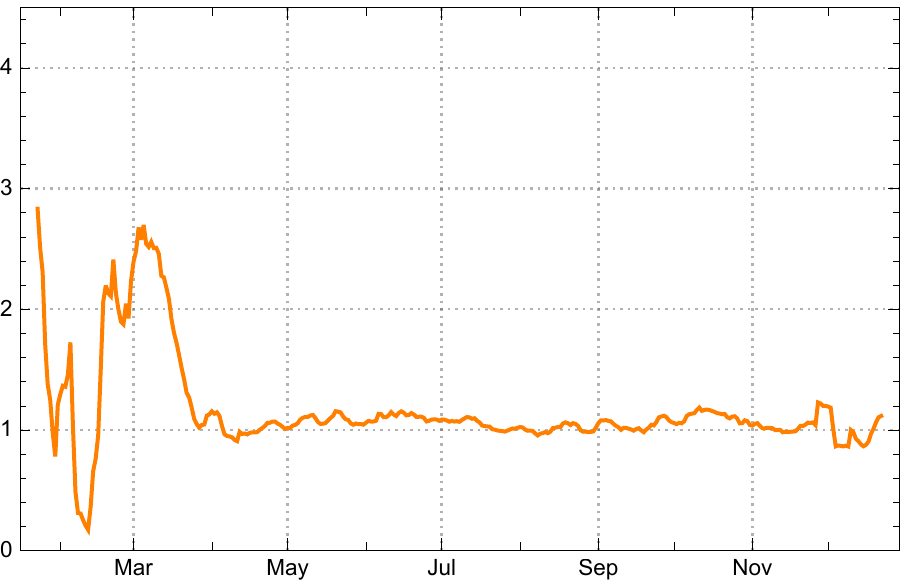} \includegraphics[scale=0.6]{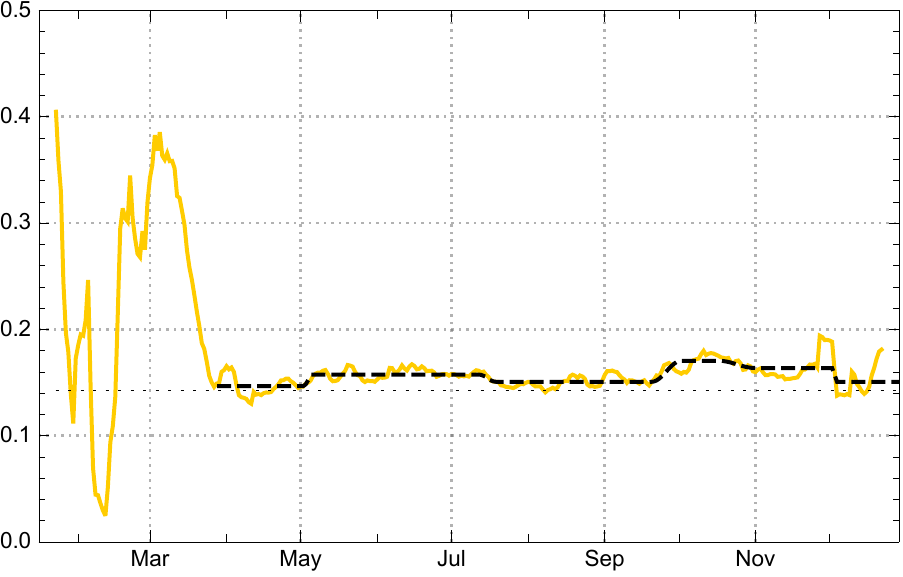} 
\caption{  Bottom, left: Empirical reproduction rates $\hat{\rho}(k)$ for the World (orange). Right: Daily strength of infection $\hat{\eta}(k)$  for the World (yellow) with model parameters $\eta_j$ in the main intervals  $J_j$ (black dashed). \label{fig rho und ak W}}
\end{figure}

These steps of steeper increase of the  daily new infections correspond to local peaks or elevated levels of the mean strength of infection and reproduction numbers. The first two peaks of the mean reproduction numbers with $\rho_{peak-1} \approx 2$ in January (China)  and $\rho_{peak-2} \approx 2.5$ in March (Europe) are followed by much lower phases of elevated levels from early May to early July, respectively in October 2020 (fig. \ref{fig rho und ak W}).

\pagebreak
We let the model start at $t_0 =$ 01/25, 2020, the  fourth day of the JHU day count, and use the following time separators for the main (constancy) intervals
$t_1= $ 03/29, $t_2= $ 05/06, $t_3= $ 07/19, $t_4= $ 10/02, $t_5= $ 11/10, $t_6=11/26 $, end of data  $t_{eod}= $  12/11, 2020. In the  count adapted to the $t_0$ chosen here the main intervals are $J_0= [1, 64], \; J_1=[65, 99],\; J_2= [103, 170], \; J_3=[177, 240]; J_4= [252, 276], \; J_5=[291, 303],\; J_6= [307, 322], ] $. The parameters $\eta_j$ and the corresponding mean reproduction numbers in these intervals are give by the following table.

\vspace{0.2cm}
\begin{center}
\begin{tabular}{|l||c|c|c|c|c|c|c|}
\hline 
\multicolumn{8}{|c|}{Model $\eta_0$ and $\eta_j$, $\rho_j$ in intervals $J_j$   for the World}\\
\hline
 & $\eta_0$ & $J_1$ & $J_2$ & $J_3$ & $J_4$ & $J_5$& $J_6$ \\
\hline 
 $\eta_j$ & 0.325  & 0.145  & 0.156 & 0.145 &  0.160  & 0.144  & 0.158    \\
 $\rho_j$  &  ---  & 1.01  & 1.09 & 1.02  & 1.12 & 1.01 &1.11\\
\hline
\end{tabular}
\end{center}
\vspace{0.5em}

\begin{figure}[h]
\includegraphics[scale=0.8]{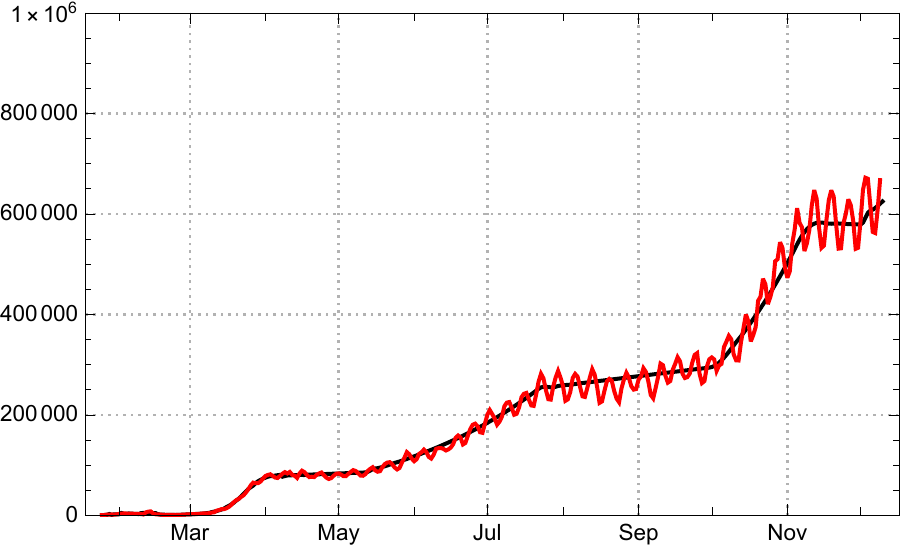}  \includegraphics[scale=0.8]{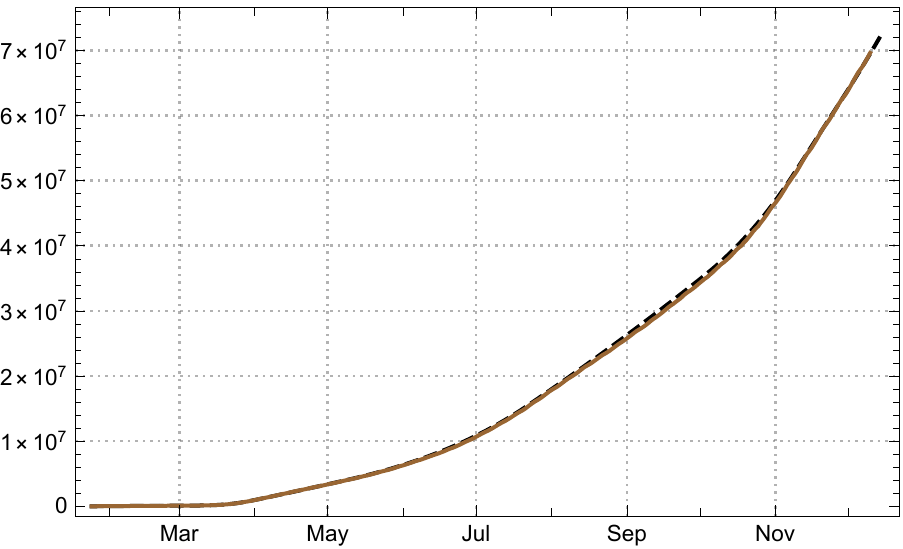} 
\caption{Left: 3-day averages of daily new reported infected for the the World  (red); empirical $\hat{A}_{new}$  solid red, model $A_{new}$ black dashed. Right: Total number of reported infected (brown);  empirical $\hat{A}_{tot}$ solid, model  $A_{tot}$ dashed. \label{fig Ineu and Itot W}}
\end{figure}

\begin{figure}[h]
\includegraphics[scale=0.6]{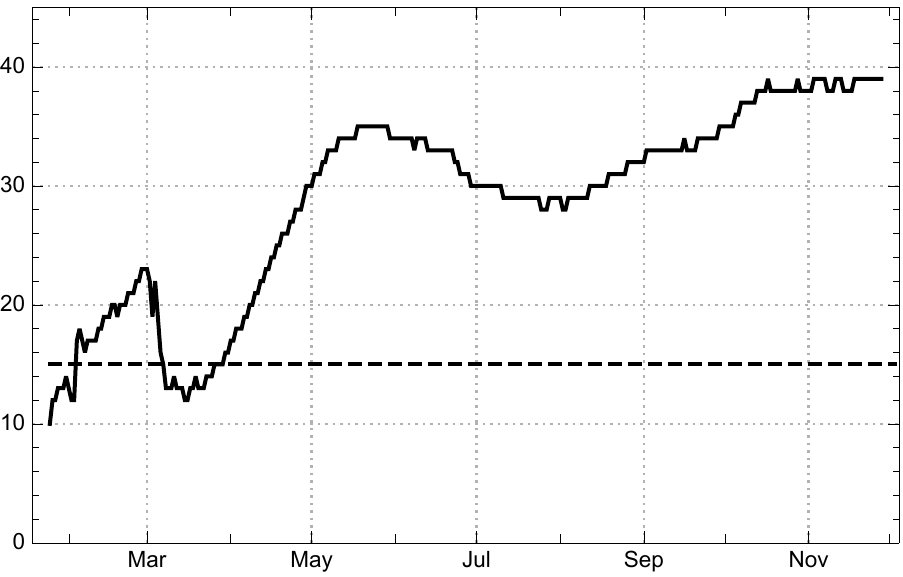} 
\includegraphics[scale=0.7]{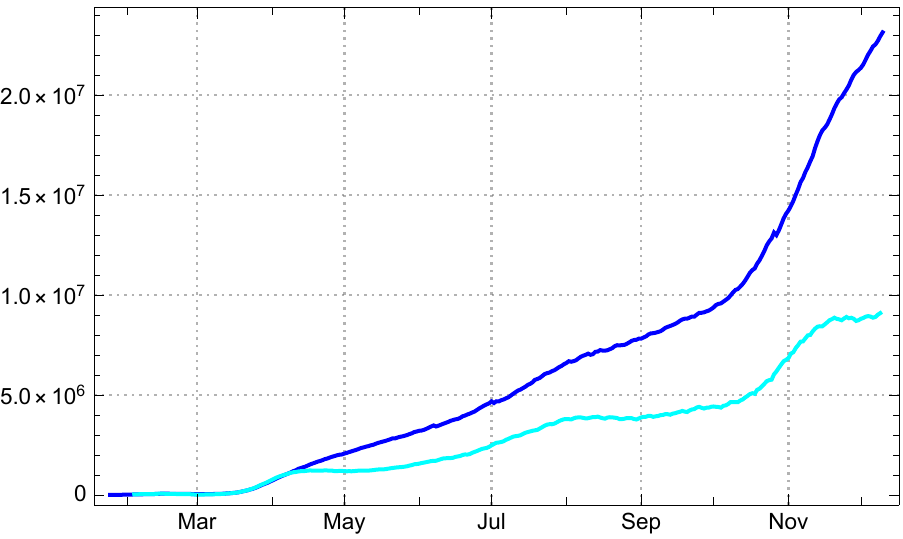}
\caption{Left: Daily values of the mean time of statistically actual   infection $\hat{q}(k)$ for  the World.  Right: Comparison of reported infected $\hat{A}$ (dark blue) and $q$-corrected number ($q=15$) of recorded actual infected $\hat{A}_q$ (bright blue) from the JHU data in the World. \label{fig q(k) W} }
\end{figure}

Because of the lack of reliable reporting for recovering dates in several countries,  among them some large ones like the USA, we cannot expect a balanced value for the sojourn in the state of actual disease, documented in the statistics. Fig. \ref{fig q(k) W}, left shows that the estimated values $\hat{q}$  keeps close to 15 or even 20 until late March. Later on  the  weight of the countries with reliable documentation of  recovering dates  is large enough to keep the mean number of $\hat{q}(k)$  between 30 and 40, even with the rise of the pandemic after May 2020 (fig. \ref{fig q(k) W}, left). Accordingly the $q$-corrected number of actually infected $\hat{A}_q$ separate from the  the ones given directly by the statistics $\hat{A}(k)$ only in mid April. Since October 2020 they difference between the two is rising progressively (same figure, right).

\begin{figure}[h]
\includegraphics[scale=0.8]{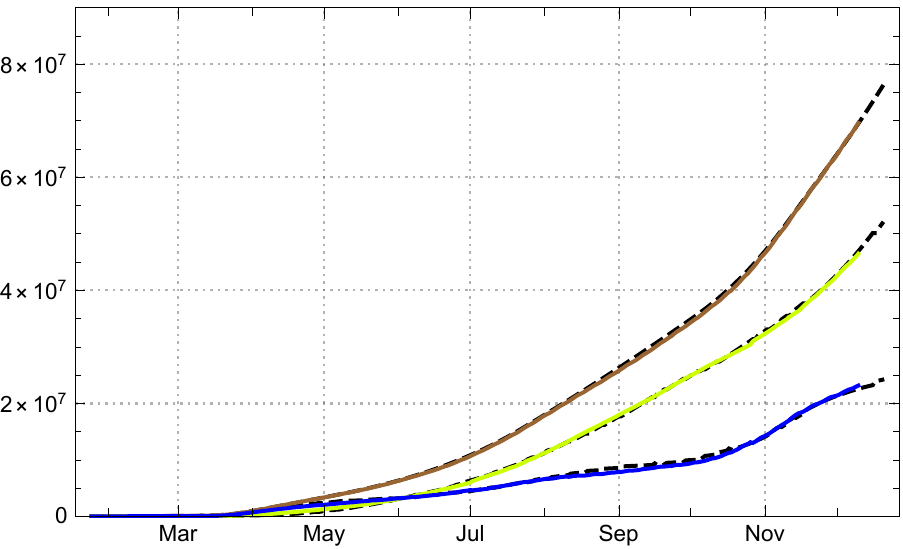}
\caption{Empirical data (solid coloured lines) and  model values (black dashed) for the World:  numbers of totally infected $\hat{A}_{tot}$ (brown), redrawn $\hat{R}$ (bright green), and   numbers of those which are  statistically displayed   as actually infected $\hat{A}$  (blue). \label{fig 3 curves W} }
\end{figure}

\pagebreak
Like in the case of those countries which have an unreliable documentation of the actual state of infected (e.g. US, Sweden, \ldots) we can here reconstruct the statistically given number $\hat{A}(k)$ by the model value $A(k)$ by using time variable durations $q(k)=\hat{q}(k)$. This is being displayed in the graph of the 3 curves of the World (fig. \ref{fig 3 curves W}). 

\newpage
\section{Discussion \label{section discussion}} 
The data evaluation in sec. \ref{section countries} shows clearly that the SEPAR$_d$ model works well for countries or territories with widely differing conditions and courses of the epidemic. For the ``tautological'' application of the  model   with daily changing coefficients of infection $\eta(k)$  this is  self-evident, while it is not so for the use of a restricted number of constancy intervals. The  examples studied in this paper show that in this mode of application  the  model is  well-behaved,  able to characterize  the  mean motion of an  epidemic  and to analyse its central dynamic. In the country studies we have shown that this is the case not only for the number of  acknowledged daily new reported, our $A_{new}(k)$  but also for data  which,   in the standard SIR approach, are not easily interpretable like  the number of actual infected persons, $A(k)$or  the $q$-normalized number $A_q(k)$.

What is the $SEPAR_d$ model good for? It is clear that it cannot predict the future. The main reason for this is that nobody knows how the contact rates are changing in the future. It allows -- though -- a prediction under assumptions. In the different countries we carried this out with different scenarios. 

The main value of the model is as a tool for analysing the development, and to learn from such an analysis.  We will discuss three such topics:
\begin{itemize}
\item[--] the role of constancy intervals
\item[--]  the role of the dark sector
\item[--]  the influence of the time between infection and quarantine
\end{itemize}

\subsubsection*{The role of constancy intervals}
The empirical values of the infection strength $\hat{\eta}(k)$ are calculated from data on reported new infected and are therefore  subject to irregularities in data taking and reporting. The most drastic consequences of this are the obvious  weekly fluctuations.  Different methods can be applied to smooth these  weekly fluctuation,  sliding 7-day averages (used here),  stochastic estimate used by the RKI (see appendix), band filter etc. Independent of the applied method there remain  effects (e.g. non- weekly reporting delays)    which  distort the calculated numbers away from being correct empirical values for the intended  quantities (e.g. $\eta(k)=\gamma \kappa(k)$). Even if they were, one would encounter day to day fluctuations resulting from the variation of intensities of contacts and of the strengths of infectiousness involved, which one is not really interested in if one wants to gain insight into the  dynamics of the epidemic. For this one needs to   distil a cross-sectional picture of  the  process. In our approach this   is achieved by constructing  {\em constancy intervals}  (main intervals) $J_j$  and model strengths of infection $\eta_j$, read off from the data, and to apply the infection recursion (\ref{eq recursion for H(k)}). \\

\subsubsection*{The role of the dark sector}
With increasing numbers of herd immunized, the influence of the dark sector on  the ratio $s(k)$ of susceptibles in the total population gains increasing weight, in particular for countries in which a high dark ratio $\delta$ may  be expected. In most of the European countries studied here we find the ratio of recorded infected at the order of magnitude of 1 \% all over the year 2020. With the non-reported ones added it can easily rise to the order of magnitude  10 \% and start to have visible effects. If our estimated values of the dark factor $\delta$ are not utterly wrong, our model calculation shows that in nearly all   countries of the study,  Germany  being the only exception,   the development of the epidemic is already noticeably   influenced  by the dark sector  at the beginning of the year 2021.  The latter contributes essentially to  turning  the tide of the reported new infected, if one assumes  constant  contact ratios $\kappa(k)$ and  mean infection strength  $\gamma$ of the virus. Of course the appearance of new mutants may change $\gamma$, and counteract the decrease of the numbers of infected predicted by the model. This seems to be the main problem for the early months of 2021.

\begin{figure}[h]
\includegraphics[scale=0.7]{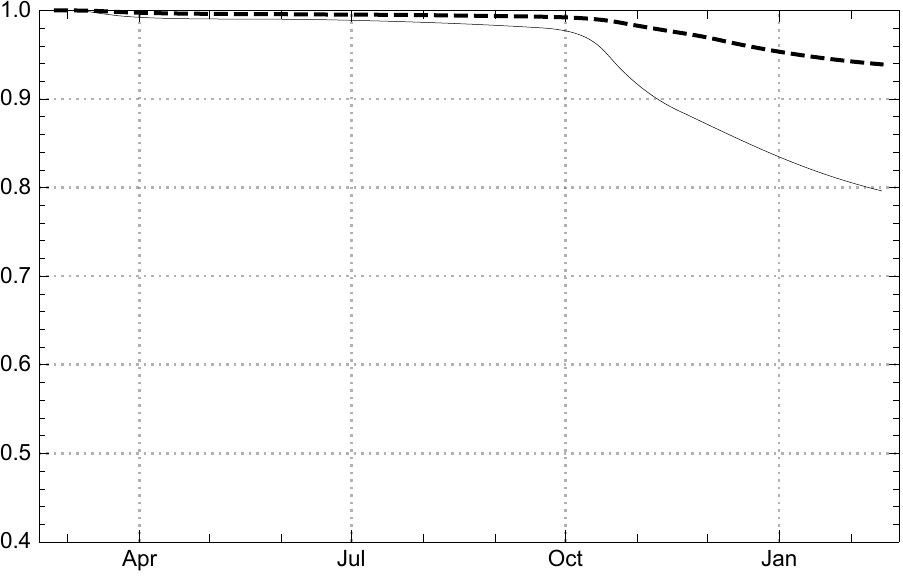} 
\caption{Ratio of susceptibles $s(k)=S(k)/N$ for Germany (dashed) and Switzerland (solid line)   at the end of the year 2020, assuming a dark ratio $\delta=2$ for Switzerland and $\delta=1$ for Germany.  \label{fig CH-D srek} }
\end{figure}

This becomes particularly succinct  by a comparing the  Swiss situation with   Germany at the end of the year 2020 (fig. \ref{fig CH-D srek}). In both countries containment measures were taken after a rise of the reproduction rate to 1.4 to 1.5 in late September /early October, although with different degrees of resoluteness and results (figs. \ref{fig rho und eta CH},  \ref{fig I-neu rho und ak D}). 

The weight of the dark sector is much stronger for the non-European countries of our study. In the case of the USA and Brazil it has started to suppress the effective reproduction number below the critical value 1, according to our model assumptions on the dark factor. But even if one would set it dowm to $\delta=1$ or 2 the effect would already occur, although a bit later and weaker. That this is not yet reflected in the numbers of newly infected may have different reasons; one of it would, of course be, that the model can no longer be trusted in this region. Others have been mentioned in the country section. And finally it could be that persons infected some months ago need not necessarily be immune against a second attack. If virologists come to this conclusion, the whole model structure would need a revision. At the moment it is too early to envisage such a drastic step.

\subsubsection*{The influence of the time between infectivity and quarantine}



A central input into the $SEPAR_d$ model is the assumption that there is a rather short period of length $p_c$, where people, who later are positively tested,  are infectious. This is closely related to the fact that people with positive test results are sent to quarantine or hospital. One can wonder what would happen, if the time between infectivity and quarantine or hospital is changed. 

It is a bit confusing, but there are two answers to this question. To explain the difference we recall the role of $p_c$ in our model. We usually  derive the  $\eta$ parameters from the data (eq. \ref{eq hat(eta)}).  For the reproduction number (\ref{eq repro number empirical})  in the simplified $SEPAR_d$-model with $p_c=p_d=p$ and a constant coefficient $\eta$ this means:
\[\rho(k)  =  p \, \eta \, s(k)  =  (p_c\,  \frac{\eta}{2}  +p_d  \frac{\eta}{2}) \, s(k)
\]
Here we assume that $p_c$ is given. This number is only  a rough estimate and may be  chosen slightly differently.   So, for each choice of the estimated number $p_c$ one gets model curves and one might ask, how much these model curves differ, in particular how much the reproduction rates would differ. The answer is: not very much. The reason is that $p_c$  enters implicitly also in the formula (\ref{eq eta}) for $\eta$, since the denominator is a sum over $p_c$ values of the daily newly infected. If we assume that this number is constant (which often is approximately the case) then in the denominator we would have the factor $p_c$ and in the formula for $\rho$ it cancels out. Thus in this situation the reconstruction of $\rho$ from $p_c$ and $\eta$ is independent of the choice of $p_c$. If the values of the newly infected changes more drastically this is not the case and one has to use the general formula (\ref{eq repro number general}), but the difference is not dramatic. So the first answer to the question is: A different estimation for $p_c$ does not have a noticeable  influence for the model curves. 

For understanding the second very different answer we have to recall that $\eta$ may be interpreted   as the product of the contact rate $\kappa$ (as measured in the model) and the strength of the infection $\gamma$. If we assume that $\gamma $ is constant, the change of $p_c$ discussed above amounts to a change of the model-$\kappa$, which does not express a changing contact behaviour. This means that our measure for the contact rate is related to our choice of $p_c$. 

Now we come to the second answer. Here we assume that the contact rate remains  the same, the contact behaviour of the society is not changed. But suppose that by some new regulations the value of $p_c$ is changed. Then, as expected, if the contact behaviour is unchanged the reproduction number changes proportionally and so the curves are different. This second answer is what we are  interested in here. Let us assume that  one finds means by which  the time until the people go to quarantine or hospital is reduced. Then less contacts take place and so the curves are flattened. This fact is well known, e.g., \citep[appendix]{Andersen_ea:2020}.\footnote{We thank S. Anderl for the hint.} 
 But how much?

For answering this question we have taken the model description for Germany,  lowering  the value of $p_c$ from $7$ to $6$ days from a certain moment on. Here we have to discuss an important point. One can only influence the time until quarantine or hospital for those who are registered, while   the infected people who end up in the dark sector behave as before. At this moment we have to give up our assumption that $p_c$ = $p_d$. So, from a  certain moment on we assume that $p_d$ is still 7 but $p_c$ is 6. 

\begin{figure}[h]
\includegraphics[scale=0.7]{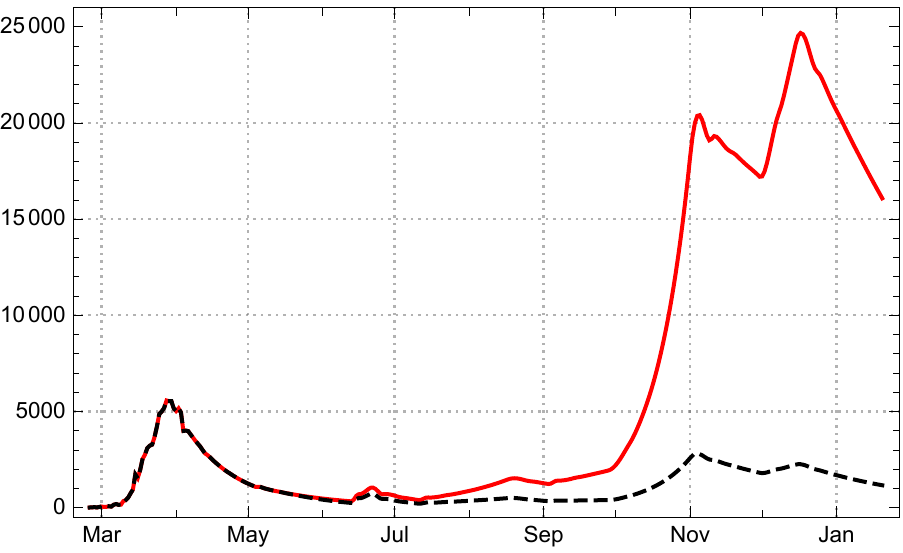} 
\hspace{0.5cm} \includegraphics[scale=0.7]{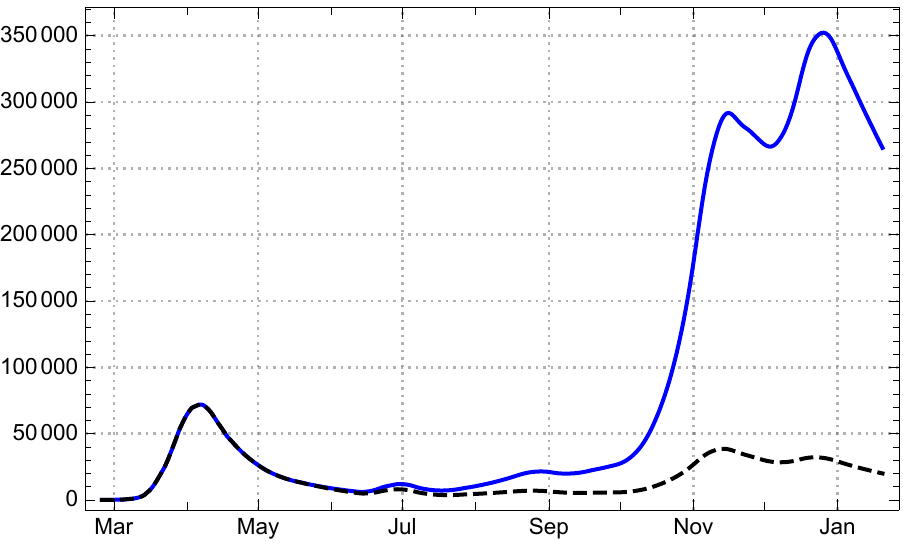}
\caption{Model calculations for reported new infected $A_{new}(k)$ (left) and reported actual infected $A(k)$ (right) for Germany. Solid lines with parameter values given in  sec. \ref{section countries}($p_c=7$ all over the year 2020). Dashed $p_c=7$ from March to May, $p_c=6$ from August onward, smooth transition in June. \label{fig reduce p D May} }
\end{figure}

\begin{figure}[h]
\includegraphics[scale=0.7]{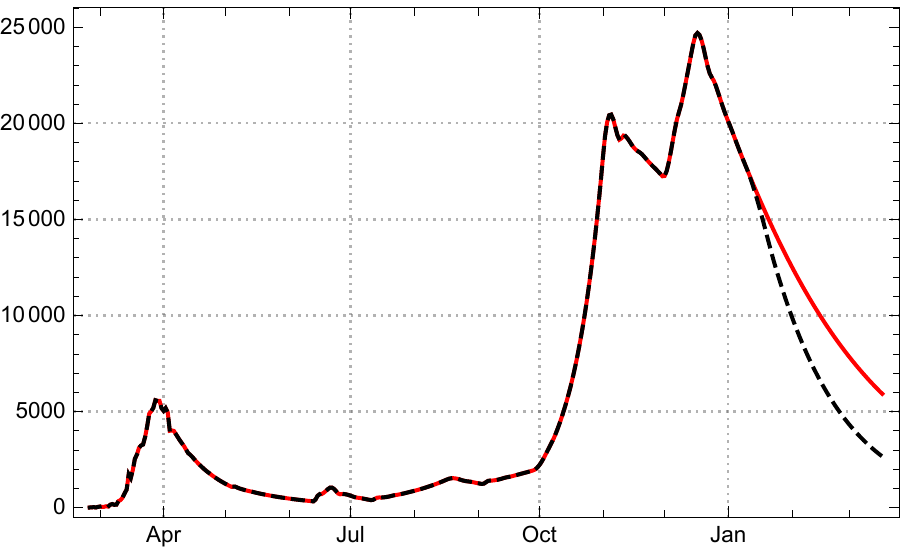} 
\hspace{0.5cm} \includegraphics[scale=0.7]{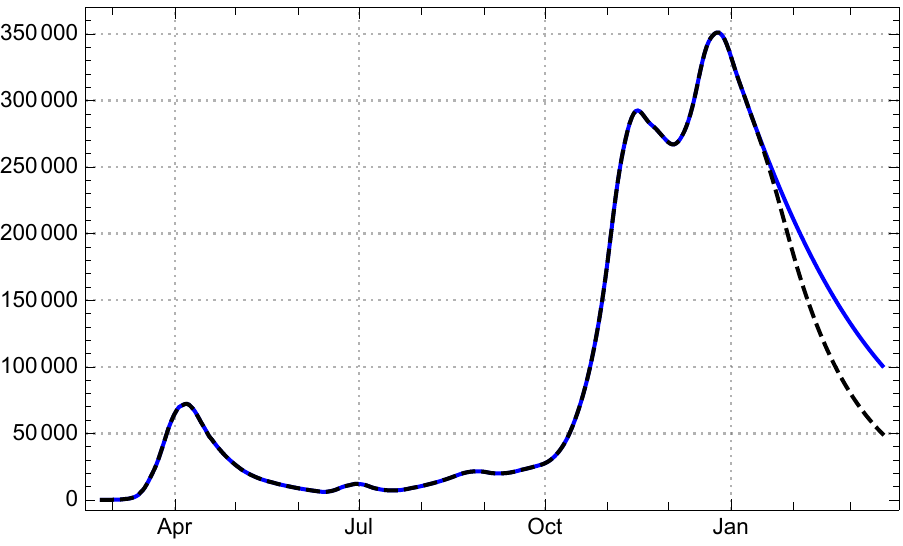}
\caption{Model calculations for reported new infected $A_{new}(k)$ (left) and reported actual infected $A(k)$ (right) for Germany (30 days prediction on the basis of data available 14 Jan 2021). Solid lines with parameter values given in the sec. \ref{section countries}, in particuilar $p_c=7$. Dashed $p_c=7$ from March 2020 to 15 Jan 2021, $p_c=6$ from February 2021 onward, smooth transition in between. \label{fig reduce p D Jan 21} }
\end{figure}

We have carried this out in two different scenarios for the expected numbers of daily new recorded infected $A_{new}(k)$ and the numbers of actual infected $A(k)$. In the first one  we compare the past development in Germany during the year 2020 with a fictitious reduction of $p_c$ from $7$ to $6$ during May 2020,  keeping $p_d = 7$ fixed (fig. \ref{fig reduce p D May}). In a second  one  we take a look into the future, perpetuating the contact rate of the last constancy interval, i.e., assuming that the contact behaviour of the population is unchanged for a while  and assume the same fictitious reduction as above in the second half of January 2021 (fig. \ref{fig reduce p D Jan 21}). This doesn't mean that we make a prediction of the future, our only aim here is to demonstrate what would happen if we could lower $p_c$ from 7 to 6. The lowering of the numbers of infected, newly recorded  and actual ones,  are very impressive.

In the past none of the regulations imposed by the German federal authorities  made an attempt to reduce the time until people got to quarantine or hospital aside from raising the number of tests.  Our considerations suggest to make a serious attempt in this direction. It has the big advantage that it does not require additional restrictions of the majority of the population and can be expected to be very effective at the same time. \\[5em]

\section*{Appendix}
\subsubsection*{Comparison with RKI reproduction numbers \label{appendix generations}}
The estimates of the reproduction numbers for Germany by the   {\em Robert Koch Institut} (RKI), Berlin, are based  on an approach using the generation time as crucial delay time. 
The {\em generation time} $\tau_g$  of an epidemic is defined as the mean time interval between a primary infection and the secondary infections induced by the first one;  similarly  the length $\tau_s$ of the {\em serial interval} as the mean time between the onset of symptoms of a primary infected and the symptom onset of secondary cases. 
There are various methods to determine time dependent effective reproduction numbers  on the basis of stochastic models for infections using both intervals. In our simplified approach with constant $e$ and $p$ these numbers correspond to $\tau_g=\tau_s=e+ \frac{p-1}{2}$. 

The RKI calculation uses a method of its own for a stochastic estimation of the numbers of newly infected, called $E(t)$, from the  raw data of newly reported cases, described in \citep{anderHeiden_ea:2020}.
The calculation of the reproduction numbers works with these $E(t)$ and assumes  constant generation time and serial intervals of equal lengths $\tau_g=\tau_s=4$
\citep{RKI:2020Erlaeuterung}.\footnote{For $e=2$ this would correspond to  $p=5$, while for  $\tau_g=\tau_s=5$ we arrive at our $p=7$.} 
Two versions of reproduction numbers are being used, a day-sharp and therefore ``sensitive'' one 
$\rho_{rki,\, 1} (t) = \frac{E(t)}{E(t-4)}$,
  and a  weekly averaged one,
\[ \rho_{rki,\, 7} (t) = \frac{\sum_{j=0}^6 E(t-j)}{\sum_{j=0}^6 E(t-4-j)} \, , 
\]
which we refer to in the following simply as $\rho_{rki}(t)$. 

The paper remarks that the  RKI  reproduction numbers (``$R$-values'') $\rho_{rki}(u)$  indexed by the date $u$ of calculation  refer to a period of infection which, after taking the incubation period $\iota$ between 4 and 6 days into account, lies between $u-16, \ldots, u-8$ (with  central day $u-12$ in the interval).  We  reformulate this redating by setting
\beq \hat{\rho}_{rki} (t-12) = \frac{\sum_{j=0}^6 E(t-j)}{\sum_{j=0}^6 E(t-4-j)} \, , , \label{eq rho-RKI}
\eeq 

\begin{figure}[h]
\includegraphics[scale=0.7]{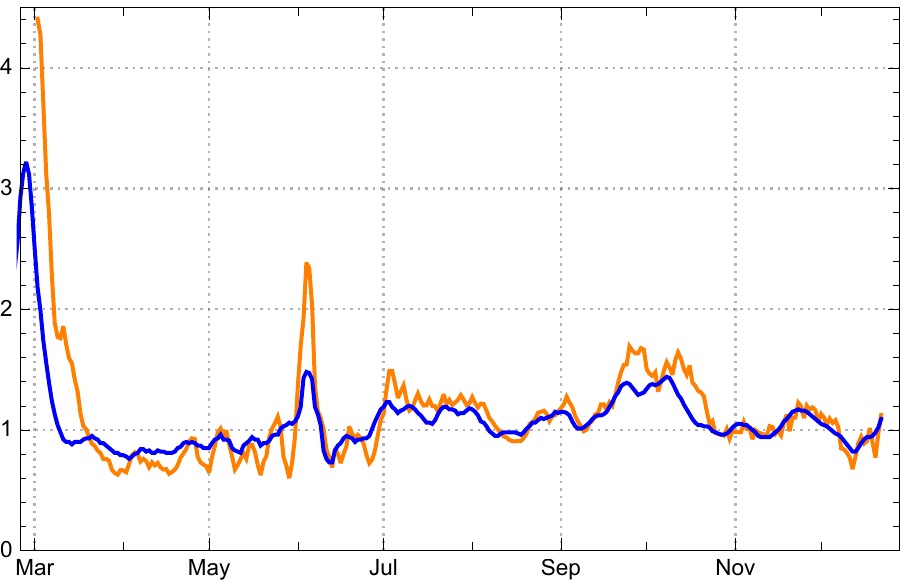} 
\caption{Empirical reproduction numbers $\hat{\rho}_7(k)$ of the SEPAR$_q$ model for Germany (orange) and reproduction numbers $\rho_{rki}(k -13)$ (7-day averages) of the RKI (blue). \label{fig rho-RKI rho7} }
\end{figure}

For a comparison with the SEPAR reproduction numbers we write (\ref{eq repro number empirical})  as 
\[\hat{ \rho} (k-(e+p+3)))= \frac{\frac{p}{7}\, \sum_{j=0}^{6}\hat{A}_{new}(k-j) }{\sum_{j=0}^{p-1} \hat{A}_{new}(k-(e+2)-j) }  \, ,
\] 
which for $e=2,\, p=7$ boils down to
\[ \hat{\rho} (k-12))= \frac{ \sum_{j=0}^{6}\hat{A}_{new}(k-j) }{\sum_{j=0}^6 \hat{A}_{new}(k-4-j) } \; .
\]
This is very close to (\ref{eq rho-RKI}).  The main differences lie in the usage of different raw data bases (RKI versus JHU) and the adjustment of the raw data (stochastic redistribution $E(t)$ versus sliding 7-day averages $\hat{A}_{new,7}$). This may explain the differences in the level of low or high plateaus shown in fig. \ref{fig rho-RKI rho7} (with 1 day additional time shift).

In this sense, our model  supports the claim of the RKI that their reproduction numbers can be used as indicators of  ``a trend analysis of the epidemic curve'' \citep[p.1]{RKI:2020Erlaeuterung}. \\[2em]

\noindent
{\bf Acknowledgements:}
 We  thank   Odo Diekmann for discussing our thoughts as non-experts at an early stage of this work; he helped us to understand compartment models better. Moroever, we  appreciate the exchange with Stephan Luckhaus, and thank Robert Schaback, Robert Fe\ss{}ler,  Jan Mohring, and  Matthias Ehrhardt for hints and discussions.  
 Calculations and graphics were made with  {\sc Mathematica} 12.

 \vspace{5em}
\small
 \bibliographystyle{apsr}

\bibliography{/home/erhard/Dropbox/Datenbanken/BibTex_files/lit_mathsci,%
/home/erhard/Dropbox/Datenbanken/BibTex_files/lit_Covid,%
/home/erhard/Dropbox/Datenbanken/BibTex_files/lit_scholz}

\end{document}